\documentclass[sigconf]{acmart}

\usepackage{enumitem} 
\usepackage{multirow,siunitx}
\usepackage{soul}
\usepackage{graphicx}
\usepackage{microtype}
\usepackage{url}

\AtBeginDocument{%
  }

\newcommand{\university}{Boston University}
\newcommand{\shortUniversity}{BU}

\newcommand{\edit}[1]{\textcolor{black}{#1}}

\copyrightyear{2026}
\acmYear{2026}
\setcopyright{cc}
\setcctype{by}
\acmConference[CHI '26]{Proceedings of the 2026 CHI Conference on Human Factors in Computing Systems}{April 13--17, 2026}{Barcelona, Spain}
\acmBooktitle{Proceedings of the 2026 CHI Conference on Human Factors in Computing Systems (CHI '26), April 13--17, 2026, Barcelona, Spain}
\acmDOI{10.1145/3772318.3791507}
\acmISBN{979-8-4007-2278-3/2026/04}

\begin{document}

\title{Surveillance, Spacing, Screaming and Scabbing: How Digital Technology Facilitates Union Busting}

\author{Frederick Reiber}
\email{freddyr@bu.edu}
\affiliation{%
  \institution{Boston University}
  \city{Boston}
  \state{Massachusetts}
  \country{USA}
}

\author{Nathan Kim}
\email{ncyk@umich.edu}
\affiliation{
    \institution{University of Michigan}
    \city{Ann Arbor}
    \state{Michigan}
    \country{USA}
}

\author{Allison McDonald}
\email{amcdon@bu.edu}
\affiliation{%
  \institution{Boston University}
  \city{Boston}
  \state{Massachusetts}
  \country{USA}
}

\author{Dana Calacci}
\email{dcalacci@psu.edu}
\affiliation{
    \institution{Pennsylvania State University}
    \city{State College}
    \state{Pennsylvania}
    \country{USA}
}


\begin{abstract}
Despite high approval ratings for unions and growing worker interest in organizing, employees in the United States still face significant barriers to securing collective bargaining agreements. A key factor is employer counter-organizing: efforts to suppress unionization through rule changes, retaliation, and disruption. Designing sociotechnical tools and strategies to resist these tactics requires a deeper understanding of the role computing technologies play in counter-organizing against unionization. In this paper, we examine three high-profile organizing efforts---at Amazon, Starbucks, and \university---using publicly available sources to identify four recurring technological tactics: surveillance, spacing, screaming and scabbing. We analyze how these tactics operate across contexts, highlighting their digital dimensions and strategic deployment. We conclude with implications for organizing in digitally-mediated workplaces, directions for future research, and emergent forms of worker resistance.
\end{abstract}




\maketitle

\section{Introduction}
A 2022 Gallup poll found that 71\% of Americans approve of labor unions~\cite{inc_labor_2024}, and workers are filing union petitions at almost double the rate recorded four years ago~\cite{nlrb_office_of_public_affairs_union_2024}. Nearly half of U.S. workers say they would vote for unionization if given the chance~\cite{kochan_worker_2019}, yet private-sector union membership is strikingly low~\cite{bureau_of_labor_statistics_union_2024}.

One key explanation for this apparent paradox of high support for unions but low participation, as documented by labor scholars in the United States, is retaliation and union busting by employers~\cite{logan_union_2006}. Despite legal protections, union elections in the U.S. are far from fair and free. Firms routinely surveil, fragment, and intimidate workers, preventing them from exercising their legal right to organize~\cite{logan_union_2006}. Recent developments in workplace technology have exacerbated these issues, shifting workplace power dynamics heavily in favor of employers and opening new avenues for tech-facilitated union busting~\cite{logan_crushing_2021, sum_its_2024, kapoor_weaving_2022, greenhouse_old-school_2023, clawson_it_2017}. 

\begin{figure*}[t]
    \centering
    \includegraphics[scale=0.26]{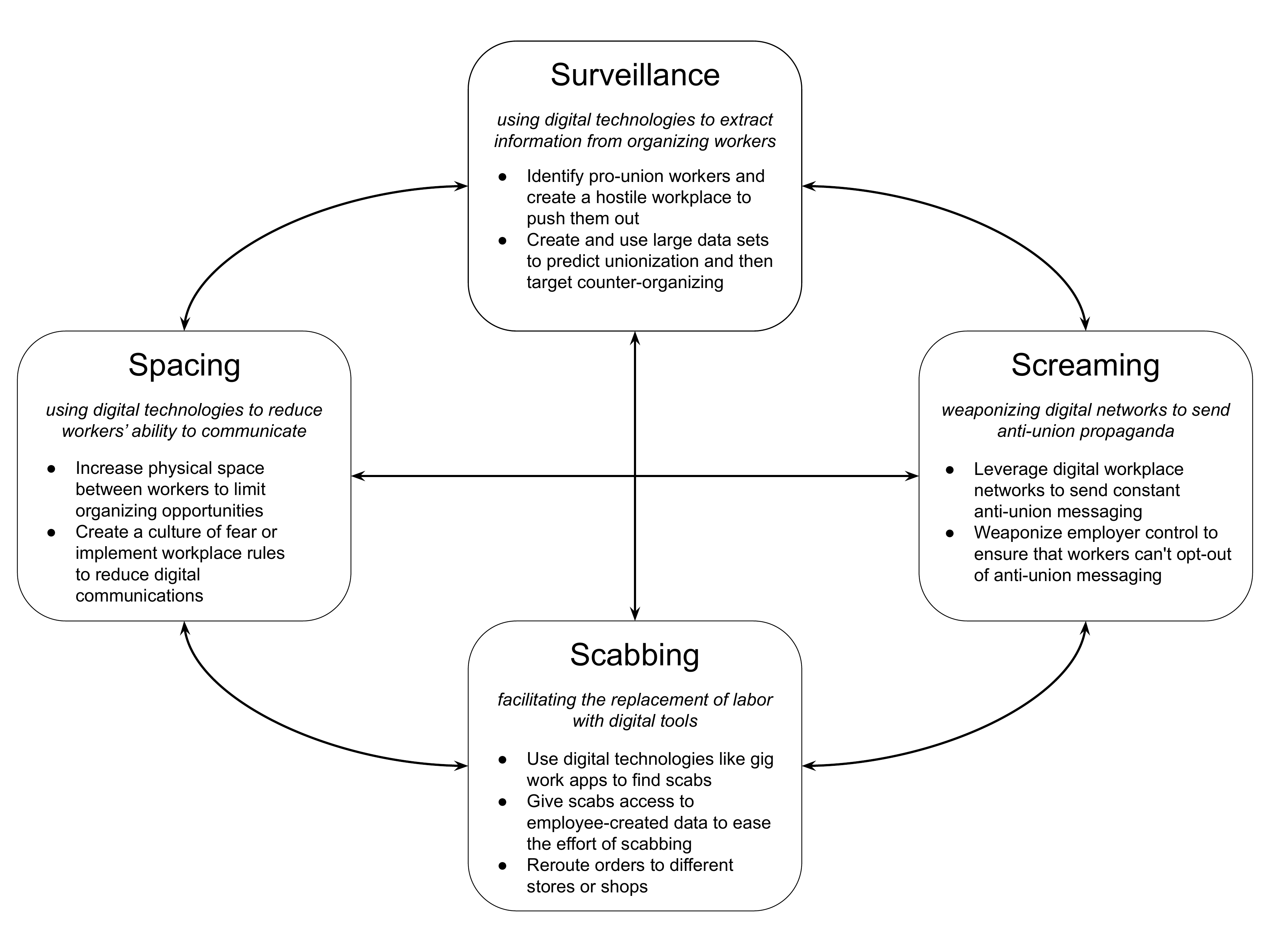}
\caption{\edit{A graphical representation and summary of our four tech-facilitated union busting tactics.}}
\Description{A graphical representation and summary of our four tech-facilitated union busting tactics.}
\label{fig:tacticsSummary}
\end{figure*}

In this paper, we examine how employers weaponize digital technologies to counter-organize---that is, to actively obstruct workers’ efforts to organize and unionize. Drawing on three recent American union campaigns at Amazon, Starbucks, and {\university} ({\shortUniversity}), we develop a set of four recurring tactics of tech-facilitated union busting:  \textit{Surveillance, Spacing, Screaming} and \textit{Scabbing}. A quick summary of our results is available in Figure \ref{fig:tacticsSummary}. Additionally, we find that these are not isolated tactics, but mutually reinforcing practices that form a broader regime of technologically-mediated counter-organizing and control~\cite{kellogg_algorithms_2020, mackenzie_marx_1984}. 

In developing an understanding of the technological tactics used in union counter-organizing, we make two main contributions to the human-computer interaction (HCI) community. First, we extend existing research on workplace technology to show how technologies in managerial contexts can become active tools of suppression in contemporary labor struggles. Second, by developing a generalized theory of how data and digital technologies facilitate union busting, we outline a design space for analyzing and contesting these tactics and for guiding future efforts to develop countermeasures and supportive infrastructure for labor organizers. Together, these contributions better position the HCI community to meet labor organizers where they are, to more fully understand the challenges of organizing under digital regimes, and to develop future technologies and practices that can strengthen collective action and protect the right to organize. 

The remainder of this paper proceeds as follows. First, we review related HCI and labor scholarship, with background on organizing and counter-organizing. We then describe how we selected our cases and developed our framework. Section~\ref{sec:tactics} presents our tactics and their interplay, providing definitions and discussions of how they are used for counter-organizing. Section~\ref{caseStruggles} details how each of the tactics was used across campaigns. To conclude, Section~\ref{sec:discussion} discusses what our results mean for design, working conditions, and the future of labor organizing.

\section{Literature Review and Background}

\subsection{HCI and Labor}

Human-computer interaction has a rich history of studying the role digital technologies play in labor and the workforce. During the initial rise of gig and platform work, scholars were quick to point out the potential for abusive systems~\cite{kittur_future_2013,irani_turkopticon_2013, irani_stories_2016, doorn_at_2020}. Recognizing the need to combat abusive systems, HCI researchers developed tools like Turkopticon~\cite{irani_turkopticon_2013} and Dynamo~\cite{salehi_we_2015} that helped correct information imbalances and provide workers with spaces for collective action. Other HCI scholars have sought to empower workers by building tools to analyze worker data~\cite{calacci_fairfare_2025, calacci_bargaining_2022}, to give workers more agency over scheduling~\cite{lee_participatory_2021}, and to promote worker well-being through alternative platforms created through participatory design~\cite{zhang_algorithmic_2022}. More recently, HCI researchers have studied the impacts of AI and automation on the workforce, seeking to understand how AI-enabled automation may present new challenges and recapitulate older forms of extraction and restructuring~\cite{lee_contrasting_2024, kang_stories_2022, lee_working_2015}. Both embedded in or explicitly discussed in these works is an understanding of power, that the workplace is often a place of conflict between employee and employer, with technology helping to mediate it~\cite{lee_contrasting_2024, anjali_anwar_watched_2021, greenbaum_back_1996, suchman_making_1995}. As Anwar, Pal, and Hui write, the ``restructuring of control through algorithmic systems in the gig economy produces information and power asymmetries that enable platforms to control workers while simultaneously obfuscating this control process''~\cite{anjali_anwar_watched_2021}.

The nature of this conflict is best illustrated by workplace surveillance issues. HCI studies in construction~\cite{kristiansen_accountability_2018}, education~\cite{lu_data_2021}, transportation~\cite{levy_contexts_2015, pritchard_how_2015}, and logistics~\cite{cheon_powerful_2023, cheon_amazon_2024,alimahomed-wilson_surveilling_2021} have sought to understand the role of digital tools in more traditional work, finding that increased surveillance often leads to unsafe working conditions and an increased workload on the worker~\cite{lu_data_2021,pritchard_how_2015,sannon_privacy_2022}. Workers also have little to no control over workplace surveillance issues, with workplace bottom lines taking precedence over workers interests, well-being, or safety~\cite{sum_its_2024, chowdhary_can_2023}. According to sociologist Christian Fuchs, ``workforce surveillance technologies are means of class struggle used by employers to try to strengthen capital's power against workers, lower wage costs, and increase absolute and relative surplus value production''~\cite{fuchs_political_2013}. The answer to combating these harms, historically, has been organized labor and labor unions~\cite{ajunwa_limitless_2017, bodie_law_2022, lee_contrasting_2024}. With their ability to leverage collective action, unions have historically been able to push back against significant workplace harms, serving as a counterbalance to the employer's control over the workplace. 

This shows the necessity of HCI research on pro-worker struc\-tures: labor unions, worker centers, and other worker advocacy organizations. We are not the first HCI researchers to make this claim---that we should target design research outside or against the logics of employment---nor to call for design work supporting organized labor. Spektor et al.~\cite{spektor_designing_2023} study the effects of an algorithmic room assignment system on hospitality workers. The work, done in partnership with UNITE HERE, focused on empowering union members in their workplace interactions with algorithmic systems. Thuppilikkat, Dhar, and Chandra~\cite{thuppilikkat_union_2024} studied the role of formal unions in platform worker struggles, highlighting the role applications like WhatsApp play in bridging the spatial and temporal gaps left by the rise of gig work. Kapoor et al.~\cite{kapoor_weaving_2022} investigate the privacy issues of tech labor organizers, finding new challenges in organizing under remote work, and Khovanskaya et al.~\cite{khovanskaya_tools_2019} draw on the historical tactics of unions to help provide tools for advocacy in platform-mediated work. Finally, Khovanskaya, Sengers, and Dombrowski~\cite{khovanskaya_bottom-up_2020} interviewed on-the-ground labor organizers, seeking to understand the challenges within organized labor data practices and ultimately calling for a shift towards more bottom-up approaches. All of these papers recognize the value of studying unions in HCI, noting both the power behind worker collective action and the need for understanding human-computer interaction within worker struggles~\cite{fox_worker-centered_2020}. 

However, understanding, designing, and ultimately supporting unionization efforts requires more than investigating and designing technology for organizers and workers. We extend this discourse by studying how employers' technology use creates barriers to unionization~\cite{clawson_it_2017, wolf_designing_2022}. By motivating our work explicitly in support of unionization, we respond to calls from other HCI researchers to incorporate political standpoints into design and HCI scholarship~\cite{ekbia_social_2016}. We also seek to situate this work as a bridge between labor-focused scholarship in HCI and direct accounts from organized labor, which constitute many of the sources we use in our case studies~\cite{guberek_keeping_2018, qiwei_sociotechnical_2024}. 

\subsection{Labor Organizing Against Technology}

Within the academic literature, there are also a number of works that seek to understand broad threats to labor unions from computational technologies, which fall into two main categories. The first focuses on how technology is being used to limit the institutional power of labor unions. In Michael Walker's work on AI, contracts, and unionization, he highlights the potential for AI to lay off unionized workers, the potential for task displacement, and the potential rise of more online forms of collective action~\cite{walker_impact_2024}. Charlotte Garden's work has highlighted the failures of labor law to respond to new technologies, studying how developments in workplace surveillance have gone beyond what labor law covers~\cite{garden_labor_2018} and arguing that labor law has failed to respond to the technology-driven fissuring of the workforce~\cite{garden_can_2024}. Others have argued that the rapid adoption of digital technologies and automation could diminish the need for unions, with the mass replacement of human labor necessitating a shift in their role~\cite{nissim_future_2021}. 

Scholars have also examined the role of platform and gig work in reducing organized labor. Tammy Katsabian argues that digital technologies are used to blur the lines of employment, with platform workers losing the legal scaffolding of the National Labor Relations Act (NLRA)~\cite{katsabian_rule_2021}. HCI scholars have pointed out data imbalances within gig and platform work, making it harder for organizers to develop an understanding of their workplace~\cite{calacci_organizing_2022}.
 
The second broad theme is that of employer control over the workplace, ensuring that workers do not question their role within the workplace or attempt to engage in collective action. The key theory for understanding this dynamic is labor process theory. Originally developed in the 1970's by sociologists and later expanded upon by HCI and communication researchers, labor process theory focuses on the everyday interactions within the labor process as a way of understanding the coercion and consent of workers~\cite{burawoy_manufacturing_1979, braverman_labor_1974, greenbaum_back_1996, edwards_contested_1979}. Building upon this, scholars in critical management studies have identified numerous ways algorithms and computational tools can be used to control the workforce, creating the six ``6 Rs'' framework---restricting, recommending, recording, rating, replacing, and rewarding~\cite{kellogg_algorithms_2020}. Other scholars have looked at the role of digital scheduling and discipline, early forms of what is now known as automated management, in preventing the building of worker resistance~\cite{ikeler_precaritys_2019, wood_despotism_2020}.

While these studies reveal how technology can be used by employers to control the workforce and how technology erodes unions' power in the workplace, they do not examine how employers weaponize technology against specific organizing efforts~\cite{bernhardt_data-driven_2023}. Some recent sociological scholarship has examined this phenomenon in single case studies, such as Teke Wiggin's work on algorithmic management during the Bessemer, Alabama, union drive~\cite{wiggin_weaponizing_2025}. We build on this work by extending our analysis across multiple case studies to identify \textit{recurring} tactics used by different employers to counter unionization, responding to the need for understanding how employers are using digital technologies to suppress worker organizing~\cite{bernhardt_data-driven_2023}.

\subsection{What it Takes to Organize}

\edit{In order to understand employer counter-organizing, we must first understand what it means to organize. We draw our organizing definition from Michael Grinthal, who defines organizing as ``the processes by which people build and exercise power by collecting and activating relationships''~\cite{grinthal_power_2011}. Accordingly, our analysis foregrounds the relational dynamics of union organizing, including relationships among workers and between workers and employers, rather than the legal infrastructure governing American unions. Counter-organizing, then, is the employer's attempts to stop the building of relationships and the exercising of their power. Our work analyzes the role of technology within counter-organizing. Shifting our focus in this way means that we are significantly less interested in the legal or financial aspects of unionization, which are often the focus of union busting literature~\cite{logan_consultants_2002, logan_corporate_2025, katsabian_rule_2021, garden_can_2024}.} 

\edit{Examples of how organizing hinges on relationship-building can be seen in many union organizing practices. For instance, workers seeking to unionize often engage in power mapping, a strategy in which organizers map out both the social and labor relationships of their workplace. Doing so requires numerous one-on-one conversations between a union organizer and their fellow workers~\cite{akhtar_out_2021, soriano-castillo_alabama_2012}. The importance of relationship-building is also visible in more direct forms of collective action. For example, when Los Angeles teachers organized a sit-down protest, they emphasized that resistance and change was only possible due to strong ties among workers and durable alliances with parents and community members~\cite{hendy_l_2011}.}

\section{Methods}

To understand the role technology plays in union busting, we adopt an exploratory case ``struggle'' (study) approach~\cite{yin_case_2009, ragin_what_1992}, using three cases of contemporary union busting---Amazon, Starbucks, and {\university}---to identify recurring patterns in the role of technology and to build a preliminary theoretical framework. To aid in replicability and validation, we provide citations to our corpus documents in the case studies section, give a full listing of all documents in our corpus within the appendix, and maintain an offline version of the corpus documents available upon request.

\subsection{Case Struggle Selection}

Our first criterion for case selection was that contemporary digital technology needed to have played a significant role in counter-organizing. This had been identified in each of our cases in recent reporting~\cite{wiggin_weaponizing_2025, vgontzas_new_2020, hanley_eyes_2021, canella_networked_2023, blanc_we_2025, press_how_2024}. Second, we wanted to pick cases that had direct pertinence to formal unionization. Because platform workers are often (mis)classified as independent contractors, they lack formal recognition and collective bargaining rights, producing dynamics beyond the scope of this study; for that reason, they were excluded from the potential cases. We also chose to focus on positions that are more service-driven, responding to current arguments that service jobs are the future of work and the critical battleground for worker organizing~\cite{dean_capitals_2025}. Our third criterion followed from other HCI research that has adopted a case study approach~\cite{shen_everyday_2021}, namely choosing cases that were (a) accessible to us via multiple data sources and (b) spanned dimensions previously identified as important when studying labor struggles, including precarity, union structure, industry, ideology and strategy~\cite{ikeler_precaritys_2019, parker_adapt_2022, hodder_essence_2015}. This led us to three cases: Amazon, Starbucks, and {\university}. We also presented our choice of cases to two active labor activists in our networks for feedback. Both supported our selection while providing critical instances for us to analyze.

\begin{table*}[t]
\setlength\extrarowheight{3pt}
\setlength\tabcolsep{0pt}
\begin{tabular*}{\textwidth}{ @{\extracolsep{\fill}} 
      l c!{\hspace{6pt}\vrule\hspace{6pt}}c c c c @{} }
\hline
\hline
\textbf{Case Struggle} & \textbf{Document Count} & \textbf{Surveillance} & \textbf{Spacing} & \textbf{Screaming} & \textbf{Scabbing} \\
\hline
\textbf{Amazon} & 58 & 16 & 12 & 10 & 9 \\
\textbf{Starbucks} & 54 & 10 & 5 & 8 & 5 \\
\textbf{\university} & 30 & 7 & 2 & 6 & 5\\
\hline
\hline
\end{tabular*}
\caption{Total counts and breakdowns of the documents reviewed along with the identified instances of technology facilitated union busting.}
\label{table:tactics counts}
\end{table*}

\subsection{Development of Cases}

To develop an understanding of these cases, we started with a small corpus of documents with which the first author was already familiar. We then expanded our corpus by following the citations in documents already included and through keyword searching. Keyword searches were conducted for each case study individually through web searches (Google and Google Scholar), popular publications that cover labor struggles (e.g., the \textit{Guardian}, \textit{Vox}, and \textit{Jacobin}), and labor publications (\textit{Labor Notes} and \textit{Working Mass}). Search terms included ``union busting,'' ``work technology,'' ``tech union busting'' and ``labor organizing'' in conjunction with keywords identifying our cases.

In general, we included any documents that added insights into the role technology played in counter-organizing or union busting. We  incorporated materials that provided background on working conditions, organizing efforts, and the broader context of our cases. We further expanded our review through National Labor Relations Board (NLRB) filings and documents, as well as articles recommended by academics or organizers knowledgeable about the cases. Much of our analysis centered on early organizing phases, from the initial public desire to unionize to attempts at securing a first collective bargaining agreement. However, we did not exclude documents that discussed behaviors outside of these phases if they highlighted a potential example of union busting by the employer and occurred recently enough to remain relevant (within 3 years of initial organizing). We also included two documents from before 2018, as they were cited by our sources and helped to provide context to the working conditions at specific employers. As we collected documents, we produced summaries and memos of each, drawing out the critical details relevant to our research questions. 

Completing this iterative search gave us a corpus of 142 relevant documents---a mix of primary and secondary sources, including firsthand accounts of the events, strategic reflections by organizers, academic analyzes of organizing drives, legal documentation from NLRB cases and \edit{a few reports of leaked Amazon documents detailing developments in their anti-union technologies.} 
In the Amazon and Starbucks cases, many of the documents focused on the same instances of counter-organizing; we excluded documents that only reiterated stories we had already seen multiple times. In the {\university} case, most of the issues were discussed in multiple corpus documents; however, our coverage of spacing instances was limited to only two documents: a worker reflection and an NLRB case file. Importantly, we did \textit{not} conduct an investigation into employer statements. While employer statements did come up---often in popular publications---we did not directly search for or include employer counter-arguments. Finally, the first author was also an \textit{active participant} in the organizing of one of our cases, {\university}. As such, some of the articles studied included reflections and reports that were either written by the first author or included quotes from the first author, a potential source of bias discussed in our positionality statement.

\subsection{Development of the Four Tactics}

After familiarizing ourselves with the cases, we set about fully answering our primary research question---what are the \textit{recurring} roles of technology in counter-organizing against unionization? We reflected on both the memos from developing our struggles and the firsthand knowledge gathered from the authors' multi-year experience in labor organizing. Doing so produced an early draft with five tactics---surveillance, spacing, screaming, stalling and scabbing---and a preliminary understanding of them. The first author then verified and refined the tactics by coding the corpus, highlighting instances of workplace technology that facilitated or assisted in union busting and categorizing them within the codebook. This left us with four consistent tactics of tech-facilitated union busting, prevalent across all three cases studied and described in the following sections. Table~\ref{table:tactics counts} summarizes the number of occurrences of each tactic through our review.

As we refined our set of tactics, we elected to remove stalling, which emerged meaningfully only in the Amazon case. Although we had initially defined this tactic as the use of digital or data technologies to slow or impede organizing, it did not appear consistently or clearly enough across our corpus to warrant inclusion. The primary example showed Amazon relying on an automated HR system to deny necessary worker accommodations---frustrating and harming workers---before later switching to a human-reviewed system in ways that disincentivized workers from unionizing~\cite{wiggin_weaponizing_2025}. We also noted an instance in which Amazon declined to participate in previously agreed-upon Zoom sessions, though this tactic’s role in \textit{counter-organizing} was minimal~\cite{logan_corporate_2025}. While stalling is indeed a common union busting tactic, it did not meaningfully intersect with workplace technologies or with counter-organizing practices in our review, and thus we chose to exclude it.

\subsection{Positionality and the Struggle of Studying Labor Struggles}

As labor organizers, researchers, and individuals with varying economic backgrounds, we are interpellated, like all political subjects, through the tactics of technology, capital, and labor. These positions have deeply influenced our thinking, both in deciding to conduct a project on tech-facilitated counter-organizing and during the process of researching. Several authors have been involved in labor activism, including organizing within their own union, organizing within local labor activist groups, and writing for a labor audience. As mentioned above, the first author was also directly involved with organizing at {\university}, and thus is subject to a potential insider or confirmation bias. Though these positions offered us firsthand experience of the tactics, we sought to overcome potential limitations in relying on personal experience by grounding our analysis in written coverage.

Additionally, as researchers within well-resourced research institutions, we are largely distanced from the precarity of contemporary work and union struggles. We also see technology not as a neutral tool but as a sociopolitical one, often driven by managerial prerogatives of control and surveillance. Our framework, heavily informed by critical HCI~\cite{ekbia_social_2016}, labor studies~\cite{vgontzas_toward_2023, logan_union_2006}, and STS~\cite{winner_artifacts_2007, green_contestation_2021, hornborg_globalised_2019}, is not neutral. Instead, this work is inherently partial, grounded in worker solidarity and framed around the tech-driven challenges workers face when attempting to organize. One way this is reflected in our work is that we have consciously chosen \textit{not} to consider employer statements, which often directly contradict narratives provided by workers. For example, Starbucks has continuously denied that the closing of stores had anything to do with unionization, despite both the NLRB finding and Starbucks Workers United claiming otherwise~\cite{wiessner_starbucks_2023, trull_addressing_2023}. We are thus highly critical of manager statements, narratives, and politics in reviewing our corpus. Importantly, this goes beyond simply disputing facts but also the narratives and politics around factual data. While this position may bias our results, we feel that this is a more accurate approach given the realities of union-drives~\cite{logan_consultants_2002, logan_union_2006, logan_crushing_2021, logan_corporate_2025}. 

\subsection{Limitations} \label{limitations}

First, our results do not describe all possible ways an employer may weaponize workplace technologies to union bust. Instead, we looked for recurring tactics, which help inform design and organizing more broadly. Second, we acknowledge that by reviewing documentation about union drives rather than conducting direct data collection, we may miss relevant incidents that were not documented publicly. While interviews could provide more information on how workers counteracted these tactics, our experience and the types of sources that were most useful in this study suggest that it would not have provided significantly more information for developing a model of tech-facilitated counter-organizing. Many of the critical data points were uncovered through legal proceedings or company leaks, as employers often have no requirement or incentive to disclose counter-organizing. Finally, all of our analyzes are situated in the United States. While there are ample examples of these tactics being used abroad---especially by global corporations---further work is needed to expand and validate these tactics in other labor contexts.

\section{The Four Tactics of Tech-Facilitated Union Busting}
\label{sec:tactics}

In this section, we detail the four recurring tactics of tech-facilitated union busting identified through our case analysis: \textbf{surveillance, spacing, screaming and scabbing}. For each of these, we provide definitions and brief examples of how these tactics are leveraged during counter-organizing. In-depth discussions of each case struggle are saved for the following section. Additionally, we situate each of these tactics within prior work, which informed our thinking in developing these tactics. Following this overview, we detail the interplay between each of these tactics, recognizing that while we have chosen to present these tactics as distinct, they operate in synthesis.

\subsection{The Four Tactics}
\subsubsection{Surveillance}

Perhaps the most well-studied tactic, \textbf{surveillance}, captures the ability of employers to track, monitor, and infer the behavior or personal traits of employees~\cite{ball_workplace_2010, ball_categorizing_2002, clawson_it_2017}. As Sum, Shi, and Fox write, ``technological advancements since the late twentieth century have led to the age of digital surveillance, making it easier for employers to constantly monitor their workers both inside and outside the workplace''~\cite{sum_its_2024}. Scholars have highlighted numerous problems with surveillance systems, including health and safety issues, an amplification of work expectations, and a decrease in the separation between work and non-work~\cite{clawson_it_2017, pritchard_how_2015,lu_data_2021}. While these direct harms to workers are themselves threats to organizing, we argue that surveillance also plays a distinct role in counter-organizing by disrupting the collective formation of workers.

One of the collective issues identified through our review is a data-imbalance~\cite{calacci_organizing_2022} in which employers are able to leverage workplace surveillance technologies to increase the accuracy and depth of information on their workforce. Employers can then use this to better plan and react to union organizing, saving resources and limiting the power of unions to create disruption through the employer's lack of information. Company leaks and reports also described employers developing systems to predict unionization. By combining both workplace and public datasets, employers likely have the ability to predict if a shop floor is at risk of unionization, even before deep organizing has begun. This significantly shifts the power towards employer counter-organizing, as it allows counter-organizing to begin before workers are organized enough to resist~\cite{logan_crushing_2021, logan_union_2006}. There is also the risk that employers will identify pro-union or organizing workers. Workers observed organizing will often face retaliation, such as being assigned increasingly difficult tasks in an attempt to prompt them to leave or provide grounds for termination~\cite{lisa_kresge_union_2020, kapoor_weaving_2022}. Employers will also surveil workers in digital spaces, with our review finding instances of employers sneaking into closed employee group chats to monitor for potential organizing and identify workers with workplace grievances.

\subsubsection{Spacing}

The second tactic observed through our review is \textbf{spacing}. Extending prior scholarship that identifies precarity and spatial dynamics as barriers to workplace resistance~\cite{ikeler_precaritys_2019, qadri_whats_2021, thuppilikkat_union_2024}, we define spacing as employers’ use of technology, or rules regulating workers' use of technology, to fracture collective formations of workers by constraining their opportunities to communicate. Through our review, we identified two primary ways this occurs. First, employers would leverage digital technologies like security cameras and management software to physically space out workers, making them unable to communicate and thus minimizing potential organizing opportunities. Workers and organizers reported being heavily isolated, sometimes confined to a single station with no contact from coworkers for the entire shift. Other times, employers weaponized their digital scheduling software to shift pro-union employees to different stores, diluting union support. Second, employers sought to limit digital communication through workplace rules and by creating a culture of fear. Examples of this include shutting down internal communications or managers telling workers not to post issues on social media. 

Importantly, our notion of spacing differs from the ``fissuring'' of the workforce identified in previous literature, wherein corporations increasingly outsource labor, liability, and accountability to third-party contractors~\cite{garden_can_2024,calacci_organizing_2022}. While fissuring and spacing similarly describe separations enforced within a workplace to curb worker power, fissuring can be seen as a management technique to restructure a workplace, removing the previous legal structures to help workers~\cite{newman_reengineering_2017}. Spacing, on the other hand, represents the direct or indirect action by an employer to curb worker's communicative power \textit{within or around} the workplace.

\subsubsection{Screaming}

Company propaganda during a union drive is nothing new~\cite{logan_consultants_2002, logan_union_2006}. Workers attempting to organize will often find themselves overwhelmed with anti-union propaganda, as employers attempt to create an atmosphere of fear, intimidation, and confusion within the workforce~\cite{logan_consultants_2002}. Examples include employer-written letters, consultant-prepared videos, and company newsletters, many of which urge workers to vote no, highlight the potential for job losses, and warn that managers may be harder on their workforce. Employers can also run captive audience meetings, where employees are forced to listen to the potential ``harms'' of unionization~\cite{masson_captive_2004}.

Digital workplace technologies heavily amplify the opportunities for employers to push anti-union messaging, essentially \textbf{screaming} at employees to avoid collective bargaining and unionization~\cite{marsh_digital_2022, baptista_digital_2020}. As observed in our review, employees attempting to unionize are met with a constant barrage of digital anti-union literature through their phones, work emails, and other digital systems of communication. Examples included workers receiving app notifications highlighting the risks of unionization or work emails from university leaders arguing that unionization was the incorrect model for academic workers. Scholars have highlighted the blurring of work and private life due to digital technologies~\cite{pauleen_making_2015}, with employees now receiving and responding to work messages outside of normal employment hours~\cite{harmer_attitude_2012}. Digital screaming weaponizes this breakdown of work and private life, with employers now able to push anti-union messaging to workers' devices outside of work, and in some cases, in the middle of the night. Unlike union communications, workers have little option to avoid employer screaming. These digital forms of anti-union propaganda are often packaged with necessary work communications, forcing employees to interact with them.

Our choice to term this tactic ``screaming'' is deliberate. In our experience, employer messaging is most effective when it seeks to overwhelm workers and drown out workers' own narratives. In this conception, messaging succeeds not through the ability to contend with worker messages but instead as a display of power that is communicated through its consistency and scope. While the multiple functions of corporate propaganda are important and deserve future study, we lean on the term ``screaming'' here to characterize the use of digital technology specifically as an amplifying tool for corporate messages in our case struggles. 

\begin{table*}[t]
  \centering 
  \begin{tabular}{>{\centering\arraybackslash}m{0.12\linewidth} | >{\raggedright\arraybackslash}m{0.41\linewidth} | >{\raggedright\arraybackslash}m{0.41\linewidth}}
    \hline
    \hline
    \textbf{Tactics (X to Y)} & \textbf{Relationship from X to Y} & \textbf{Relationship from Y to X} \\
    \hline
    \edit{Surveillance \& Spacing} & Digital surveillance systems are often used to enforce physical spacing. \textbf{Example:} Amazon leveraged its workplace surveillance systems to space organizers and workers. & \edit{A physically spaced out workforce may turn to digital channels, but doing so presents new opportunities for employer surveillance.} \textbf{Example}: Starbucks Workers United utilized social media \edit{to help organize across their spaced out workplace, allowing their employers to digitally surveil them.} \\
    \hline
    \edit{Surveillance \& Screaming} & Systems like algorithmic management provide new dimensions for employers to scream at their workforce. \textbf{Example:} Amazon's A to Z app, used for employee management, would send anti-union messaging to workers. & Digital screaming can help implement surveillance systems by having workers self-report or attest on working conditions. \textbf{Example:} Workers at {\shortUniversity} were screamed at through workplace emails about the attestation form system. \\ 
    \hline
    \edit{Surveillance \& Scabbing} & Surveillance systems make it easier for workers to scab, as the information needed to replace the worker is available digitally. \textbf{Example:} {\shortUniversity}'s learning management system gave them access to much of the course related data of striking workers. & Scabs hired through gig work apps are easier to control through the app's built-in surveillance systems. \textbf{Example:} Numerous academic works have argued gig work amplifies control, including~\cite{wood_good_2019, sannon_privacy_2022}. \\
    \hline
    \edit{Spacing \& Screaming} & \edit{A spaced out workforce has fewer opportunities for communication with pro-union sources, which may increase the power of employer screaming.} \textbf{Example:} {\university} would send anti-union emails to workers who were not receiving union emails. & Employers will send propaganda designed to scare workers, attempting to discourage talking to union organizers. \textbf{Example:} Amazon claimed that union organizers would share personal data with bad actors.  \\ 
    \hline
    \edit{Spacing \& Scabbing} & Workers are easier to replace when spaced out, as employers don't need to worry about social connections when hiring a scab. \textbf{Example:} Prior work on striking argues that relationships are critical for participation and to deter worker scabbing~\cite{akkerman_solidarity_2013}, but technological spacing lowers the potential for these relationships. & \edit{As gig work platforms are used to replace union labor, their inherently dispersed labor process can increase workplace spacing.} \textbf{Example:} \edit{Amazon utilized its gig work app, Amazon Flex, to replace striking workers likely without their awareness and reducing opportunities for workers to communicate.} \\ 
    \hline
    \edit{Screaming \& Scabbing} & Employers will leverage their digital networks to argue they are not hiring scabs or replacing labor. \textbf{Example:} Starbucks used its website to claim that they are not closing stores because of unionization, disputing NLRB claims. & The ease with which employers can replace labor or reorganize around strikes is often used in screaming to discourage workers from taking disruptive actions. \textbf{Example:} Amazon claimed that striking workers would have little impact due to their ability to reroute orders. \\
    \hline
    \hline
  \end{tabular}
  \caption{Summary of the interplay between our identified tactics with an example from our case review or the literature.}
  \label{table:interplay}
\end{table*}

\subsubsection{Scabbing}

The final tactic of our review on tech-facilitated union busting is \textbf{scabbing}. Drawing inspiration from colloquial uses of the term to describe workers who strike-break~\cite{akkerman_solidarity_2013, dimaggio_victory_2017}, we define \textbf{scabbing} as the use of computational technologies to facilitate the replacement of union labor. Rather than understanding technology as an inevitable harbinger of automation, independent of the political aims of management, technology as scabbing captures the contingent and multifaceted processes by which management replaces workers and curbs worker power. 

Within our review, we identified three ways that digital technologies have facilitated scabbing. First, employers leveraged digital tools to aid in finding scabs, using information and communication tools to hire replacements for pro-union or organizing workers. Second, employers used computational tools to reorganize their workforce, rerouting orders to different store or shop locations and thereby limiting the impact of striking workers. Third, employers used technologies to lower the workload of replacement workers. As we will see in our case struggles, scabs were able to leverage information and data created by striking workers to more easily replace them, thus lowering the impact of the strike. 

\subsection{Mutually-Reinforcing Tactics: Towards a Regime of Counter-Organizing}

Critical to our analysis is recognizing that these tactics should not be understood separately; instead, they constitute a broader apparatus of tech-facilitated union busting. Take, for example, spacing and its interplay with the other tactics. When workers are spaced out and unable to communicate, the effectiveness of employer screaming is potentially increased. Now, organizing workers are unable to inoculate other workers against company propaganda, essentially making it so that the employer is the only source of union-related information and thus positioning them as the sole source of credible information~\cite{logan_consultants_2002}. This then allows the employer to create a culture of fear around unionization, further amplifying employee spacing as workers become afraid to talk to organizers. Workplace digital surveillance allows the employer to better target  spacing, more easily identifying which workers are organizing and thus which workers to remove from the broader population. Additionally, digital surveillance is often the primary way of enforcing physical spacing, with workers' locations being actively tracked on the shop floor. Spacing also makes it easier to scab striking workers, as employers have an easier time replacing workers who don't rely on social connections at work, reducing potential backlash from scabbing.

This dynamic, in which individual tactics reinforce each other and thus their own power, is not unique to spacing. As we detail in Table~\ref{table:interplay}, each of our counter-organizing tactics interweaves with the other tactics, forming a broader system of tech-facilitated union busting that is incredibly difficult to overcome. Spacing, then, illustrates how each tactic gains strength from the others, operating not in isolation but as part of a tightly interwoven apparatus of digital counter-organizing. Recognizing this systemic character is crucial not only for our analysis but also for designing strategies of worker resistance.

\section{Case Struggles}\label{caseStruggles}

We now turn to our case struggle analysis. Here we more deeply analyze the role of digital workplace technologies in union busting. Across all three cases, employers were consistent in their use of surveillance, spacing, screaming, and scabbing, although each applied these tactics in different ways. A summary of our results is available in Table~\ref{tab:example}.

\begin{table*}
\centering
\begin{tabular}{>{\centering\arraybackslash}m{2cm}|
                >{\raggedright\arraybackslash}m{3.5cm}|
                >{\raggedright\arraybackslash}m{3.5cm}|
                >{\raggedright\arraybackslash}m{3.5cm}|
                >{\raggedright\arraybackslash}m{3.5cm}}
\hline
\hline
\textbf{Case Struggle} & \textbf{Surveillance} & \textbf{Spacing} & \textbf{Screaming} & \textbf{Scabbing} \\
\hline
\textbf{Amazon} & Workers are surveilled with security cameras and are monitored in private digital channels like Facebook group chats. & Organizers are physical separated through management software and cameras, and workers are made to be afraid of speaking to organizers through employer communications. & Workers are blasted with anti-union messaging through phone applications, emails, and at the workplace. & Amazon is easily able to re-route orders through it digital fulfillment system or hire scabs through its gig work platform. \\
\hline
\textbf{Starbucks} & Starbucks' Media Relations Department was monitoring social media and public discussions on union related posts. & Workers are told not to discuss organizing or workplace issues online and union supporters are moved around to different stores. & Workers are sent mass texts and memos about the harms of unionization often linked to an anti-union website. & Starbucks uses its digital hiring system to more easily find replacement workers. \\
\hline
\textbf{\university} & Workers not on strike were required to attest to working through an online digital survey. & BU failed to provide the union with necessary digital contact information including work emails. & Workers received constant emails from the employer arguing against unionization or putting the union in a bad light. & Learning management software designed to help organize courses aided in replacing striking instructors. \\
\hline
\hline
\end{tabular}
\caption{High level summary of how each tactic played out in all three case struggles.}
\label{tab:example}
\end{table*}

\subsection{Amazon}

Amazon is currently the second-largest employer in the United States, with an estimated 1.6 million workers in 2022. The company presents itself as embodying the ``future of work,'' advertising its expansive automation systems and arguing for their increased implementation~\cite{stefano_perago_embracing_2025}. Reality and workers, however, tell a different story. In this case, we primarily focus on two instances of organizing at Amazon, while also \edit{drawing on leaks from Amazon's internal anti-union infrastructure}. First, we focus on a union campaign in Bessemer, Alabama, in which fulfillment center workers attempted to join the Retail, Wholesale and Department Store Union. At the time of writing, workers have rejected unionization twice: once in 2021 and a second time in 2022, with a third court-mandated vote currently unscheduled~\cite{selyukh_amazon_2024}. Second, we review the organizing efforts at JFK8, a warehouse in Staten Island, NYC. Here, workers were able to win recognition from the National Labor Relations Board, but have gone without a first contract since their election on April 1, 2022. Since then, the workers have affiliated themselves with the International Brotherhood of Teamsters, which is attempting to organize workers within the delivery operation~\cite{reese_teamsters_2022, greenhouse_old-school_2023, logan_corporate_2025}. In general, these drives have relied on more traditional tactics like physical media and door-knocking, with Amazon both heavily counter-organizing and weaponizing the legal system~\cite{jaffe_itll_2021, logan_crushing_2021, logan_corporate_2025}. \edit{Finally, we note that for all of the leaks discussed, it is unclear which sectors of Amazon's business are being target as many of them come from Amazon's global intelligence projects~\cite{rey_leaked_2020}; however we are aware that these tools have been deployed against Amazon-owned Whole Foods which is discussed below.}

\subsubsection{Surveilling those who drive and predicting a union drive}
Part of the reason Amazon has evaded collective bargaining agreements is that Amazon doesn’t wait for organizing to gain traction. In 2020, \textit{Business Insider} leaked internal documents from Amazon, highlighting a data science tool used to \textbf{surveil} and predict which Whole Foods stores were at risk for unionization~\cite{peterson_amazon-owned_2020}. By combining both internal and external data, including employee turnover, calls to human resources, and violations recorded by the Occupational Safety and Health Administration, Amazon ranked stores in a ``heat-map'' system, warning the employer of which stores were at risk of unionization~\cite{peterson_amazon-owned_2020}. Further reports of the leaked memo detailed Amazon's desire for more funding and staffing to help optimize and combine different sources of surveillance data~\cite{rey_leaked_2020, kessler_companies_2020}. The ability to predict which stores are at risk of unionization makes organizing incredibly difficult for organizers, as Amazon can better target specific stores for union busting, allowing them to lower their financial costs. For example, Amazon may hire and fire workers at a faster rate, preventing organizers from building a base, or Amazon may provide short-term marginal relief over workplace issues, only to later roll them back after union risk is lower.

Our review also revealed numerous reports and leaks of \textbf{surveillance} practices, including job listings for anti-union ``intelligence analysts''~\cite{palmer_how_2020}, the surveillance of internal employee communications~\cite{jason_del_rey_amazon_2020}, and employers spying on private worker Facebook group chats~\cite{cox_inside_2020}. All of these digital surveillance practices enable Amazon to union bust in ways that current labor laws or organizing practices are unable to overcome. Workers who wish to leverage their legal right to unionize risk identification, even in channels thought to be private. Once identified, Amazon can leverage their almost unilateral control of the workplace to ensure that organizers never build momentum. Even if individual organizers do manage to avoid being targeted, the collective privacy issues around group prediction allow Amazon to union bust at the store level, making it no surprise that Amazon rarely faces a union election. 

\subsubsection{Screaming and spacing when workers try to organize}
Despite these challenges, workers have tried to organize. Reflections of the Bessemer, Alabama drive described it as a former steel town, highlighting its predominant Black population, decreasing public sector opportunities, and an eroded welfare system~\cite{jaffe_itll_2021, macgillis_lessons_2021}. Opened during the onset of the pandemic, Amazon's Bessemer facility was originally met with applause from city leaders, as it represented new sources of income and city revenue for the stagnating town~\cite{anonymous_what_2022}. The jobs, however, were less than ideal. Pro-worker outlets described the already isolating and exhausting conditions as becoming even more extreme under the heightened micromanagement imposed during the pandemic~\cite{peter_olney_bamazon_2020}. Rightly frustrated, employees reached out to the Retail, Wholesale and Department Store Union (RWDSU) for help in a union drive. After some initial card collecting in the summer of 2020, RWDSU sent organizers to the warehouse, heavily amplifying the ongoing union battle~\cite{logan_crushing_2021}.

One of Amazon's earliest tactics was to send a barrage of anti-union propaganda, \textbf{screaming} at their employees. Academic accounts detailed the numerous digital avenues used to scream at organizing workers. At the warehouse, Amazon placed what academic John Logan referred to as ``acid screens''---large digital monitors constantly warning workers about the dangers of signing a union card~\cite{logan_consultants_2002}. Workers found themselves bombarded with work emails, text messages, and app notifications containing anti-union messaging, often with a link to the anti-union website that Amazon had set up~\cite{logan_crushing_2021}. Many of these constant blasts occurred on the employees' personal devices, like phones, as Amazon's A to Z app---a required tool for work---would send push notifications to workers. As one worker put it, ``they were constantly sending us anti-union propaganda in the middle of the night, super early in the morning, while we were working''~\cite{wiggin_weaponizing_2025}. This is a canonical example of screaming, weaponizing the digital network of the workplace to heavily push anti-union messaging.

When looking at Amazon warehouses broadly, we found numerous examples of \textbf{spacing} in our corpus. Reports from \textit{Vox}, \textit{The Atlantic}, and \textit{The Verge} detailed an oppressive regime of spacing~\cite{evans_ruthless_2019, ghaffary_real_2020, dzieza_robots_2020}. During the pandemic, a relatively high instance of Amazon organizing, Amazon leveraged an artificial intelligence and camera system to enforce worker spacing during the labor process. However, as \textit{Vox} and the \textit{Open Markets Institute} describe, these spacing tools were really used to union bust~\cite{ghaffary_real_2020, hanley_eyes_2020}. As detailed in the Open Markets Institute report, ``the corporation uses its surveillance infrastructure to move around employees whom management suspects of collectively organizing. `They
spread the workers out,' said Mohamed, adding that `you cannot talk to your colleagues{'}''~\cite{hanley_eyes_2020}.

Important here, again, is the overlap between the four tactics; the physical \textbf{spacing} of the workforce was facilitated by \textbf{surveillance} technologies. This pattern---of worker spacing being shaped by our other tactics---also occurred with screaming. As anthropologist Orin Starn noted in his write up of Amazon's union busting, Amazon would \textbf{scream} to their employees that union organizers might share personal data with bad actors~\cite{orin_starn_inside_2024}. This is a combination of screaming and spacing, with the employer screaming at employees to be afraid of workplace organizers, thus manufacturing space between workers. 

In Bessemer, Amazon went so far as to lobby city officials to change the light timer outside the warehouse, preventing organizers from reaching workers on their way to work, suggesting that organizers had little---if any---ability to communicate with workers at the facility~\cite{brandom_amazon_2021}. Thus, Amazon weaponized its workplace digital technologies to enforce \textbf{spacing} between workers and to \textbf{scream} at its employees, ensuring that organizing was unsuccessful. For Bessemer, these challenges coalesced into a lost first election: 1798 votes against and 738 for, although the status of this election is still contested~\cite{selyukh_its_2021, palmer_amazon_2020, selyukh_amazon_2024}.

\subsubsection{Click here to hire a scab}
Amazon has also continued to contest its other recent election results. As stated above, Amazon currently has one unionized warehouse, JFK8. Amazon is refusing to bargain with them, instead attempting to challenge the constitutionality of the NLRB~\cite{gurley_amazon_2022, nerkar_amazon_2024}.

In 2024, in an attempt to get Amazon to recognize its employees' rights, workers at five delivery stations and three fulfillment centers---including JFK8---launched a holiday season strike, with reportedly thousands of workers participating. The goal was to slow down holiday packages and affect holiday deliveries~\cite{scheiber_amazon_2024, sam_gindin_was_2025}.

In our corpus review, we read claims from Amazon---published by popular outlets---that assured packages would not be delayed, highlighting their ability to shift orders from one hub to another~\cite{scheiber_amazon_2024, nerkar_amazon_2024}. We also reviewed an academic account of a similar instance, in which striking workers in Germany were unable to disrupt Amazon's business. As sociologist and labor scholar Nantina Vgontzas notes, Amazon's fulfillment operating system provides them with added flexibility when dealing with economic disruptions~\cite{vgontzas_new_2020}. As Amazon's orders are primarily digital, Amazon can simply pause orders at striking facilities or reroute orders to non-striking facilities, distributing the struck labor. The effect is that an individual shop floor's strength is weakened, effectively \textbf{scabbing} the struck work. This, in turn, weakens workers' economic strength as they now need to organize across fulfillment centers, further shifting the balance of power in favor of the employer.

\textit{The New York Times} also reported that Amazon hired \textbf{scabs} through its gig work app, Amazon Flex, which allows people to deliver packages in their own cars~\cite{scheiber_amazon_2024}, explicitly leveraging digital technology to union bust and scab. Amazon---likely---did not inform the gig workers that they were scabbing. While we were unable to find documentation of the calls for work, previous incidents of gig work apps being used to scab---such as the UNITE HERE Local 11 strikes---did \textit{not} include such information, with workers only becoming aware after accepting and arriving for the job~\cite{brown_hotel_2023}. Thus, the holiday strike offers a critical lesson: traditional tactics falter against the flexible, tech- and data-driven approach of Amazon. For any meaningful disruption to occur, workers may need to rethink how to organize against a company that can reroute both packages and people with the tap of a button.

\subsection{Starbucks}

Starbucks, unlike Amazon, has had a less traditional path to organizing. Described as a worker-to-worker model of organizing, workers at Starbucks relied heavily on digital spaces for organizing, subverting the physical spacing created by the Starbucks business model~\cite{blanc_we_2025}. Our main focus is the ongoing Starbucks Workers United (SBWU) campaign, which seeks to unionize workers across the United States. Originating in Buffalo and Rochester, New York, Starbucks Workers United came out of a group of salts---workers who get a job with the primary goal of organizing a union---backed by the Rochester Regional 
Joint Board of Workers United~\cite{gurley_starbucks_2021, jenny_brown_salts_2025}. After an initial filing and vote in late 2022, baristas at the Elmwood store in Buffalo, New York voted to unionize, becoming the first officially recognized union shop out of Starbucks's 9,000 company owned locations in the U.S.~\cite{scheiber_starbucks_2021, amelia_lucas_starbucks_2021}. Since the initial drive, baristas nationwide have voted to join SBWU, which now represents approximately 550 stores in at least 40 states. However, at the time of writing, they are without a first contract. In our work, we primarily focus on this ongoing struggle for both union recognition and to secure a first contract.

\subsubsection{Smothering a digital union drive}

Critical for understanding organizing at Starbucks is the campaign’s reliance on social media and the eruption of activity following the initial union vote. Reading through news coverage, we found a rapidly increasing number of unionized stores as the campaign progressed~\cite{gurley_starbucks_2021, gurley_starbucks_2022, gurley_starbucks_2021-2}. A significant factor in this rapid rise, as detailed in academic coverage, was the role of digital technologies. In his book chapter on Starbucks Workers United, labor relations scholar Eric Blanc describes how workers leveraged social media and digital communications to build a nationwide movement. For example, workers often posted clips of themselves striking, dealing with bad managers, or facing illegal firings on platforms like TikTok. On other occasions, workers would join other digital protest movements, like the online pro-Palestine movement, leading to company-wide boycotts. As one worker recalled, ``We knew from the beginning that to bring the company to the table we'd have to damage their image enough to make them scared they'd lose a whole generation of Gen Z customers, [...] all media was important, but the campaign's reach was widest online''~\cite{blanc_we_2025}. Other academic work reinforced this theme, highlighting how the use of Instagram and TikTok gave workers better control over their union's narrative, as the more direct means of communication allowed them to dispute the progressive image of Starbucks~\cite{canella_networked_2023}. However, this reliance on digital tools for organizing has also opened new opportunities for union busting and employee surveillance.

Unlike Amazon, our review did not find reports of Starbucks employees needing to deal with constant electronic surveillance within the labor process. In fact, we found the opposite, with a news article highlighting that the \textit{lack} of direct workplace surveillance was helpful~\cite{gurley_starbucks_2022-1}. Instead, workers faced \textbf{surveillance} through many of the digital tools used in organizing. Reviewed NLRB documents confirmed that workers were being surveilled through their social media, with the Starbucks media relations department monitoring and tracking the likes, shares, and comments on union-related posts~\cite{national_labor_relations_board_summary_2023}. Popular articles described how workers who appeared on television~\cite{blest_starbucks_2022-2} or signed public petitions~\cite{blest_starbucks_2022-4} were targeted for retaliation. Other reports revealed that managers pressured employees not to discuss workplace issues publicly, arguing that it was in the best interest of the partners---Starbucks's term for its employees---to keep such matters internal. Managers also monitored workers' Instagram stories, requiring employees to carefully consider what they posted online~\cite{lauren_kaori_gurley_its_2021, canella_networked_2023}. Additionally, these instances serve as examples of \textbf{spacing}, as they demonstrate how employers use both workplace pressure and digital surveillance to limit the ability of workers to organize and communicate through digital platforms.

Starbucks also leveraged their own internal digital network to \textbf{scream} at employees. Articles from a PhD student covering the drive highlighted how Starbucks had been sending mass texts to workers, delivering anti-union material and propaganda~\cite{faith_bennett_starbucks_2021}. Other coverage detailed memos sent to workers~\cite{lauren_kaori_gurley_its_2021} and employers' fabrication of tweets~\cite{gurley_starbucks_2022}. These screams would often urge workers to vote no on unionization and point them to their union busting website \textit{One.Starbucks}~\cite{blest_lateness_2022}. The website currently contains numerous anti-union talking points, highlighting how a union could ``prohibit your store manager from working directly with you on employment concerns'' or ``require you to pay union dues''~\cite{starbucks_we_2025}. As an interviewed worker for \textit{In These Times} put it, Starbucks had created a ``fear culture,'' with workers being afraid to discuss unionization efforts for fear of losing their jobs, and thus their income and health insurance~\cite{sainato_starbucks_2022, hamilton_nolan_buffalo_2021}.

\subsubsection{The digital storefront as digital control}
We also found Starbucks weaponizing shop floor digital technologies, allowing them to target specific stores in their counter-organizing. One of the most covered topics in the news was credit card tipping~\cite{danielle_wiener-bronner_starbucks_2022}. One of SBWU's main asks at the bargaining table was for employees to be able to receive tips through digital transactions. Before 2022, customers could tip only with cash or through Starbucks-managed platforms like the Starbucks app or Starbucks Card. In an attempt to dissuade workers from unionizing, Starbucks began rolling out support for credit card tips at non-unionized stores only, essentially trying to bribe workers not to unionize~\cite{alex_n_press_starbucks_2024, zakarin_exclusive_2023}. 

Another well-documented action in our corpus was the ``churn-and-burn'' of the Starbucks workforce. We reviewed numerous reports of Starbucks closing stores in response to union organizing, essentially burning their workforce. For example, Starbucks closed every store in Ithaca after it became the first U.S. city in which all locations were unionized~\cite{alex_n_press_starbucks_2022}. In Buffalo, two of the original stores that had attempted to unionize were closed, one for a remodel and the other to become a training store~\cite{gurley_starbucks_2021-2}. In total, the NLRB found that Starbucks had closed 23 stores to counteract union organizing~\cite{wiessner_starbucks_2023}. At other, more crucial stores, Starbucks would lean heavily into the churn. In Seattle, the birthplace of Starbucks, workers found themselves being shifted around three different shops after one store petitioned for an election. Here, the strategy, as argued by \textit{Jacobin}, was to mix organized and non-organized employees, reducing the likelihood of a successful election~\cite{john_logan_starbucks_2022, alex_n_press_starbucks_2024, saurav_sarkar_across_2022}. 

We observe that the Starbucks churn-and-burn was largely facilitated by digital technologies. Starbucks's hiring platform is digital~\cite{khaleda_rahman_gen-z_2022}, as is their scheduling platform~\cite{kantor_working_2014}, allowing them to more easily find new employees or weaponize scheduling to union bust~\cite{greenhouse_will_2023}. For us, these are prime examples of \textbf{scabbing} and \textbf{spacing}. Organizing workers who are fired by Starbucks are quickly replaced through a digital hiring system---or, in Seattle, shifted and spaced across stores through a digital scheduling system. Credit card tips, however, are a bit more complicated. The weaponization of credit card tips does not neatly fit within one of our individual components but rather represents a synthesis of our tactics. \textbf{Surveillance} logics around digital transactions shift control in favor of the employer, giving employers more tools to weaponize against union-organizing. The employer then does so through targeted reorganization of the labor process, which also functions as a \textbf{scream}, telling workers not to unionize.

\subsection{\university}

We now turn to our third case study, graduate worker organizing at {\university}. Over the past decade, graduate workers have emerged as a growing force in U.S. labor organizing and are now one of the biggest organizing waves the U.S. labor movement has seen in decades~\cite{valentina_luketa_how_2024}. While universities have historically engaged in less aggressive union busting than the employers in our first two cases, workers at {\university} still faced similar challenges, morphed by the blurred boundaries between labor and education~\cite{nicolette_manglos-weber_whats_2024}. Much like our other two cases, we focus on organizing after a public announcement of unionization up to the securing of a first contract. However, unlike the other two cases, {\shortUniversity} workers were not up against a multi-national corporation, but instead a private research university. This likely made it much easier for {\shortUniversity} workers to successfully organize an employer-wide strike. Workers at {\university} are also significantly more protected than workers in our other two case struggles, as the university---unlike Amazon or Starbucks---has a direct incentive to keep students in their programs and maintain their dual-roles.   

\subsubsection{A spaced workforce while screaming at faculty}

Organizing at a university presents unique spatial challenges. Universities are incredibly spread out as academic workers are scattered across dozens of buildings, floors, and offices. This dispersion makes digital communication tools---email, Slack, and similar platforms---essential for both work and organizing. Yet, these systems are often controlled by the university, raising concerns over employer \textbf{surveillance} and worker privacy. Workers at {\university} also recognized this, as we read reports of workers developing their own systems of communication, mostly through a worker-run Slack~\cite{freddy_reiber_opinion_2025}. These channels, however, were far from frictionless. Challenges in creating networks outside of the workplace lead to uneven participation and constrained communication, leaving certain departments difficult to reach and workers with little information about ongoing organizing or union activity~\cite{freddy_reiber_opinion_2025}. 

These conflicts were also amplified by the university \textbf{screaming} at the broader community. In our review, we found numerous examples of the university provost’s office sending out anti-union emails to the broader community~\cite{camille_bugayong_bugwu_2024, maya_mitchell_students_2024}. Examples of these included arguments that graduate work is not ``labor,'' the need to avoid bringing in a third party (a union), or highlighting lower turnout in specific union-related votes~\cite{jean_morrison_message_2022, ryan_quinn_boston_2024}. Workers attempted to organize against these emails with an in-person walkout, indicating they lacked effective \textit{digital} tools to counter workplace screaming. This was reinforced by {\shortUniversity}’s control over digital contact lists. As of 2014, the NLRB requires employers to provide unions with employee contact information so they are better able to reach and communicate with the employees they represent. However, NLRB documents highlighted how the University refused to provide any information, forcing the union to file an unfair labor practice for access~\cite{national_labor_relations_board_trustees_2023}. Refusing to share employee contact information limits organizers' ability to reach workers and further amplifies the effectiveness of employer messaging. As scholar John Logan notes in his article on corporate union busting, denying access or even providing outdated information is a common union busting tactic, and our review suggests that this strategy has extended into the digital realm, another example of \textbf{spacing}~\cite{logan_corporate_2025}.

\subsubsection{Surveilling yourself on strike}

After months of bargaining, graduate workers decided to walk off the job, recognizing the need to put economic pressure on the university to secure a contract~\cite{vanessa_bartlett_striking_2024, city_council_council_2024}. {\shortUniversity} retaliated immediately. A new tactic was the creation of an attestation form, which required non-striking workers to attest to working through a digital web-based survey in order to receive a paycheck. After protests over the university’s original plan---having advisors and administrators report which of their students were striking---the university quickly switched to an online digital survey that required workers themselves to attest to fulfilling job duties for the week~\cite{abigail_hassan_office_2024}. The almost instant implementation of these forms sent workers and organizers into confusion, as workers found the messaging and guidance around self reporting to be ``really inconsistent''~\cite{abigail_hassan_boston_2024, abigail_hassan_office_2024}.

Digitization of this process alters the power dynamics in two key ways. First, the use of these digital forms required workers to interface with the university directly, representing a form of digital \textbf{surveillance}. No longer was employer surveillance managed by a human who could potentially exert solidarity and ``forget'' to report. Instead, it was managed by a computational tool that directly serves the employer, allowing them to better identify workers who are striking and to more accurately withhold pay. The attestation forms also helped to provide the employer with significantly better data. As mentioned earlier, the union members reported failing to have accurate data on organizing~\cite{freddy_reiber_opinion_2025}, but with a mandatory attestation system, the university now had week-by-week, department-level intelligence on strike participation. One of the potential ways to disrupt production is by creating information disruptions, preventing employers from knowing who is striking or where a strike is likely to occur. These attestation forms likely gave management a strategic advantage in targeting counter-organizing efforts, allowing them to better predict the strike’s trajectory or timing for bargaining concessions. We also observed reports of these attestation forms serving as a form of \textbf{screaming}, with workers reporting emails about the self-reporting ``two to four times a day''~\cite{truman_dickerson_bu_2024} and one instance of the University publicly emailing out the names of people who did not complete the attestation form~\cite{adora_brown_bu_2024}.

\subsubsection{When software is scabbing}

Finally, we observed two instances in which software was used, or threatened to be used, to facilitate \textbf{scabbing}. The best-publicized case stemmed from a leaked email by university administrators, which suggested using ``generative AI [to] give feedback or facilitate `discussion' on readings or assignments''~\cite{tony_ho_tran_boston_2025}. This email prompted immediate community backlash, after which {\shortUniversity} denied any intention to use AI as a scab. While we found no evidence that the university actually deployed generative AI in this way, the incident highlights how generative AI may pose a threat to organized labor~\cite{samantha_genzer_bu_2024, lauren_coffey_boston_2024}. However, as researchers, we want to be cautious in making broad claims about the potential for AI-enabled scabbing specifically. As noted previously, universities have strong incentives to keep graduate workers enrolled and within the academic system, making employee terminations significantly less likely. In most other labor disputes, employers who \textit{could} replace striking workers with AI would probably do so well before a walkout.

A more concrete case involved the learning management system (LMS). According to reports from \textit{Boston.com}, some striking graduate instructors were locked out of their course sites---including the university’s LMS, Blackboard, and the system for submitting final grades~\cite{adora_brown_bu_2024}. We also reviewed accounts of graduate instructors removing their own materials from the LMS out of fear that struck work would be continued by a human replacement instructor in their courses~\cite{johanna_alonso_undergrads_2024, freddy_reiber_opinion_2025}. Although specific details remain unclear---likely because replacement instructors did not want to publicly reveal their process---we are confident that the university’s LMS was used to help facilitate \textbf{scabbing}. It is important to recognize the distinction between a digital course space, which is controlled by the university, and an analog course managed directly by an instructor. In a digital system, workers are much more easily replaced, since course materials, grades, and attendance records are all stored on a server, with access that can be quickly reassigned by a university administrator. In contrast with an analog system, most---if not all---of the organization and content can be withheld by the striking workers, as they are not compelled to respond to employer orders. This episode illustrates how employer-controlled technologies---even those ostensibly designed to support workers---can be leveraged against them. While LMS developers likely did not intend to undermine labor actions, design practices that consolidate institutional control can, when activated during a strike, weaken workers’ bargaining power and ability to advocate for fair conditions.

\section{Discussion}\label{sec:discussion}

Through our case struggle analysis, we have identified four consistent tactics that help characterize the ways in which technology is being leveraged to union bust against workers attempting to organize their workplaces. We now discuss the implications of these results in two main ways. To start, we discuss the implications for labor and labor organizing, contextualizing our results within contemporary debates around the labor process, privacy, and organizing. We then engage in a discussion of research and design, arguing for a deeper understanding of technologies for organizing, for designing around shop-floor conflicts, and for further work in this space.

\subsection{Labor Implications}

\subsubsection{\edit{Digital technologies more explicitly embed union busting within the labor process}}

\edit{Employers have a long history of developing production technologies in ways that counter worker autonomy and collective power. In \textit{Labor and Monopoly Capital}, Harry Braverman argues that industrial technologies are designed to maximize managerial control through deskilling, which he famously describes as the separation of conception from execution~\cite{braverman_labor_1974}. David Noble similarly shows how General Electric’s postwar investments in shop-floor automation were an explicit response to the union militancy of the 1940s, redistributing control over production in order to curb labor power~\cite{noble_forces_1986}. Joan Greenbaum extends this tradition by theorizing the rationalization of work under digital systems, demonstrating how managerial uses of computing technologies reorganize labor in ways that devalue skills and degrade working conditions~\cite{greenbaum_windows_2004}. Our findings point to the continuation of these dynamics---for example, Starbucks’ use of scheduling software to keep workers separated---but they also suggest an important shift. Rather than weakening unions indirectly through transformations of the labor process alone, digital workplace technologies increasingly function as more \textit{explicit} instruments of union busting embedded directly in workers’ everyday labor.}

\edit{Workers in our cases were required to navigate digital tools that were simultaneously essential for performing their jobs and functioning as direct tools of counter-organizing. For example, Amazon's workplace applications not only coordinated labor but also propagated anti-union messaging, directly communicating employer opposition to organizing efforts. In this respect, digital workplace technologies more closely resemble concrete anti-union interventions, such as captive audience meetings, rather than the more indirect forms of labor control emphasized in classic labor process accounts~\cite{logan_union_2006, logan_consultants_2002}. Noble’s analysis helps clarify this distinction. In his accounts, the implementation of automation aimed to redistribute power by reconfiguring production, transferring control over work from a unionized workforce to a non-unionized workforce~\cite{noble_forces_1986}. In our accounts, workers faced not only this redistribution---through the scabbing tactic---but also direct and explicit union busting with tools that disseminated propaganda or predicted unionization. Workers need not only contend with labor process technology at the macro level but also at the micro level, as employers are able to counter-organize in more direct, precise, and explicit ways.}

\subsubsection{Traditional workplaces are facing similar challenges to gig workers}

This deeper embedding of counter-organizing capacity within the labor process also presents theoretical implications for how scholars understand power in digitally-mediated workplaces. The shift suggests that workplace control increasingly operates through infrastructural and technological mechanisms rather than through physical displays of authority or force, much more similar to the way scholars understand the limited power of gig workers~\cite{doorn_at_2020, khovanskaya_tools_2019}. Similar to earlier analyses, workers in our cases lacked control over the data they produced. For example, graduate workers at {\university} were forced out of course spaces they had created. For unions---especially those rooted in ``bread-and-butter'' models of collective bargaining---this transformation necessitates a rethinking of strategy. Organizing cannot focus solely on wages and scheduling or even on traditional forms of employer opposition; it must also grapple with the technological systems that structure workers’ interactions, obscure managerial decision-making, and reshape possibilities for collective action. Doing so, however, is likely to be a challenge, as other work has shown that workers' direct employers are not always in charge of their labor conditions~\cite{katsabian_rule_2021}. To build durable workplace power, organizers must attend simultaneously to the conditions of work and to the digital infrastructures that increasingly mediate them, treating both as sites of contestation rather than neutral background features of employment.

\subsubsection{Workers need privacy at the workplace}

One of the other critical themes in our results is the need for workplace privacy. Unlike more historic forms of union busting, workers under digital regimes often face increasingly opaque monitoring~\cite{sum_its_2024, ajunwa_limitless_2017}, with the workers in our case struggles being surveilled in private or outside of work channels. Thus, workplace organizers need to be critically aware of their digital trail. Workers should assume that all digital communications on workplace platforms, including private messages on these platforms like Slack, are accessible to the employer; thus, workers should avoid organizing through them. Organizers should also attempt to increase the digital distance between themselves and their workplace. This includes not installing company software on personal devices when possible and avoiding the use of workplace-managed technology for personal or union functions. Additionally, we see opportunities for union staffers to help provide digital-organizing training, equipping workplace organizers with the necessary tools to understand potential tech harms and to better organize in workplaces augmented by technology. Many of these challenges could be solved by strong federal privacy regulation; however, as Sum, Shi, and Fox point out, large tech companies have significant lobbying power, and thus meaningful policy change remains a challenge despite this well-documented need~\cite{sum_its_2024}.

We also wish to stress the importance of collective privacy. Our results, especially those regarding surveillance, suggest that our definition of meaningful privacy needs to be extended beyond the individual. As Calacci and Stein argue, current data protection laws are likely to fail as they protect the privacy of the individual data subject, and thus allow for harms through the lack of collective privacy~\cite{calacci_access_2023}. Our results bolster this argument, as workers at Amazon faced collective prediction in addition to individual surveillance. These harms cannot be avoided through individualized security and privacy strategies.
To rectify this, we echo the call from a recent report from \textit{Data \& Society}, which argues for legal protections that ``eliminate the employer surveillance prerogative'' with a ban on the electronic monitoring of workers~\cite{minsu_longiaru_privacy_2025}. Although such a ban would not halt all counter-organizing tactics discussed here, we believe that it would meaningfully mitigate barriers to organizing, given that many of the other tactics are, as our results indicate, deeply intertwined with surveillance.

\subsubsection{Deep organizing is even more critical under technological regimes}

Finally, we also want to stress the importance of deep organizing, especially in tech-dominated workplaces~\cite{mcalevey_no_2016}. Employer screaming and spacing add new levels to the obfuscation of working conditions, surveillance makes it much easier to single out organizing workers, and scabbing reduces the power of potential strikes. Thus, we strongly suggest workers not only develop models of the social connections within their workplace but also of the technical systems. Our results indicate that these sociotechnical systems are deeply impacting union drives, and thus we recommend workers be pro-active in understanding potential tech harms or threats. One strong tool for this analysis is the \textit{workers' inquiry}. Originally developed by Marx in the 1880s, the \textit{workers' inquiry} is a set of questions that seeks to expand upon our understanding and points of leverage within the workplace. Recently expanded upon by scholars of digital economies, the digital workers' inquiry seeks to build knowledge about the workplace from the bottom-up~\cite{the_capacitor_collective_notes_2025}, leveraging the knowledge workers develop through their labor~\cite{woodcock_marx_2019, woodcock_workers_2014, woodcock_towards_2021}. While currently applied primarily by data, platform, and AI workers~\cite{weizenbaum-institut_data_2024, woodcock_towards_2021}, we see opportunities for more traditional workers to apply similar methods, especially as similar technology enters their shop floors.

\subsection{Design and Research Implications}

\subsubsection{\edit{Recognizing the co-option of technology within union busting}}

In doing this work, we have become deeply cynical about most contemporary discussions of technology design when they intersect with worker struggles. Across all three cases, employers readily weaponized available technologies to union bust, doing so even in the face of public pressure. This dynamic is, of course, not new. In their 2022 paper, ``Designing within Capitalism,'' Wolf, Asad, and Dombrowski found that none of the design projects aimed at helping workers address wage theft succeeded. Highlighting the constraints of designing within a hostile political economy, they conclude that there is a systemic barrier to pro-worker social-computing projects~\cite{wolf_designing_2022}. Although we have not conducted a parallel analysis, focusing instead on the harms of design, we advance a similar point: attempts to design interventions for organizing under technologically saturated regimes are likely doomed to fail.

Employers have little---if any---incentive to refrain from weaponizing workplace technologies to counter-organize, especially given the limited power of labor law in the face of new digital tools~\cite{garden_can_2024}. Our results contain numerous examples of ostensibly ``non-political'' technologies---digital hiring platforms, internal communication tools, scheduling systems---becoming sources of counter-organizing. These cases suggest that design interventions that do not explicitly engage with or shift employee power are readily co-opted for union busting. Much of this vulnerability stems from the political economy of workplace technology in the United States. Unlike other nations, U.S. corporate law has long been hostile to arrangements that constrain owners’ or managers’ exclusive discretion~\cite{jager_codetermination_2022}, rendering it nearly impossible to meaningfully correct the power imbalances embedded in work technologies. Other design-oriented approaches---such as digital workerism---while effective for advocacy and research, remain structurally outside workplace and shop floor conflicts, limiting their ability to address the harms we identify~\cite{calacci_organizing_2022, calacci_fairfare_2025}. Moreover, the tactics we identify---surveillance, spacing, screaming, and scabbing---do not directly impede the legal ability to unionize; rather, they strike at the core of unionization itself: the ability to \textit{organize}.

If workers are to meaningfully exercise their legal rights, we must expand our understanding of what it takes to unionize a workplace \edit{and seek to combat not only formal legal barriers but the sociotechnical conditions that erode collective capacity. From a design perspective, this requires treating co-option not as an untended consequence but as a predictable outcome of employer control. Rather than asking how technologies might ``support'' organizing in the abstract, designers must grapple with how tools are embedded within relations of control, ownership, and enforcement that overwhelmingly favor employers. This suggests a shift away from artifact-centered interventions and towards strategies that explicitly align with worker power even when such approaches sit uneasily with dominant paradigms of technological design. Fully realizing this means \textit{not} engaging with design approaches premised on neutral or comprehensive stakeholder analysis; we give an example in Section~\ref{sabotage}. Absent such commitments, efforts to design for organizing risk reproducing the very asymmetries they seek to contest, offering new surfaces for managerial control rather than durable resources for collective struggle.}

\subsubsection{Designing for sabotage}
\label{sabotage}

Following Khovanskaya et al., we also turn to historical union tactics~\cite{khovanskaya_tools_2019} as a means of grounding pro-worker struggle design. In her 1917 article, ``Sabotage: The Conscious Withdrawal of the Workers' Industrial Efficiency,'' American labor leader and feminist Elizabeth Gurley Flynn argues for sabotage. She writes, ``Sabotage means either to slacken up and interfere with the quantity, or to botch in your skill and interfere with the quality of capitalist production, or to give poor service. Sabotage is not physical violence, sabotage is an internal, industrial process''~\cite{elizabeth_gurley_flynn_sabotage_1917}. As design researchers and organizers, we see significant value in deepening our understanding of resistance and strongly advocate embracing Flynn's notions of sabotage.

Workers have already started to engage in this work. Through our review, we identified grass-roots developers who have created a fake job application tool to help combat Starbucks's hiring of scabs. As designers and activists, we see opportunities to aid in this work, helping to provide technological scaffolding for workers to better engage in sabotage. HCI researchers already have a deep history of designing against harmful technologies in gig work, recognizing data imbalances and pushing back against them~\cite{irani_turkopticon_2013, calacci_fairfare_2025}. We advocate for expanding this work, pushing beyond correcting data imbalances and moving to more direct forms of resistance facilitated through technologies. Examples of this could include apps that inform gig workers if they are being hired as a scab \textit{before} accepting the job, systems that help coordinate workplace slowdowns and boycott overtime, or data science tools to aid in identifying economic weak points for striking. 

One in-depth example of this is leveraging technology to organize a work-to-rule ``strike.'' As described by \textit{Labor Notes}' editorial board, a work-to-rule attack involves following the rules of the company handbook or union contract to the letter, skipping all the daily shortcuts and extras on which company production is dependent~\cite{labor_notes_editorial_board_ways_2019}. This includes things like delaying truck routes for a required 20-minute safety check, ensuring that \textit{all} necessary equipment is available, completing paperwork in detail, or calling managers about anything slightly tricky. Doing so while the threat of a strike is active allows workers to keep their paychecks, slow down production, and ensure that employers keep their expensive strike contingency plan in place. The challenge with these ``strikes'' is that employers will often label this as a partial strike or a slowdown. Workers organizing a work-to-rule campaign must be incredibly careful not to provide their employer with evidence of union coordination. We see opportunities for security and privacy technology to help facilitate this kind of slowdown. Anonymous messaging applications that don't identify users or organizers could be used to coordinate or disseminate information while ensuring the employer is unable to connect it to the union. Workers could also develop digital literature highlighting how to work to the rule, providing workers with easy checklists detailing the necessary safety regulations, and---if done securely---avoiding employers becoming aware of a coordinated work slowdown.     

\subsubsection{Uncovering the invisible work of organizing}

Finally, we see research opportunities in furthering this work. Digital work technologies can make labor, laborers, and the conditions of labor invisible~\cite{irani_turkopticon_2013, gruszka_out_2022, ming_i_2023, ming_labor_2024, star_layers_1999}. Counter-organizing during labor is no different, with many of the strategies identified being shrouded in tech-obfuscation. As researchers with a strong technical background, it is imperative that we continue to find and document these challenges. Finding cases with enough documentation to ground our analysis was more challenging than expected, as labor organizers and scholars tend to focus on organizing strategies rather than the barriers to it. Researchers familiar with both computational technology design and qualitative methods are thus uniquely positioned to help build out this work. We see opportunities for HCI researchers to utilize their skills to engage in research from below, helping to deepen collective understanding of tech organizing harms and empower workers. For example, interviews could be conducted to understand how workers overcame and resisted the tactics identified within our results. The development of this work could also lead to data leverage for the workplace, helping to identify points of contestation over computational technologies~\cite{vincent_data_2021}.

\section{Conclusion}

Workers want unions---the numbers are clear---yet many face significant challenges in their ability to organize or fight back. One reason is workplace technology, which is increasingly weaponized to facilitate counter-organizing. Through our review of three case struggles, we identified four recurring tactics: surveillance, spacing, screaming and scabbing, often interwoven in practice. In doing so, we have offered one of the first studies, to our knowledge, on the role technology plays in undermining unionization. This work not only helps inform strategies for organizing and designing against oppressive technologies, but also aims to spark new discussions on how best to support worker resistance and dignity in the face of evolving technological barriers.

\begin{acks}
We are deeply grateful for the labor organizers, activists, and writers we have spoken to who have helped in the preparation of this work, including Adrienne Williams, Cella Sum, Jeff Uehlinger, Lucia Vilallonga, Jacksyn Baekberg, and Rachel McCleery. We are also indebted to all of our fantastic reviewers who have provided stellar feedback at each step of the process, the Political Economy and Algorithms Collective, and to all of the workers who continue to organize, unionize, and fight.

\end{acks}

\bibliographystyle{ACM-Reference-Format}
\bibliography{references,laborCorpus}

@misc{orin_starn_inside_2024,
	title = {Inside {Amazon}’s {Union}-{Busting} {Tactics}},
	url = {https://www.sapiens.org/culture/amazon-union-busting-anthropologist/},
	language = {en-US},
	urldate = {2025-09-03},
	journal = {Sapiens},
	author = {{Orin Starn}},
	month = aug,
	year = {2024},
}

@misc{gurley_amazon_2022,
	title = {Amazon {Cracks} {Down} on {Organizing} {After} {Historic} {Union} {Win}},
	url = {https://www.vice.com/en/article/amazon-cracks-down-on-organizing-after-historic-union-win/},
	language = {en-US},
	urldate = {2025-09-03},
	journal = {VICE},
	author = {Gurley, Lauren Kaori},
	month = apr,
	year = {2022},
}

@misc{peter_olney_bamazon_2020,
	title = {{BAmazon} {Union}: {Anticipating} the {Battle} in {Bessemer}, {Alabama}},
	shorttitle = {{BAmazon} {Union}},
	url = {https://labornotes.org/2020/12/bamazon-union-anticipating-battle-bessemer-alabama},
	language = {en},
	urldate = {2025-09-03},
	journal = {Labor Notes},
	author = {{Peter Olney} and {Rand Wilson}},
	month = dec,
	year = {2020},
}

@misc{rey_leaked_2020,
	title = {Leaked: {Confidential} {Amazon} memo reveals new software to track unions},
	url = {https://www.vox.com/recode/2020/10/6/21502639/amazon-union-busting-tracking-memo-spoc},
	language = {en-US},
	urldate = {2025-09-03},
	journal = {Vox},
	author = {Rey, Jason Del and {Shirin Ghaffary}},
	month = oct,
	year = {2020},
}

@techreport{hanley_eyes_2020,
	address = {Rochester, NY},
	type = {{SSRN} {Scholarly} {Paper}},
	title = {Eyes {Everywhere}: {Amazon}'s {Surveillance} {Infrastructure} and {Revitalizing} {Worker} {Power}},
	shorttitle = {Eyes {Everywhere}},
	url = {https://papers.ssrn.com/abstract=4089862},
	language = {en},
	number = {4089862},
	urldate = {2025-09-03},
	institution = {Open Markets Institue},
	author = {Hanley, Daniel and Hubbard, Sally},
	month = sep,
	year = {2020},
}

@misc{evans_ruthless_2019,
	title = {Ruthless {Quotas} at {Amazon} {Are} {Maiming} {Employees}},
	url = {https://www.theatlantic.com/technology/archive/2019/11/amazon-warehouse-reports-show-worker-injuries/602530/},
	language = {en},
	urldate = {2025-09-03},
	journal = {The Atlantic},
	author = {Evans, Will},
	month = nov,
	year = {2019},
	note = {Section: Technology},
}

@misc{alex_n_press_starbucks_2024,
	title = {The {Starbucks} {Workers}’ {Union} {Has} {Finally} {Broken} {Through}},
	url = {https://jacobin.com/2024/02/starbucks-workers-united-master-contract-bargaining},
	language = {en-US},
	urldate = {2025-09-03},
	journal = {Jacobin},
	author = {{Alex N. Press}},
	month = feb,
	year = {2024},
}

@misc{truman_dickerson_bu_2024,
	title = {{BU} graduate workers reflect on living conditions, union demands as strike carries on},
	url = {https://dailyfreepress.com/04/10/22/203734/bu-graduate-workers-reflect-on-living-conditions-union-demands-as-strike-carries-on/},
	language = {en, sv},
	urldate = {2025-09-03},
	journal = {The Daily Free Press},
	author = {{Truman Dickerson}},
	month = apr,
	year = {2024},
	note = {Section: Community},
}

@misc{maya_mitchell_students_2024,
	title = {Students, administration uncertain about {BUGWU} strike as semester ends},
	url = {https://dailyfreepress.com/04/29/02/204322/students-administration-uncertain-about-bugwu-strike-as-semester-ends/},
	language = {en, sv},
	urldate = {2025-09-03},
	journal = {The Daily Free Press},
	author = {{Maya Mitchell} and {Abigail Hassan}},
	month = apr,
	year = {2024},
	note = {Section: Campus},
}

@misc{national_labor_relations_board_trustees_2023,
	title = {Trustees of {Boston} {University} {\textbar} {National} {Labor} {Relations} {Board}},
	url = {https://www.nlrb.gov/case/01-CA-315655},
	urldate = {2025-09-03},
	journal = {NLRB},
	author = {{National Labor Relations Board}},
	month = apr,
	year = {2023},
}

@misc{abigail_hassan_office_2024,
	title = {Office of {Provost} changes payroll policy amid {BUGWU} strike},
	url = {https://dailyfreepress.com/04/04/21/203500/office-of-provost-changes-payroll-policy-amid-bugwu-strike/},
	language = {en, sv},
	urldate = {2025-09-03},
	journal = {The Daily Free Press},
	author = {{Abigail Hassan} and {Maya Mitchell}},
	month = apr,
	year = {2024},
	note = {Section: Campus},
}

@misc{camille_bugayong_bugwu_2024,
	title = {{BUGWU} calls undergraduates, faculty, staff to join walkout in response to {Provost} emails},
	url = {https://dailyfreepress.com/05/02/14/204393/bugwu-calls-undergraduates-faculty-staff-to-join-walkout-in-response-to-provost-emails/},
	language = {en, sv},
	urldate = {2025-09-03},
	journal = {The Daily Free Press},
	author = {{Camille Bugayong}},
	month = may,
	year = {2024},
	note = {Section: Campus},
}

@misc{samantha_genzer_bu_2024,
	title = {{BU} community reacts to {CAS} dean suggestion to replace striking graduate teaching assistants with {AI} tools},
	url = {https://dailyfreepress.com/04/05/20/203574/bu-community-reacts-to-cas-dean-suggestion-to-replace-striking-graduate-teaching-assistants-with-ai-tools/},
	language = {en, sv},
	urldate = {2025-09-03},
	journal = {The Daily Free Press},
	author = {{Samantha Genzer}},
	month = apr,
	year = {2024},
	note = {Section: Campus},
}

@misc{adora_brown_bu_2024,
	title = {{BU} faces two unfair labor charges amidst graduate student strike},
	url = {https://www.boston.com/news/local-news/2024/05/08/bu-faces-two-unfair-labor-charges-amidst-graduate-student-strike/},
	language = {en-US},
	urldate = {2025-09-03},
	journal = {Boston.com},
	author = {{Adora Brown}},
	month = may,
	year = {2024},
}

@misc{vanessa_bartlett_striking_2024,
	title = {Striking {BU} {Graduate} {Workers} {Demand} {More} for {Parents} and {Families}},
	url = {https://working-mass.com/2024/04/04/striking-bu-graduate-workers-demand-more-for-parents-and-families/},
	language = {en-US},
	urldate = {2025-09-03},
	journal = {Working Mass},
	author = {{Vanessa Bartlett}},
	month = apr,
	year = {2024},
}

@misc{jean_morrison_message_2022,
	title = {A {Message} to {BU} {Graduate} {Students}: {Our} {Commitment} to {Progress} {Together} {\textbar} {Office} of the {Provost}},
	url = {https://www.bu.edu/provost/2022/09/23/a-message-to-bu-graduate-students-our-commitment-to-progress-together/},
	urldate = {2025-09-03},
	journal = {BU Office of the Provost},
	author = {{Jean Morrison}},
	month = sep,
	year = {2022},
}

@misc{tony_ho_tran_boston_2025,
	title = {Boston {University} {Suggests} {Replacing} {Striking} {Grad} {Students} {With} {AI}},
	url = {https://www.thedailybeast.com/boston-university-suggests-replacing-striking-grad-students-with-ai/},
	urldate = {2025-09-03},
	journal = {Daily Beast},
	author = {{Tony Ho Tran}},
	month = jan,
	year = {2025},
}

@misc{city_council_council_2024,
	title = {Council {Supports} {Boston} {University} {Graduate} {Workers} {\textbar} {Boston}.gov},
	url = {https://www.boston.gov/news/council-supports-boston-university-graduate-workers},
	language = {en},
	urldate = {2025-09-03},
	journal = {City of Boston},
	author = {{City Council}},
	month = mar,
	year = {2024},
}

@misc{freddy_reiber_opinion_2025,
	title = {Opinion – {Reflections} {From} an {Organizer} in the {Longest} {Grad} {Student} {Strike} in ({Recent}) {History}},
	url = {https://working-mass.com/2025/01/24/opinion-reflections-from-an-organizer-in-the-longest-grad-student-strike-in-recent-history/},
	language = {en-US},
	urldate = {2025-09-03},
	journal = {Working Mass},
	author = {{Freddy Reiber}},
	month = jan,
	year = {2025},
}

@misc{johanna_alonso_undergrads_2024,
	title = {Undergrads {Suffer} the {Impact} of {BU}’s {Graduate} {Student} {Strike}},
	url = {https://www.insidehighered.com/news/students/academics/2024/05/13/boston-u-strike-plagues-undergraduate-writing-students},
	language = {en},
	urldate = {2025-09-03},
	journal = {Inside Higher Ed},
	author = {{Johanna Alonso}},
	month = may,
	year = {2024},
}

@misc{lauren_coffey_boston_2024,
	title = {Boston {University} {Denies} {It} {Would} {Use} {AI} to {Replace} {Striking} {Teaching} {Assistants}},
	url = {https://www.insidehighered.com/news/quick-takes/2024/04/01/boston-university-denies-it-would-replace-striking-tas-ai},
	language = {en},
	urldate = {2025-09-03},
	journal = {Inside Higher Ed},
	author = {{Lauren Coffey}},
	month = apr,
	year = {2024},
}

@misc{ryan_quinn_boston_2024,
	title = {Boston {U} {Grad} {Worker} {Strike} {Now} {Longest} in a {Decade}},
	url = {https://www.insidehighered.com/news/faculty-issues/labor-unionization/2024/08/23/boston-u-grad-worker-strike-now-longest-decade},
	language = {en},
	urldate = {2025-09-03},
	journal = {Inside Higher Ed},
	author = {{Ryan Quinn}},
	month = aug,
	year = {2024},
}

@misc{nicolette_manglos-weber_whats_2024,
	title = {What’s behind the grad student strike at {Boston} {University}?},
	url = {https://www.christiancentury.org/features/what-s-behind-grad-student-strike-boston-university},
	language = {en},
	urldate = {2025-09-03},
	journal = {The Christian Century},
	author = {{Nicolette Manglos-Weber}},
	month = jul,
	year = {2024},
}

@misc{kantor_working_2014,
	title = {Working {Anything} but 9 to 5},
	url = {https://www.nytimes.com/interactive/2014/08/13/us/starbucks-workers-scheduling-hours.html, https://www.nytimes.com/interactive/2014/08/13/us/starbucks-workers-scheduling-hours.html},
	language = {en-US},
	urldate = {2025-09-03},
	journal = {The New York Times},
	author = {Kantor, Jodi},
	month = aug,
	year = {2014},
}

@misc{jenny_brown_salts_2025,
	title = {Salts and {Peppers} {Build} a {Union} at {Starbucks}},
	url = {https://www.labornotes.org/blogs/2025/08/salts-and-peppers-build-union-starbucks},
	language = {en},
	urldate = {2025-09-03},
	journal = {Labor Notes},
	author = {{Jenny Brown}},
	month = aug,
	year = {2025},
}

@misc{danielle_wiener-bronner_starbucks_2022,
	title = {Starbucks union organizers wanted credit-card tipping. {Now} they’re being left out {\textbar} {CNN} {Business}},
	url = {https://www.cnn.com/2022/12/16/business/starbucks-tipping},
	language = {en},
	urldate = {2025-09-03},
	journal = {CNN},
	author = {{Danielle Wiener-Bronner}},
	month = dec,
	year = {2022},
}

@misc{gurley_starbucks_2022,
	title = {The {Starbucks} {Unionization} {Effort} {Is} {Now} {An} '{Unprecedented}' {Nationwide} {Movement}},
	url = {https://www.vice.com/en/article/the-starbucks-unionization-effort-is-now-an-unprecedented-nationwide-movement/},
	language = {en-US},
	urldate = {2025-09-03},
	journal = {VICE},
	author = {Gurley, Lauren Kaori},
	month = jan,
	year = {2022},
}

@misc{amelia_lucas_starbucks_2021,
	title = {Starbucks will have at least one unionized cafe in {Buffalo}, {New} {York} — a {U}.{S}. first for the chain},
	url = {https://www.cnbc.com/2021/12/09/starbucks-employees-at-a-buffalo-cafe-vote-to-unionize-a-first-for-the-coffee-chain-in-the-us.html},
	language = {en},
	urldate = {2025-09-03},
	journal = {CNBC},
	author = {{Amelia Lucas} and {Kate Rogers}},
	month = dec,
	year = {2021},
	note = {Section: Restaurants},
}

@misc{scheiber_starbucks_2021,
	title = {Starbucks {Faces} {Rare} {Union} {Challenge} as {Buffalo} {Workers} {Seek} {Vote}},
	url = {https://www.nytimes.com/2021/08/30/business/starbucks-coffee-buffalo-union.html},
	language = {en-US},
	urldate = {2025-09-03},
	journal = {The New York Times},
	author = {Scheiber, Noam},
	month = aug,
	year = {2021},
}

@misc{national_labor_relations_board_summary_2023,
	title = {Summary of {NLRB} {Decisions} for {Week} of {February} 13 - 17, 2023 {\textbar} {National} {Labor} {Relations} {Board}},
	url = {https://www.nlrb.gov/cases-decisions/weekly-summaries-decisions/summary-of-nlrb-decisions-for-week-of-february-13-17-0},
	urldate = {2025-09-03},
	journal = {NLRB},
	author = {{National Labor Relations Board}},
	month = feb,
	year = {2023},
}

@misc{khaleda_rahman_gen-z_2022,
	title = {Gen-{Z} {Activists} {Flood} {Starbucks} {With} {Fake} {Job} {Applications} {Over} {Firings}},
	url = {https://www.newsweek.com/gen-z-activists-flood-starbucks-fake-job-applications-over-firings-1681786},
	language = {en},
	urldate = {2025-09-03},
	journal = {Newsweek},
	author = {{Khaleda Rahman}},
	month = feb,
	year = {2022},
	note = {Section: U.S.},
}

@misc{alex_n_press_starbucks_2022,
	title = {Starbucks {Workers} in {Ithaca} {Say} the {Company} {Is} {Trying} to {Crush} {Them}},
	url = {https://jacobin.com/2022/07/starbucks-workers-ithaca-new-york-union-busting-unfair-labor-practice-law},
	language = {en-US},
	urldate = {2025-09-03},
	journal = {Jacobin},
	author = {{Alex N. Press}},
	month = jul,
	year = {2022},
}

@misc{hamilton_nolan_buffalo_2021,
	title = {Buffalo {Starbucks} {Workers} {Say} {They} {Will} {Unionize} {One} {Store} {At} a {Time}},
	url = {https://inthesetimes.com/article/buffalo-starbucks-workers-union-labor-india-walton-colectivo},
	language = {en},
	urldate = {2025-09-03},
	journal = {In These Times},
	author = {{Hamilton Nolan}},
	month = aug,
	year = {2021},
}

@misc{saurav_sarkar_across_2022,
	title = {Across the {Country}, {Starbucks}’s {Anti}-{Union} {Push} {Is} {Getting} {Worse}},
	url = {https://jacobin.com/2022/09/starbucks-workers-united-anti-union-busting},
	language = {en-US},
	urldate = {2025-09-03},
	journal = {Jacobin},
	author = {{Saurav Sarkar}},
	month = sep,
	year = {2022},
}

@misc{zakarin_exclusive_2023,
	title = {{EXCLUSIVE}: {Starbucks} {Illegally} {Withheld} {Raises} \& {Tips} from {Union} {Workers}, {NLRB} {Says}},
	url = {https://perfectunion.us/starbucks-nlrb-credit-card-tip-complaint/},
	language = {en-US},
	urldate = {2025-09-03},
	journal = {More Perfect Union},
	author = {Zakarin, Jordan},
	month = mar,
	year = {2023},
}

@misc{john_logan_starbucks_2022,
	title = {Starbucks’ {Howard} {Schultz} {Should} {Win} an {Award} for the {Nation}’s {Most} {Flagrant} {Union} {Buster}},
	url = {https://jacobin.com/2022/07/starbucks-howard-schultz-union-busting-firings-store-closures},
	language = {en-US},
	urldate = {2025-09-03},
	journal = {Jacobin},
	author = {{John Logan}},
	month = jul,
	year = {2022},
}

@misc{faith_bennett_starbucks_2021,
	title = {Starbucks {Workers} {Are} {Organizing} — and {Management} {Is} {Worried}},
	url = {https://jacobin.com/2021/10/starbucks-workers-united-buffalo-union-drive-organizing-coffee-shop-industry-labor},
	language = {en-US},
	urldate = {2025-09-03},
	journal = {Jacobin},
	author = {{Faith Bennett}},
	month = oct,
	year = {2021},
}

@misc{gurley_starbucks_2021,
	title = {Starbucks {Is} {Blocking} {Union} {Activists} and {Workers} on {Twitter}},
	url = {https://www.vice.com/en/article/starbucks-is-blocking-union-activists-and-workers-on-twitter/},
	language = {en-US},
	urldate = {2025-09-03},
	journal = {VICE},
	author = {Gurley, Lauren Kaori},
	month = oct,
	year = {2021},
}

@misc{blest_starbucks_2022-2,
	title = {Starbucks {CEO} {Howard} {Schultz} {Says} {Companies} {Are} {Being} ‘{Assaulted}’ by {Unions}},
	url = {https://www.vice.com/en/article/starbucks-union-howard-schultz/},
	language = {en-US},
	urldate = {2025-09-03},
	journal = {VICE},
	author = {Blest, Paul},
	month = apr,
	year = {2022},
}

@misc{blest_lateness_2022,
	title = {Lateness, {Cursing}, a {Broken} {Sink}: {Starbucks} {Keeps} {Firing} {Pro}-{Union} {Employees}},
	shorttitle = {Lateness, {Cursing}, a {Broken} {Sink}},
	url = {https://www.vice.com/en/article/alleged-starbucks-union-busting/},
	language = {en-US},
	urldate = {2025-09-03},
	journal = {VICE},
	author = {Blest, Paul},
	month = apr,
	year = {2022},
}

@misc{lauren_kaori_gurley_its_2021,
	title = {'{It}'s {Almost} {Comical}:' {Starbucks} {Is} {Blatantly} {Trying} to {Crush} {Its} {Union}},
	shorttitle = {'{It}'s {Almost} {Comical}},
	url = {https://www.vice.com/en/article/its-almost-comical-starbucks-is-blatantly-trying-to-crush-its-union/},
	language = {en-US},
	urldate = {2025-09-03},
	journal = {VICE},
	author = {{Lauren Kaori Gurley}},
	month = sep,
	year = {2021},
}

@misc{gurley_starbucks_2022-1,
	title = {The {Starbucks} {Union} {Movement} {Is} ‘{Unstoppable}’},
	url = {https://www.vice.com/en/article/the-starbucks-union-movement-is-unstoppable/},
	language = {en-US},
	urldate = {2025-09-03},
	journal = {VICE},
	author = {Gurley, Lauren Kaori},
	month = apr,
	year = {2022},
}

@misc{sainato_starbucks_2022,
	title = {Starbucks creating ‘culture of fear’ as it fires dozens involved in union efforts},
	url = {https://www.theguardian.com/us-news/2022/aug/25/starbucks-union-employees-fired},
	language = {en-GB},
	urldate = {2025-09-03},
	journal = {The Guardian},
	author = {Sainato, Michael},
	month = aug,
	year = {2022},
}

@misc{blest_starbucks_2022-4,
	title = {The {Starbucks} {Union} {Push} {Is} {So} {Successful} {It} {Could} {Start} a {Turf} {War}},
	url = {https://www.vice.com/en/article/wisconsin-starbucks-unions-ufcw/},
	language = {en-US},
	urldate = {2025-09-03},
	journal = {VICE},
	author = {Blest, Paul},
	month = may,
	year = {2022},
}

@misc{sam_gindin_was_2025,
	title = {Was the {Teamsters}’ {Amazon} {Strike} a {Success}?},
	url = {https://jacobin.com/2025/03/teamsters-amazon-strike-union-strategy},
	language = {en-US},
	urldate = {2025-09-03},
	journal = {Jacobin},
	author = {{Sam Gindin}},
	month = mar,
	year = {2025},
}

@misc{scheiber_amazon_2024,
	title = {Amazon {Delivery} {Drivers} at {Seven} {Hubs} {Walk} {Out}},
	url = {https://www.nytimes.com/2024/12/19/business/economy/amazon-teamsters-strike.html},
	language = {en-US},
	urldate = {2025-09-03},
	journal = {The New York Times},
	author = {Scheiber, Noam and Nerkar, Santul},
	month = dec,
	year = {2024},
}

@misc{nerkar_amazon_2024,
	title = {Amazon {Warehouse} {Workers} in {New} {York} {City} {Join} {Protest}},
	url = {https://www.nytimes.com/2024/12/21/business/amazon-warehouse-staten-island-strike.html},
	language = {en-US},
	urldate = {2025-09-03},
	journal = {The New York Times},
	author = {Nerkar, Santul and Scheiber, Noam},
	month = dec,
	year = {2024},
}

@misc{selyukh_its_2021,
	title = {It's {A} {No}: {Amazon} {Warehouse} {Workers} {Vote} {Against} {Unionizing} {In} {Historic} {Election}},
	shorttitle = {It's {A} {No}},
	url = {https://www.npr.org/2021/04/09/982139494/its-a-no-amazon-warehouse-workers-vote-against-unionizing-in-historic-election},
	language = {en},
	urldate = {2025-09-03},
	journal = {NPR},
	author = {Selyukh, Alina},
	month = apr,
	year = {2021},
}

@misc{selyukh_amazon_2024,
	title = {Amazon ordered to let workers vote on unionizing -- for the 3rd time},
	url = {https://www.npr.org/2024/11/06/nx-s1-5087567/amazon-union-warehouse-bessemer-alabama},
	language = {en},
	urldate = {2025-09-03},
	journal = {NPR},
	author = {Selyukh, Alina},
	month = nov,
	year = {2024},
}

@misc{palmer_amazon_2020,
	title = {Amazon moves closer to facing its first unionization vote in six years},
	url = {https://www.cnbc.com/2020/12/22/amazon-moves-closer-to-facing-its-first-unionization-vote-in-six-years.html},
	language = {en},
	urldate = {2025-09-03},
	journal = {CNBC},
	author = {Palmer, Annie},
	month = dec,
	year = {2020},
	note = {Section: Technology},
}

@misc{anonymous_what_2022,
	title = {What happens when {Amazon} comes to town},
	url = {https://www.economist.com/united-states/what-happens-when-amazon-comes-to-town/21808308},
	urldate = {2025-09-03},
	journal = {The Economist},
	author = {{Anonymous}},
	month = mar,
	year = {2022},
}

@misc{macgillis_lessons_2021,
	title = {Lessons {From} {Bessemer}: {What} {Amazon}’s {Union} {Defeat} {Means} for the {American} {Labor} {Movement}},
	shorttitle = {Lessons {From} {Bessemer}},
	url = {https://www.propublica.org/article/lessons-from-bessemer-what-amazons-union-defeat-means-for-the-american-labor-movement},
	language = {en},
	urldate = {2025-09-03},
	journal = {ProPublica},
	author = {MacGillis, Alec},
	month = apr,
	year = {2021},
}

@misc{abigail_hassan_boston_2024,
	title = {Boston {University} faculty, staff create successful petition against {Provost}’s strike pay policies},
	url = {https://dailyfreepress.com/03/29/12/203314/boston-university-faculty-staff-create-successful-petition-against-provosts-strike-pay-policies/},
	language = {en, sv},
	urldate = {2025-09-04},
	journal = {The Daily Free Press},
	author = {{Abigail Hassan}},
	month = mar,
	year = {2024},
	note = {Section: Campus},
}

@misc{starbucks_we_2025,
	title = {We are all {Starbucks} {Partners}},
    author = {Starbucks},
	url = {https://one.starbucks.com/},
	language = {en-US},
	urldate = {2025-09-04},
	journal = {One.Starbucks},
    year = {2025},
}

@misc{gurley_starbucks_2021-2,
	title = {Starbucks {Temporarily} {Closes} 2 {Stores} {That} {Are} {Trying} to {Unionize}},
	url = {https://www.vice.com/en/article/starbucks-temporarily-closes-two-stores-that-are-trying-to-unionize/},
	language = {en-US},
	urldate = {2025-09-04},
	journal = {VICE},
	author = {Gurley, Lauren Kaori},
	month = oct,
	year = {2021},
}

@misc{stefano_perago_embracing_2025,
	title = {Embracing {Automation}: {AI}, {Robotics}, and the {Future} of {Work}},
	shorttitle = {Embracing {Automation}},
	url = {https://www.aboutamazon.eu/embracing-automation-ai-robotics-and-the-future-of-work},
	language = {en},
	urldate = {2025-09-04},
	journal = {EU About Amazon},
	author = {{Stefano Perago}},
	month = may,
	year = {2025},
	note = {Section: Policy},
}

@misc{greenhouse_will_2023,
	title = {Will {Starbucks}’ union-busting stifle a union rebirth in the {US}?},
	url = {https://www.theguardian.com/us-news/2023/aug/28/will-starbucks-union-busting-stifle-a-union-rebirth-in-the-us},
	language = {en-GB},
	urldate = {2025-01-22},
	journal = {The Guardian},
	author = {Greenhouse, Steven},
	month = aug,
	year = {2023},
}

@misc{greenhouse_old-school_2023,
	title = {‘{Old}-school union busting’: how {US} corporations are quashing the new wave of organizing},
	shorttitle = {‘{Old}-school union busting’},
	url = {https://www.theguardian.com/us-news/2023/feb/26/amazon-trader-joes-starbucks-anti-union-measures},
	language = {en-GB},
	urldate = {2025-01-21},
	journal = {The Guardian},
	author = {Greenhouse, Steven},
	month = feb,
	year = {2023},
}

@misc{hendy_l_2011,
	title = {L.{A}. {Teachers} {Use} {Privatization} {Fight} to {Build} {Community} {Power}},
	url = {https://labornotes.org/2011/03/la-teachers-use-privatization-fight-build-community-power},
	language = {en},
	urldate = {2025-08-18},
	journal = {Labor Notes},
	author = {Hendy, Sherlett and Lippe-Klein, Noah},
	month = mar,
	year = {2011},
}

@misc{labor_notes_editorial_board_ways_2019,
	title = {Ways to {Not} {Quite} {Strike}},
	url = {https://labornotes.org/2019/10/ways-not-quite-strike},
	language = {en},
	urldate = {2025-08-18},
	journal = {Labor Notes},
	author = {{Labor Notes Editorial Board}},
	month = oct,
	year = {2019},
}

@misc{soriano-castillo_alabama_2012,
	title = {In {Alabama} {Poultry} {Workers} {Victory}, {A} {Vote} to {Stick} {Together}},
	url = {https://labornotes.org/blogs/2012/06/alabama-poultry-workers-victory-vote-stick-together},
	language = {en},
	urldate = {2025-08-18},
	journal = {Labor Notes},
	author = {Soriano-Castillo, Eduardo},
	month = jun,
	year = {2012},
}

@misc{wiessner_starbucks_2023,
	title = {Starbucks closed 23 stores to deter unionizing, {US} agency says},
	url = {https://www.reuters.com/business/starbucks-closed-23-us-stores-deter-unionizing-agency-claims-2023-12-14/},
	language = {en},
	urldate = {2025-08-15},
	journal = {Reuters},
	author = {Wiessner, Daniel and Wiessner, Daniel},
	month = dec,
	year = {2023},
}

@book{yin_case_2009,
	address = {Thousand Oaks, California, USA},
	title = {Case {Study} {Research}: {Design} and {Methods}},
	isbn = {978-1-4129-6099-1},
	shorttitle = {Case {Study} {Research}},
	language = {en},
	publisher = {SAGE},
	author = {Yin, Robert K.},
	year = {2009},
	note = {Google-Books-ID: FzawIAdilHkC},
}

@incollection{winner_artifacts_2007,
	address = {Milton Park, Abingdon-on-Thames, Oxfordshire, England, UK},
	title = {Do {Artifacts} {Have} {Politics}?},
	isbn = {978-1-315-25969-7},
	booktitle = {Computer {Ethics}},
	publisher = {Routledge},
	author = {Winner, Langdon},
	year = {2007},
	note = {Num Pages: 16},
}

@book{woodcock_marx_2019,
	address = {Chicago, Illinois, U.S.},
	title = {Marx at the {Arcade}: {Consoles}, {Controllers}, and {Class} {Struggle}},
	isbn = {978-1-60846-867-6},
	shorttitle = {Marx at the {Arcade}},
	language = {en},
	publisher = {Haymarket Books},
	author = {Woodcock, Jamie},
	month = jun,
	year = {2019},
	note = {Google-Books-ID: 2vOPDwAAQBAJ},
}

@book{wood_despotism_2020,
	address = {Ithaca, New York, USA},
	title = {Despotism on {Demand}: {How} {Power} {Operates} in the {Flexible} {Workplace}},
	isbn = {978-1-5017-4890-5},
	shorttitle = {Despotism on {Demand}},
	language = {en},
	publisher = {Cornell University Press},
	author = {Wood, Alex J.},
	month = may,
	year = {2020},
	note = {Google-Books-ID: xHW4DwAAQBAJ},
}

@incollection{walker_impact_2024,
	address = {Cheltenham, England},
	title = {The impact of {AI} on contracts and unionisation},
	isbn = {978-1-80088-997-2},
	url = {https://www.elgaronline.com/edcollchap/book/9781800889972/book-part-9781800889972-27.xml},
	language = {eng},
	urldate = {2024-12-05},
	booktitle = {Handbook of {Artificial} {Intelligence} at {Work}},
	publisher = {Edward Elgar Publishing},
	author = {Walker, Michael},
	month = feb,
	year = {2024},
	note = {Section: Handbook of Artificial Intelligence at Work},
	pages = {356--370},
}

@book{ragin_what_1992,
	address = {Cambridge, England},
	title = {What {Is} a {Case}?: {Exploring} the {Foundations} of {Social} {Inquiry}},
	isbn = {978-1-316-10174-2},
	shorttitle = {What {Is} a {Case}?},
	language = {en},
	publisher = {Cambridge University Press},
	author = {Ragin, Charles C. and Becker, Howard Saul},
	month = jul,
	year = {1992},
	note = {Google-Books-ID: XmduBAAAQBAJ},
}

@book{edwards_contested_1979,
	address = {New York City, USA},
	title = {Contested {Terrain}},
	isbn = {978-0-465-01412-5},
	language = {en},
	publisher = {Basic Books},
	author = {Edwards, Richard},
	month = apr,
	year = {1979},
	note = {Google-Books-ID: 1B9HAAAAMAAJ},
}

@book{dean_capitals_2025,
	address = {Brooklyn, New York, USA},
	title = {Capital's {Grave}: {Neofeudalism} and the {New} {Class} {Struggle}},
	isbn = {978-1-80429-521-2},
	shorttitle = {Capital's {Grave}},
	language = {en},
	publisher = {Verso Books},
	author = {Dean, Jodi},
	month = mar,
	year = {2025},
	note = {Google-Books-ID: voVDEQAAQBAJ},
}

@book{burawoy_manufacturing_1979,
	address = {Chicago, Illinois, U.S.},
	title = {Manufacturing {Consent}: {Changes} in the {Labor} {Process} {Under} {Monopoly} {Capitalism}},
	isbn = {978-0-226-08037-6},
	shorttitle = {Manufacturing {Consent}},
	language = {en},
	publisher = {University of Chicago Press},
	author = {Burawoy, Michael},
	year = {1979},
	note = {Google-Books-ID: uDu9QgAACAAJ},
}

@book{braverman_labor_1974,
	address = {New York City, USA},
	title = {Labor and {Monopoly} {Capital}: {The} {Degradation} of {Work} in the {Twentieth} {Century}},
	isbn = {978-0-85345-340-6},
	shorttitle = {Labor and {Monopoly} {Capital}},
	language = {en},
	publisher = {Monthly Review Press},
	author = {Braverman, Harry},
	year = {1974},
	note = {Google-Books-ID: YEocaPj7PtUC},
}

@incollection{ball_categorizing_2002,
	address = {Milton Park, Abingdon-on-Thames, Oxfordshire, England, UK},
	title = {Categorizing the workers: {Electronic} surveillance and social ordering in the call center},
	isbn = {978-0-203-99488-7},
	shorttitle = {Categorizing the workers},
	booktitle = {Surveillance as {Social} {Sorting}},
	publisher = {Routledge},
	author = {Ball, Kirstie},
	year = {2002},
	note = {Num Pages: 25},
}

@book{the_capacitor_collective_notes_2025,
	address = {Brooklyn, NY, USA},
	title = {Notes {Toward} a {Digital} {Workers}' {Inquiry}},
	publisher = {Common Notions},
	author = {{The Capacitor Collective}},
	month = nov,
	year = {2025},
}

@misc{akhtar_out_2021,
	title = {Out of the {Pandemic}: {UC} {Student} {Researchers} {Emerge} {United}},
	shorttitle = {Out of the {Pandemic}},
	url = {https://labornotes.org/blogs/2021/05/out-pandemic-uc-student-researchers-emerge-united},
	language = {en},
	urldate = {2025-08-18},
	journal = {Labor Notes},
	author = {Akhtar, Ahmed and Banks, Jess},
	month = may,
	year = {2021},
}

@misc{valentina_luketa_how_2024,
	title = {How {Tens} of {Thousands} of {Grad} {Workers} {Are} {Organizing} {Themselves}},
	url = {https://labornotes.org/2024/05/how-tens-thousands-grad-workers-are-organizing-themselves},
	language = {en},
	urldate = {2025-09-10},
	journal = {Labor Notes},
	author = {{Valentina Luketa}},
	month = may,
	year = {2024},
}

@misc{bureau_of_labor_statistics_union_2024,
	title = {Union {Members} {Summary} - 2023 {A01} {Results}},
	url = {https://www.bls.gov/news.release/union2.nr0.htm},
	language = {en},
	urldate = {2025-01-21},
	journal = {Bureau of Labor Statistics},
	author = {{Bureau of Labor Statistics}},
	month = jan,
	year = {2024},
}

@article{logan_union_2006,
	title = {The {Union} {Avoidance} {Industry} in the {United} {States}},
	volume = {44},
	issn = {1467-8543},
	url = {https://onlinelibrary.wiley.com/doi/abs/10.1111/j.1467-8543.2006.00518.x},
	doi = {10.1111/j.1467-8543.2006.00518.x},
	language = {en},
	number = {4},
	urldate = {2026-01-27},
	journal = {British Journal of Industrial Relations},
	author = {Logan, John},
	year = {2006},
	pages = {651--675},
}

@book{noble_forces_1986,
	address = {Oxford, New York},
	title = {Forces of {Production}: {A} {Social} {History} of {Industrial} {Automation}},
	isbn = {978-0-19-504046-3},
	shorttitle = {Forces of {Production}},
	publisher = {Oxford University Press},
	author = {Noble, David F.},
	month = feb,
	year = {1986},
}

@article{calacci_access_2023,
	title = {From access to understanding: {Collective} data governance for workers},
	volume = {14},
	issn = {2031-9525},
	shorttitle = {From access to understanding},
	url = {https://doi.org/10.1177/20319525231167981},
	doi = {10.1177/20319525231167981},
	language = {en},
	number = {2},
	urldate = {2025-01-21},
	journal = {European Labour Law Journal},
	publisher = {SAGE Publications},
	author = {Calacci, Dana and Stein, Jake},
	month = jun,
	year = {2023},
	pages = {253--282},
}

@article{calacci_bargaining_2022,
	title = {Bargaining with the {Black}-{Box}: {Designing} and {Deploying} {Worker}-{Centric} {Tools} to {Audit} {Algorithmic} {Management}},
	volume = {6},
	shorttitle = {Bargaining with the {Black}-{Box}},
	url = {https://dl.acm.org/doi/10.1145/3570601},
	doi = {10.1145/3570601},
	number = {CSCW2},
	urldate = {2025-04-24},
	journal = {Proc. ACM Hum.-Comput. Interact.},
	author = {Calacci, Dana and Pentland, Alex},
	month = nov,
	year = {2022},
	pages = {428:1--428:24},
}

@inproceedings{calacci_organizing_2022,
	address = {New York, NY, USA},
	series = {{CHIWORK} '22},
	title = {Organizing in the {End} of {Employment}: {Information} {Sharing}, {Data} {Stewardship}, and {Digital} {Workerism}},
	isbn = {978-1-4503-9655-4},
	shorttitle = {Organizing in the {End} of {Employment}},
	url = {https://dl.acm.org/doi/10.1145/3533406.3533424},
	doi = {10.1145/3533406.3533424},
	urldate = {2025-05-12},
	booktitle = {Proceedings of the 1st {Annual} {Meeting} of the {Symposium} on {Human}-{Computer} {Interaction} for {Work}},
	publisher = {Association for Computing Machinery},
	author = {Calacci, Dana},
	month = jun,
	year = {2022},
	pages = {1--9},
}

@article{jager_codetermination_2022,
	title = {Codetermination and {Power} in the {Workplace} {Special} {Issue}: {Not} {So} {Free} to {Contract}: {How} {Unequal} {Workplace} {Power} {Undercuts} the "{Freedom} of {Contract}" {Framework}: {Section} {B}: {Assessing} {Economic} {Claims} in {Philosophy} and {Employment} {Law}},
	volume = {3},
	shorttitle = {Codetermination and {Power} in the {Workplace} {Special} {Issue}},
	url = {https://heinonline.org/HOL/P?h=hein.journals/jlolwadpl3&i=205},
	language = {eng},
	number = {1},
	urldate = {2025-12-02},
	journal = {Journal of Law and Political Economy},
	author = {Jager, Simon and Noy, Shakked and Schoefer, Benjamin},
	year = {2022},
	pages = {[i]--224},
}

@techreport{minsu_longiaru_privacy_2025,
	type = {Policy {Brief}},
	title = {The "{Privacy} {Trap}: {How} "{Privacy}-{Preserving} {AI} {Techniques}" {Mask} the {New} {Worker} {Surveillance} and {Datafication}},
	url = {https://datasociety.net/wp-content/uploads/2025/10/The-Privacy-Trap.pdf},
	institution = {Data \& Society},
	author = {{Minsu Longiaru} and {Wilneida Negron} and {Brian J. Chen} and {Aiha Nguyen} and {Seema N. Patel} and {Dana Calacci}},
	month = oct,
	year = {2025},
}

@inproceedings{wolf_designing_2022,
	address = {New York, NY, USA},
	series = {{DIS} '22},
	title = {Designing within {Capitalism}},
	isbn = {978-1-4503-9358-4},
	url = {https://dl.acm.org/doi/10.1145/3532106.3533559},
	doi = {10.1145/3532106.3533559},
	urldate = {2025-07-03},
	booktitle = {Proceedings of the 2022 {ACM} {Designing} {Interactive} {Systems} {Conference}},
	publisher = {Association for Computing Machinery},
	author = {Wolf, Christine T. and Asad, Mariam and Dombrowski, Lynn S.},
	month = jun,
	year = {2022},
	pages = {439--453},
}

@misc{lisa_kresge_union_2020,
	title = {Union {Collective} {Bargaining} {Agreement} {Strategies} in {Response} to {Technology}},
	url = {https://laborcenter.berkeley.edu/union-collective-bargaining-agreement-strategies-in-response-to-technology/},
	language = {en-US},
	urldate = {2024-12-10},
	journal = {UC Berkeley Labor Center},
	author = {{Lisa Kresge}},
	month = dec,
	year = {2020},
}

@article{grinthal_power_2011,
	title = {Power {With}: {Practice} {Models} for {Social} {Justice} {Lawyering}},
	volume = {15},
	shorttitle = {Power {With}},
	url = {https://heinonline.org/HOL/Page?handle=hein.journals/hybrid15&id=27&div=&collection=},
	journal = {University of Pennsylvania Journal of Law and Social Change},
	author = {Grinthal, Michael},
	year = {2011},
	pages = {25},
}

@techreport{greenbaum_windows_2004,
	title = {Windows on the {Workplace}},
	language = {en},
	institution = {Monthly Review Press},
	author = {Greenbaum, Joan M.},
	month = jun,
	year = {2004},
	note = {Google-Books-ID: mV9YAAAAYAAJ},
}

@article{wood_good_2019,
	title = {Good {Gig}, {Bad} {Gig}: {Autonomy} and {Algorithmic} {Control} in the {Global} {Gig} {Economy}},
	volume = {33},
	issn = {0950-0170},
	shorttitle = {Good {Gig}, {Bad} {Gig}},
	url = {https://pmc.ncbi.nlm.nih.gov/articles/PMC6380453/},
	doi = {10.1177/0950017018785616},
	number = {1},
	urldate = {2025-11-17},
	journal = {Work, Employment \& Society},
	author = {Wood, Alex J and Graham, Mark and Lehdonvirta, Vili and Hjorth, Isis},
	month = feb,
	year = {2019},
	pages = {56--75},
}

@misc{dzieza_robots_2020,
	title = {Robots aren’t taking our jobs — they’re becoming our bosses},
	url = {https://www.theverge.com/2020/2/27/21155254/automation-robots-unemployment-jobs-vs-human-google-amazon},
	language = {en},
	urldate = {2024-07-10},
	journal = {The Verge},
	author = {Dzieza, Josh},
	month = feb,
	year = {2020},
}

@misc{jason_del_rey_amazon_2020,
	title = {Amazon white-collar employees are fuming over management targeting a fired warehouse worker},
	url = {https://www.vox.com/recode/2020/4/5/21206385/amazon-fired-warehouse-worker-christian-smalls-employee-backlash-david-zapolsky-coronavirus},
	language = {en-US},
	urldate = {2025-09-03},
	journal = {Vox},
	author = {{Jason Del Rey} and {Shirin Ghaffary}},
	month = apr,
	year = {2020},
}

@misc{cox_inside_2020,
	title = {Inside {Amazon}’s {Secret} {Program} to {Spy} {On} {Workers}’ {Private} {Facebook} {Groups}},
	url = {https://www.vice.com/en/article/amazon-is-spying-on-its-workers-in-closed-facebook-groups-internal-reports-show/},
	language = {en-US},
	urldate = {2025-09-03},
	journal = {VICE},
	author = {Cox, Lauren Kaori Gurley {and} Joseph},
	month = sep,
	year = {2020},
}

@article{gruszka_out_2022,
	title = {Out of sight, out of mind? ({In})visibility of/in platform-mediated work},
	volume = {24},
	issn = {1461-4448},
	shorttitle = {Out of sight, out of mind?},
	url = {https://doi.org/10.1177/1461444820977209},
	doi = {10.1177/1461444820977209},
	language = {EN},
	number = {8},
	urldate = {2025-09-02},
	journal = {New Media \& Society},
	publisher = {SAGE Publications},
	author = {Gruszka, Katarzyna and Böhm, Madeleine},
	month = aug,
	year = {2022},
	pages = {1852--1871},
}

@article{ming_i_2023,
	title = {"{I} {Go} {Beyond} and {Beyond}" {Examining} the {Invisible} {Work} of {Home} {Health} {Aides}},
	volume = {7},
	url = {https://dl.acm.org/doi/10.1145/3579492},
	doi = {10.1145/3579492},
	number = {CSCW1},
	urldate = {2025-09-02},
	journal = {Proc. ACM Hum.-Comput. Interact.},
	author = {Ming, Joy and Kuo, Elizabeth and Go, Katie and Tseng, Emily and Kallas, John and Vashistha, Aditya and Sterling, Madeline and Dell, Nicola},
	month = apr,
	year = {2023},
	pages = {59:1--59:21},
}

@article{star_layers_1999,
	title = {Layers of {Silence}, {Arenas} of {Voice}: {The} {Ecology} of {Visible} and {Invisible} {Work}},
	volume = {8},
	issn = {1573-7551},
	shorttitle = {Layers of {Silence}, {Arenas} of {Voice}},
	url = {https://doi.org/10.1023/A:1008651105359},
	doi = {10.1023/A:1008651105359},
	language = {en},
	number = {1},
	urldate = {2025-09-02},
	journal = {Computer Supported Cooperative Work (CSCW)},
	author = {Star, Susan Leigh and Strauss, Anselm},
	month = mar,
	year = {1999},
	pages = {9--30},
}

@misc{elizabeth_gurley_flynn_sabotage_1917,
	title = {Sabotage: {The} {Conscious} {Withdrawal} of the {Workers}' {Industrial} {Efficiency}},
	url = {https://www.marxists.org/subject/women/authors/flynn/1917/sabotage.htm},
	urldate = {2025-08-28},
	publisher = {IWW Publishing Bureau},
	author = {{Elizabeth Gurley Flynn}},
	year = {1917},
}

@misc{weizenbaum-institut_data_2024,
	title = {The {Data} {Workers}’ {Inquiry} - {Unveiling} the {Hidden} {Labor} {Behind} {AI}},
	url = {https://www.youtube.com/watch?v=tAMqrXlEPDI},
	urldate = {2025-08-28},
	author = {{Weizenbaum-Institut}},
	month = may,
	year = {2024},
}

@article{woodcock_towards_2021,
	title = {Towards a {Digital} {Workerism}: {Workers}’ {Inquiry}, {Methods}, and {Technologies}},
	volume = {15},
	issn = {1871-4765},
	shorttitle = {Towards a {Digital} {Workerism}},
	url = {https://doi.org/10.1007/s11569-021-00384-w},
	doi = {10.1007/s11569-021-00384-w},
	language = {en},
	number = {1},
	urldate = {2025-08-28},
	journal = {NanoEthics},
	author = {Woodcock, Jamie},
	month = apr,
	year = {2021},
	pages = {87--98},
}

@article{woodcock_workers_2014,
	title = {The workers’ inquiry from {Trotskyism} to {Operaismo}: {A} political methodology for investigating the workplace},
	volume = {14},
	issn = {1473-2866},
	shorttitle = {The workers’ inquiry from {Trotskyism} to {Operaismo}},
	url = {http://www.ephemerajournal.org/contribution/workers%E2%80%99-inquiry-trotskyism-operaismo-political-methodology-investigating-workplace},
	language = {en},
	number = {3},
	urldate = {2025-08-28},
	journal = {Ephemera: theory \& politics in organizations},
	author = {Woodcock, Jamie},
	month = aug,
	year = {2014},
	pages = {493--513},
}

@article{ikeler_precaritys_2019,
	title = {Precarity’s {Prospect}: {Contingent} {Control} and {Union} {Renewal} in the {Retail} {Sector}},
	volume = {45},
	issn = {0896-9205},
	shorttitle = {Precarity’s {Prospect}},
	url = {https://doi.org/10.1177/0896920517749706},
	doi = {10.1177/0896920517749706},
	language = {EN},
	number = {4-5},
	urldate = {2025-07-07},
	journal = {Critical Sociology},
	publisher = {SAGE Publications Ltd},
	author = {Ikeler, Peter},
	month = jul,
	year = {2019},
	pages = {501--516},
}

@article{dimaggio_victory_2017,
	title = {Victory at {Verizon}: {The} {Anatomy} of a {Strike}},
	volume = {26},
	issn = {1095-7960},
	shorttitle = {Victory at {Verizon}},
	url = {https://doi.org/10.1177/1095796016681545},
	doi = {10.1177/1095796016681545},
	language = {EN},
	number = {1},
	urldate = {2025-08-23},
	journal = {New Labor Forum},
	publisher = {SAGE Publications Inc},
	author = {DiMaggio, Daniel},
	month = jan,
	year = {2017},
	pages = {28--35},
}

@article{akkerman_solidarity_2013,
	title = {Solidarity, {Strikes}, and {Scabs}: {How} {Participation} {Norms} {Affect} {Union} {Members}’ {Willingness} to {Strike}},
	volume = {40},
	issn = {0730-8884},
	shorttitle = {Solidarity, {Strikes}, and {Scabs}},
	url = {https://doi.org/10.1177/0730888413481481},
	doi = {10.1177/0730888413481481},
	language = {EN},
	number = {3},
	urldate = {2025-08-23},
	journal = {Work and Occupations},
	publisher = {SAGE Publications Inc},
	author = {Akkerman, Agnes and Born, Marieke J. and Torenvlied, René},
	month = aug,
	year = {2013},
	pages = {250--280},
}

@article{harmer_attitude_2012,
	title = {Attitude, aptitude, ability and autonomy: the emergence of ‘offroaders’, a special class of nomadic worker},
	volume = {31},
	issn = {0144-929X},
	shorttitle = {Attitude, aptitude, ability and autonomy},
	url = {https://doi.org/10.1080/0144929X.2010.489117},
	doi = {10.1080/0144929X.2010.489117},
	number = {5},
	urldate = {2025-08-23},
	journal = {Behaviour \& Information Technology},
	publisher = {Taylor \& Francis},
	author = {Harmer, Brian M. and Pauleen, David J.},
	month = may,
	year = {2012},
	note = {\_eprint: https://doi.org/10.1080/0144929X.2010.489117},
	pages = {439--451},
}

@article{pauleen_making_2015,
	title = {Making {Sense} of {Mobile} {Technology}: {The} {Integration} of {Work} and {Private} {Life}},
	volume = {5},
	issn = {2158-2440},
	shorttitle = {Making {Sense} of {Mobile} {Technology}},
	url = {https://doi.org/10.1177/2158244015583859},
	doi = {10.1177/2158244015583859},
	language = {EN},
	number = {2},
	urldate = {2025-08-23},
	journal = {SAGE Open},
	publisher = {SAGE Publications},
	author = {Pauleen, David and Campbell, John and Harmer, Brian and Intezari, Ali},
	month = apr,
	year = {2015},
	pages = {2158244015583859},
}

@article{newman_reengineering_2017,
	title = {Reengineering {Workplace} {Bargaining}: {How} {Big} {Data} {Drives} {Lower} {Wages} and {How} {Reframing} {Labor} {Law} {Can} {Restore} {Information} {Equality} in the {Workplace}},
	volume = {85},
	shorttitle = {Reengineering {Workplace} {Bargaining}},
	url = {https://heinonline.org/HOL/P?h=hein.journals/ucinlr85&i=713},
	language = {eng},
	number = {3},
	urldate = {2025-08-23},
	journal = {University of Cincinnati Law Review},
	author = {Newman, Nathan},
	year = {2017},
	pages = {693--760},
}

@article{alimahomed-wilson_surveilling_2021,
	chapter = {Work in the Global Economy},
	title = {Surveilling {Amazon}’s warehouse workers: racism, retaliation, and worker resistance amid the pandemic},
	volume = {1},
	shorttitle = {Surveilling {Amazon}’s warehouse workers},
	url = {https://bristoluniversitypressdigital.com/view/journals/wge/1/1-2/article-p55.xml},
	doi = {10.1332/273241721X16295348549014},
	language = {en},
	number = {1-2},
	urldate = {2025-08-21},
	journal = {Work in the Global Economy},
	publisher = {Bristol University Press},
	author = {Alimahomed-Wilson, Jake and Reese, Ellen},
	month = oct,
	year = {2021},
	pages = {55--73},
}

@article{cheon_powerful_2023,
	title = {Powerful {Futures}: {How} a {Big} {Tech} {Company} {Envisions} {Humans} and {Technologies} in the {Workplace} of the {Future}},
	volume = {7},
	shorttitle = {Powerful {Futures}},
	url = {https://dl.acm.org/doi/10.1145/3610103},
	doi = {10.1145/3610103},
	number = {CSCW2},
	urldate = {2025-08-21},
	journal = {Proc. ACM Hum.-Comput. Interact.},
	author = {Cheon, EunJeong},
	month = oct,
	year = {2023},
	pages = {312:1--312:35},
}

@article{anjali_anwar_watched_2021,
	title = {Watched, but {Moving}: {Platformization} of {Beauty} {Work} and {Its} {Gendered} {Mechanisms} of {Control}},
	volume = {4},
	shorttitle = {Watched, but {Moving}},
	url = {https://dl.acm.org/doi/10.1145/3432949},
	doi = {10.1145/3432949},
	number = {CSCW3},
	urldate = {2025-08-21},
	journal = {Proc. ACM Hum.-Comput. Interact.},
	author = {Anjali Anwar, Ira and Pal, Joyojeet and Hui, Julie},
	month = jan,
	year = {2021},
	pages = {250:1--250:20},
}

@inproceedings{greenbaum_back_1996,
	address = {New York, NY, USA},
	series = {{CSCW} '96},
	title = {Back to labor: returning to labor process discussions in the study of work},
	isbn = {978-0-89791-765-0},
	shorttitle = {Back to labor},
	url = {https://dl.acm.org/doi/10.1145/240080.240259},
	doi = {10.1145/240080.240259},
	urldate = {2025-08-21},
	booktitle = {Proceedings of the 1996 {ACM} conference on {Computer} supported cooperative work},
	publisher = {Association for Computing Machinery},
	author = {Greenbaum, Joan},
	month = nov,
	year = {1996},
	pages = {229--237},
}

@article{suchman_making_1995,
	title = {Making work visible},
	volume = {38},
	issn = {0001-0782},
	url = {https://dl.acm.org/doi/10.1145/223248.223263},
	doi = {10.1145/223248.223263},
	number = {9},
	urldate = {2025-08-21},
	journal = {Commun. ACM},
	author = {Suchman, Lucy},
	month = sep,
	year = {1995},
	pages = {56--64},
}

@inproceedings{thuppilikkat_union_2024,
	address = {San José, Costa Rica},
	series = {Volume 8, {Issue} {CSCW2}},
	title = {Union {Makes} {Us} {Strong}: {Space}, {Technology}, and {On}-{Demand} {Ridesourcing} {Digital} {Labour} {Platforms} {\textbar} {Proceedings} of the {ACM} on {Human}-{Computer} {Interaction}},
	volume = {Volume 8},
	url = {https://dl.acm.org/doi/10.1145/3687002},
	doi = {https://doi.org/10.1145/3687002},
	urldate = {2025-04-30},
	booktitle = {Proceedings of the {ACM} on {Human}-{Computer} {Interaction}, {Volume} 8, {Issue} {CSCW2}},
	publisher = {ACM},
	author = {Thuppilikkat, Ashique Ali and Dhar, Dipsita and Chandra, Priyank},
	month = nov,
	year = {2024},
	pages = {1--36},
}

@article{bodie_law_2022,
	title = {The {Law} of {Employee} {Data}: {Privacy}, {Property}, {Governance}},
	volume = {97},
	shorttitle = {The {Law} of {Employee} {Data}},
	url = {https://heinonline.org/HOL/P?h=hein.journals/indana97&i=735},
	language = {eng},
	number = {2},
	urldate = {2025-08-20},
	journal = {Indiana Law Journal},
	author = {Bodie, Matthew T.},
	year = {2022},
	pages = {707--754},
}

@article{garden_can_2024,
	title = {Can {Labor} {Law} {Still} {Protect} {Concerted} {Activity}? {The} {Piper} {Lecture}},
	volume = {99},
	shorttitle = {Can {Labor} {Law} {Still} {Protect} {Concerted} {Activity}?},
	url = {https://heinonline.org/HOL/P?h=hein.journals/chknt99&i=516},
	language = {eng},
	number = {2},
	urldate = {2025-08-20},
	journal = {Chicago-Kent Law Review},
	author = {Garden, Charlotte},
	year = {2024},
	pages = {527--550},
}

@article{katsabian_rule_2021,
	title = {The {Rule} of {Technology}: {How} {Technology} {Is} {Used} to {Disturb} {Basic} {Labor} {Law} {Protections}},
	volume = {25},
	shorttitle = {The {Rule} of {Technology}},
	url = {https://heinonline.org/HOL/P?h=hein.journals/lewclr25&i=927},
	language = {eng},
	number = {3},
	urldate = {2025-08-20},
	journal = {Lewis \& Clark Law Review},
	author = {Katsabian, Tammy},
	year = {2021},
	pages = {895--952},
}

@misc{trull_addressing_2023,
	title = {Addressing a {NLRB} {Complaint} involving store closures in 2022},
	url = {https://one.starbucks.com/get-the-facts/complaint-involving-store-closures-in-2022/},
	language = {en-US},
	urldate = {2025-08-15},
	journal = {One.Starbucks},
	author = {Trull, Andrew},
	month = dec,
	year = {2023},
}

@article{press_how_2024,
	title = {How the {U}.{S}. {Labor} {Movement} {Is} {Confronting} {AI}},
	volume = {33},
	issn = {1095-7960},
	url = {https://doi.org/10.1177/10957960241276516},
	doi = {10.1177/10957960241276516},
	language = {EN},
	number = {3},
	urldate = {2025-08-14},
	journal = {New Labor Forum},
	publisher = {SAGE Publications Inc},
	author = {Press, Alex},
	month = sep,
	year = {2024},
	pages = {15--22},
}

@article{shen_everyday_2021,
	title = {Everyday {Algorithm} {Auditing}: {Understanding} the {Power} of {Everyday} {Users} in {Surfacing} {Harmful} {Algorithmic} {Behaviors}},
	volume = {5},
	shorttitle = {Everyday {Algorithm} {Auditing}},
	url = {https://dl.acm.org/doi/10.1145/3479577},
	doi = {10.1145/3479577},
	number = {CSCW2},
	urldate = {2025-08-14},
	journal = {Proc. ACM Hum.-Comput. Interact.},
	author = {Shen, Hong and DeVos, Alicia and Eslami, Motahhare and Holstein, Kenneth},
	month = oct,
	year = {2021},
	pages = {433:1--433:29},
}

@article{green_contestation_2021,
	title = {The {Contestation} of {Tech} {Ethics}: {A} {Sociotechnical} {Approach} to {Technology} {Ethics} in {Practice}},
	volume = {2},
	issn = {2688-5255},
	shorttitle = {The {Contestation} of {Tech} {Ethics}},
	url = {https://ieeexplore.ieee.org/document/9684741},
	doi = {10.23919/JSC.2021.0018},
	number = {3},
	urldate = {2025-08-15},
	journal = {Journal of Social Computing},
	author = {Green, Ben},
	month = sep,
	year = {2021},
	pages = {209--225},
}

@misc{hornborg_globalised_2019,
	title = {A globalised solar-powered future is wholly unrealistic – and our economy is the reason why},
	url = {http://theconversation.com/a-globalised-solar-powered-future-is-wholly-unrealistic-and-our-economy-is-the-reason-why-118927},
	language = {en-US},
	urldate = {2025-08-15},
	journal = {The Conversation},
	author = {Hornborg, Alf},
	month = sep,
	year = {2019},
}

@article{hodder_essence_2015,
	title = {The essence of trade unions: understanding identity, ideology and purpose},
	volume = {29},
	issn = {0950-0170},
	shorttitle = {The essence of trade unions},
	url = {https://doi.org/10.1177/0950017014568142},
	doi = {10.1177/0950017014568142},
	language = {EN},
	number = {5},
	urldate = {2025-08-14},
	journal = {Work, Employment and Society},
	publisher = {SAGE Publications Ltd},
	author = {Hodder, Andy and Edwards, Paul},
	month = oct,
	year = {2015},
	pages = {843--854},
}

@article{canella_networked_2023,
	title = {Networked {Unionism}: {Reframing} the {Labour} {Movement} and {Starbucks} {Workers} {United}’s {Hybrid} {Organising} {Practices}},
	volume = {21},
	issn = {1726-670X},
	shorttitle = {Networked {Unionism}},
	url = {https://www.triple-c.at/index.php/tripleC/article/view/1358},
	doi = {10.31269/triplec.v21i1.1358},
	language = {en-US},
	number = {1},
	urldate = {2025-04-24},
	journal = {tripleC: Communication, Capitalism \& Critique. Open Access Journal for a Global Sustainable Information Society},
	author = {Canella, Gino},
	month = jan,
	year = {2023},
	pages = {1--17},
}

@inproceedings{kristiansen_accountability_2018,
	address = {New York, NY, USA},
	series = {{CHI} '18},
	title = {Accountability in the {Blue}-{Collar} {Data}-{Driven} {Workplace}},
	isbn = {978-1-4503-5620-6},
	url = {https://dl.acm.org/doi/10.1145/3173574.3173906},
	doi = {10.1145/3173574.3173906},
	urldate = {2025-08-05},
	booktitle = {Proceedings of the 2018 {CHI} {Conference} on {Human} {Factors} in {Computing} {Systems}},
	publisher = {Association for Computing Machinery},
	author = {Kristiansen, Kristian Helbo and Valeur-Meller, Mathias A. and Dombrowski, Lynn and Holten Moller, Naja L..},
	month = apr,
	year = {2018},
	pages = {1--12},
}

@article{logan_consultants_2002,
	title = {Consultants, lawyers, and the ‘union free’ movement in the {USA} since the 1970s},
	volume = {33},
	copyright = {Blackwell Publishers Ltd. 2002},
	issn = {1468-2338},
	url = {https://onlinelibrary.wiley.com/doi/abs/10.1111/1468-2338.00230},
	doi = {10.1111/1468-2338.00230},
	language = {en},
	number = {3},
	urldate = {2025-07-24},
	journal = {Industrial Relations Journal},
	author = {Logan, John},
	year = {2002},
	note = {\_eprint: https://onlinelibrary.wiley.com/doi/pdf/10.1111/1468-2338.00230},
	pages = {197--214},
}

@inproceedings{chowdhary_can_2023,
	address = {New York, NY, USA},
	series = {{FAccT} '23},
	title = {Can {Workers} {Meaningfully} {Consent} to {Workplace} {Wellbeing} {Technologies}?},
	isbn = {979-8-4007-0192-4},
	url = {https://dl.acm.org/doi/10.1145/3593013.3594023},
	doi = {10.1145/3593013.3594023},
	urldate = {2025-04-28},
	booktitle = {Proceedings of the 2023 {ACM} {Conference} on {Fairness}, {Accountability}, and {Transparency}},
	publisher = {Association for Computing Machinery},
	author = {Chowdhary, Shreya and Kawakami, Anna and Gray, Mary L and Suh, Jina and Olteanu, Alexandra and Saha, Koustuv},
	month = jun,
	year = {2023},
	pages = {569--582},
}

@inproceedings{kittur_future_2013,
	address = {New York, NY, USA},
	series = {{CSCW} '13},
	title = {The future of crowd work},
	isbn = {978-1-4503-1331-5},
	url = {https://dl.acm.org/doi/10.1145/2441776.2441923},
	doi = {10.1145/2441776.2441923},
	urldate = {2025-07-07},
	booktitle = {Proceedings of the 2013 conference on {Computer} supported cooperative work},
	publisher = {Association for Computing Machinery},
	author = {Kittur, Aniket and Nickerson, Jeffrey V. and Bernstein, Michael and Gerber, Elizabeth and Shaw, Aaron and Zimmerman, John and Lease, Matt and Horton, John},
	month = feb,
	year = {2013},
	pages = {1301--1318},
}

@inproceedings{pritchard_how_2015,
	address = {New York, NY, USA},
	series = {{CHI} '15},
	title = {How to {Drive} a {London} {Bus}: {Measuring} {Performance} in a {Mobile} and {Remote} {Workplace}},
	isbn = {978-1-4503-3145-6},
	shorttitle = {How to {Drive} a {London} {Bus}},
	url = {https://dl.acm.org/doi/10.1145/2702123.2702307},
	doi = {10.1145/2702123.2702307},
	urldate = {2025-07-22},
	booktitle = {Proceedings of the 33rd {Annual} {ACM} {Conference} on {Human} {Factors} in {Computing} {Systems}},
	publisher = {Association for Computing Machinery},
	author = {Pritchard, Gary W. and Briggs, Pam and Vines, John and Olivier, Patrick},
	month = apr,
	year = {2015},
	pages = {1885--1894},
}

@article{lu_data_2021,
	title = {Data {Work} in {Education}: {Enacting} and {Negotiating} {Care} and {Control} in {Teachers}' {Use} of {Data}-{Driven} {Classroom} {Surveillance} {Technology}},
	volume = {5},
	shorttitle = {Data {Work} in {Education}},
	url = {https://dl.acm.org/doi/10.1145/3479596},
	doi = {10.1145/3479596},
	number = {CSCW2},
	urldate = {2025-07-22},
	journal = {Proc. ACM Hum.-Comput. Interact.},
	author = {Lu, Alex Jiahong and Dillahunt, Tawanna R. and Marcu, Gabriela and Ackerman, Mark S.},
	month = oct,
	year = {2021},
	pages = {452:1--452:26},
}

@article{masson_captive_2004,
	title = {"{Captive} {Audience}" {Meetings} in {Union} {Organizing} {Campaigns}: {Free} {Speech} or {Unfair} {Advantage}?},
	volume = {56},
	issn = {0017-8322{\textless}br /{\textgreater}© Copyright University of California, College of the Law San Francisco},
	shorttitle = {"{Captive} {Audience}" {Meetings} in {Union} {Organizing} {Campaigns}},
	url = {https://repository.uclawsf.edu/hastings_law_journal/vol56/iss1/4},
	number = {1},
	journal = {UC Law Journal},
	author = {Masson, Elizabeth},
	month = jan,
	year = {2004},
	pages = {169},
}

@article{wiggin_weaponizing_2025,
	title = {Weaponizing the {Workplace}: {How} {Algorithmic} {Management} {Shaped} {Amazon}’s {Antiunion} {Campaign} in {Bessemer}, {Alabama}},
	volume = {11},
	issn = {2378-0231},
	shorttitle = {Weaponizing the {Workplace}},
	url = {https://doi.org/10.1177/23780231251318389},
	doi = {10.1177/23780231251318389},
	language = {EN},
	urldate = {2025-04-02},
	journal = {Socius},
	publisher = {SAGE Publications},
	author = {Wiggin, Teke},
	month = apr,
	year = {2025},
	pages = {23780231251318389},
}

@article{qiwei_sociotechnical_2024,
	title = {The {Sociotechnical} {Stack}: {Opportunities} for {Social} {Computing} {Research} in {Non}-{Consensual} {Intimate} {Media}},
	volume = {8},
	shorttitle = {The {Sociotechnical} {Stack}},
	url = {https://dl.acm.org/doi/10.1145/3686914},
	doi = {10.1145/3686914},
	number = {CSCW2},
	urldate = {2025-03-26},
	journal = {Proc. ACM Hum.-Comput. Interact.},
	author = {Qiwei, Li and McDonald, Allison and Haimson, Oliver L. and Schoenebeck, Sarita and Gilbert, Eric},
	month = nov,
	year = {2024},
	pages = {375:1--375:21},
}

@inproceedings{lee_participatory_2021,
	address = {New York, NY, USA},
	series = {{AIES} '21},
	title = {Participatory {Algorithmic} {Management}: {Elicitation} {Methods} for {Worker} {Well}-{Being} {Models}},
	isbn = {978-1-4503-8473-5},
	shorttitle = {Participatory {Algorithmic} {Management}},
	url = {https://dl.acm.org/doi/10.1145/3461702.3462628},
	doi = {10.1145/3461702.3462628},
	urldate = {2024-11-20},
	booktitle = {Proceedings of the 2021 {AAAI}/{ACM} {Conference} on {AI}, {Ethics}, and {Society}},
	publisher = {Association for Computing Machinery},
	author = {Lee, Min Kyung and Nigam, Ishan and Zhang, Angie and Afriyie, Joel and Qin, Zhizhen and Gao, Sicun},
	month = jul,
	year = {2021},
	pages = {715--726},
}

@book{mcalevey_no_2016,
	address = {Oxford, United Kingdom},
	title = {No {Shortcuts}: {Organizing} for {Power} in the {New} {Gilded} {Age}},
	isbn = {978-0-19-062471-2},
	shorttitle = {No {Shortcuts}},
	language = {en},
	publisher = {Oxford University Press},
	author = {McAlevey, Jane},
	year = {2016},
	note = {Google-Books-ID: CNfeDAAAQBAJ},
}

@inproceedings{cheon_amazon_2024,
	address = {New York, NY, USA},
	series = {{DIS} '24},
	title = {Amazon {Z} to {A}: {Speculative} {Design} to {Understand} the {Future} of {Labor}-{Intensive} {Workplaces}},
	isbn = {979-8-4007-0583-0},
	shorttitle = {Amazon {Z} to {A}},
	url = {https://dl.acm.org/doi/10.1145/3643834.3661561},
	doi = {10.1145/3643834.3661561},
	urldate = {2025-03-18},
	booktitle = {Proceedings of the 2024 {ACM} {Designing} {Interactive} {Systems} {Conference}},
	publisher = {Association for Computing Machinery},
	author = {Cheon, EunJeong and Khovanskaya, Vera},
	month = jul,
	year = {2024},
	pages = {314--327},
}

@inproceedings{lee_contrasting_2024,
	address = {New York, NY, USA},
	series = {{CHI} '24},
	title = {Contrasting {Perspectives} of {Workers}: {Exploring} {Labor} {Relations} in {Workplace} {Automation} and {Potential} {Interventions}},
	isbn = {979-8-4007-0330-0},
	shorttitle = {Contrasting {Perspectives} of {Workers}},
	url = {https://dl.acm.org/doi/10.1145/3613904.3642907},
	doi = {10.1145/3613904.3642907},
	urldate = {2025-03-26},
	booktitle = {Proceedings of the 2024 {CHI} {Conference} on {Human} {Factors} in {Computing} {Systems}},
	publisher = {Association for Computing Machinery},
	author = {Lee, Hee Rin},
	month = may,
	year = {2024},
	pages = {1--17},
}

@article{garden_labor_2018,
	title = {Labor {Organizing} in the {Age} of {Surveillance}},
	volume = {63},
	issn = {0036-3030},
	url = {https://scholarship.law.slu.edu/lj/vol63/iss1/5},
	number = {1},
	journal = {Saint Louis University Law Journal},
	author = {Garden, Charlotte},
	month = jan,
	year = {2018},
	pages = {55},
}

@article{kellogg_algorithms_2020,
	title = {Algorithms at {Work}: {The} {New} {Contested} {Terrain} of {Control}},
	volume = {14},
	copyright = {© Academy of Management Annals},
	shorttitle = {Algorithms at {Work}},
	url = {https://journals.aom.org/doi/abs/10.5465/annals.2018.0174},
	doi = {10.5465/annals.2018.0174},
	language = {EN},
	number = {1},
	urldate = {2025-01-06},
	journal = {Academy of Management Annals},
	publisher = {Academy of Management AnnalsBriarcliff Manor, NY},
	author = {Kellogg, Katherine C. and Valentine, Melissa A. and Christin, Angèle},
	month = jan,
	year = {2020},
	pages = {366--410},
}

@misc{parker_adapt_2022,
	title = {Adapt or {Resist}? {Unions} and the {Political} {Economy} of {Automation}},
	url = {https://gsipe-workshop.github.io/files/AMParker__Unions_and_the_Political_Economy_of_Automation_220829.pdf},
	language = {en},
	urldate = {2025-05-12},
	author = {Parker, Adam M},
	month = aug,
	year = {2022},
}

@incollection{vgontzas_new_2020,
	address = {London, N6},
	series = {Amazon in the {Global} {Economy}},
	title = {A {New} {Industrial} {Working} {Class}? {Challenges} in {Disrupting} {Amazon}’s {Fulfillment} {Process} in {Germany}},
	isbn = {978-0-7453-4148-4},
	shorttitle = {A {New} {Industrial} {Working} {Class}?},
	url = {https://www.jstor.org/stable/j.ctv16zjhcj.14},
	doi = {10.2307/j.ctv16zjhcj.14},
	urldate = {2025-03-25},
	booktitle = {The {Cost} of {Free} {Shipping}},
	publisher = {Pluto Press},
	author = {Vgontzas, Nantina},
	editor = {Alimahomed-Wilson, Jake and Reese, Ellen},
	year = {2020},
	pages = {116--128},
}

@book{blanc_we_2025,
	address = {Oakland, CA, USA},
	title = {We {Are} the {Union}: {How} {Worker}-to-{Worker} {Organizing} {Is} {Revitalizing} {Labor} and {Winning} {Big}},
	isbn = {978-0-520-39491-9},
	shorttitle = {We {Are} the {Union}},
	language = {en},
	publisher = {Univ of California Press},
	author = {Blanc, Eric},
	month = feb,
	year = {2025},
	note = {Google-Books-ID: Us0uEQAAQBAJ},
}

@inproceedings{sannon_privacy_2022,
	address = {New York, NY, USA},
	series = {{CHI} '22},
	title = {Privacy, {Surveillance}, and {Power} in the {Gig} {Economy}},
	isbn = {978-1-4503-9157-3},
	url = {https://dl.acm.org/doi/10.1145/3491102.3502083},
	doi = {10.1145/3491102.3502083},
	urldate = {2025-04-28},
	booktitle = {Proceedings of the 2022 {CHI} {Conference} on {Human} {Factors} in {Computing} {Systems}},
	publisher = {Association for Computing Machinery},
	author = {Sannon, Shruti and Sun, Billie and Cosley, Dan},
	month = apr,
	year = {2022},
	pages = {1--15},
}

@article{qadri_whats_2021,
	title = {What's in a {Network}? {Infrastructures} of {Mutual} {Aid} for {Digital} {Platform} {Workers} during {COVID}-19},
	volume = {5},
	shorttitle = {What's in a {Network}?},
	url = {https://dl.acm.org/doi/10.1145/3479563},
	doi = {10.1145/3479563},
	number = {CSCW2},
	urldate = {2025-05-06},
	journal = {Proc. ACM Hum.-Comput. Interact.},
	author = {Qadri, Rida},
	month = oct,
	year = {2021},
	pages = {419:1--419:20},
}

@inproceedings{khovanskaya_bottom-up_2020,
	address = {New York, NY, USA},
	series = {{CHI} '20},
	title = {Bottom-{Up} {Organizing} with {Tools} from {On} {High}: {Understanding} the {Data} {Practices} of {Labor} {Organizers}},
	isbn = {978-1-4503-6708-0},
	shorttitle = {Bottom-{Up} {Organizing} with {Tools} from {On} {High}},
	url = {https://dl.acm.org/doi/10.1145/3313831.3376185},
	doi = {10.1145/3313831.3376185},
	urldate = {2025-04-24},
	booktitle = {Proceedings of the 2020 {CHI} {Conference} on {Human} {Factors} in {Computing} {Systems}},
	publisher = {Association for Computing Machinery},
	author = {Khovanskaya, Vera and Sengers, Phoebe and Dombrowski, Lynn},
	month = apr,
	year = {2020},
	pages = {1--13},
}

@inproceedings{salehi_we_2015,
	address = {New York, NY, USA},
	series = {{CHI} '15},
	title = {We {Are} {Dynamo}: {Overcoming} {Stalling} and {Friction} in {Collective} {Action} for {Crowd} {Workers}},
	isbn = {978-1-4503-3145-6},
	shorttitle = {We {Are} {Dynamo}},
	url = {https://dl.acm.org/doi/10.1145/2702123.2702508},
	doi = {10.1145/2702123.2702508},
	urldate = {2025-04-24},
	booktitle = {Proceedings of the 33rd {Annual} {ACM} {Conference} on {Human} {Factors} in {Computing} {Systems}},
	publisher = {Association for Computing Machinery},
	author = {Salehi, Niloufar and Irani, Lilly C. and Bernstein, Michael S. and Alkhatib, Ali and Ogbe, Eva and Milland, Kristy and Clickhappier},
	month = apr,
	year = {2015},
	pages = {1621--1630},
}

@inproceedings{irani_stories_2016,
	address = {New York, NY, USA},
	series = {{CHI} '16},
	title = {Stories {We} {Tell} {About} {Labor}: {Turkopticon} and the {Trouble} with "{Design}"},
	isbn = {978-1-4503-3362-7},
	shorttitle = {Stories {We} {Tell} {About} {Labor}},
	url = {https://dl.acm.org/doi/10.1145/2858036.2858592},
	doi = {10.1145/2858036.2858592},
	urldate = {2025-04-24},
	booktitle = {Proceedings of the 2016 {CHI} {Conference} on {Human} {Factors} in {Computing} {Systems}},
	publisher = {Association for Computing Machinery},
	author = {Irani, Lilly C. and Silberman, M. Six},
	month = may,
	year = {2016},
	pages = {4573--4586},
}

@inproceedings{ming_labor_2024,
	address = {New York, NY, USA},
	series = {{CSCW} {Companion} '24},
	title = {Labor, {Visibility}, and {Technology}: {Weaving} {Together} {Academic} {Insights} and {On}-{Ground} {Realities}},
	isbn = {979-8-4007-1114-5},
	shorttitle = {Labor, {Visibility}, and {Technology}},
	url = {https://dl.acm.org/doi/10.1145/3678884.3681827},
	doi = {10.1145/3678884.3681827},
	urldate = {2025-04-28},
	booktitle = {Companion {Publication} of the 2024 {Conference} on {Computer}-{Supported} {Cooperative} {Work} and {Social} {Computing}},
	publisher = {Association for Computing Machinery},
	author = {Ming, Joy and Pei, Lucy and Varanasi, Rama Adithya and Kawakami, Anna and Verdezoto, Nervo and Cheon, EunJeong},
	month = nov,
	year = {2024},
	pages = {708--711},
}

@inproceedings{zhang_algorithmic_2022,
	address = {New York, NY, USA},
	series = {{CHI} '22},
	title = {Algorithmic {Management} {Reimagined} {For} {Workers} and {By} {Workers}: {Centering} {Worker} {Well}-{Being} in {Gig} {Work}},
	isbn = {978-1-4503-9157-3},
	shorttitle = {Algorithmic {Management} {Reimagined} {For} {Workers} and {By} {Workers}},
	url = {https://dl.acm.org/doi/10.1145/3491102.3501866},
	doi = {10.1145/3491102.3501866},
	urldate = {2025-04-28},
	booktitle = {Proceedings of the 2022 {CHI} {Conference} on {Human} {Factors} in {Computing} {Systems}},
	publisher = {Association for Computing Machinery},
	author = {Zhang, Angie and Boltz, Alexander and Wang, Chun Wei and Lee, Min Kyung},
	month = apr,
	year = {2022},
	pages = {1--20},
}

@inproceedings{spektor_designing_2023,
	address = {New York, NY, USA},
	series = {{DIS} '23},
	title = {Designing for {Wellbeing}: {Worker}-{Generated} {Ideas} on {Adapting} {Algorithmic} {Management} in the {Hospitality} {Industry}},
	isbn = {978-1-4503-9893-0},
	shorttitle = {Designing for {Wellbeing}},
	url = {https://dl.acm.org/doi/10.1145/3563657.3596018},
	doi = {10.1145/3563657.3596018},
	urldate = {2025-04-28},
	booktitle = {Proceedings of the 2023 {ACM} {Designing} {Interactive} {Systems} {Conference}},
	publisher = {Association for Computing Machinery},
	author = {Spektor, Franchesca and Fox, Sarah E and Awumey, Ezra and Riordan, Christine A. and Rho, Hye Jin and Kulkarni, Chinmay and Martinez-Lopez, Marlen and Stringam, Betsy and Begleiter, Ben and Forlizzi, Jodi},
	month = jul,
	year = {2023},
	pages = {623--637},
}

@inproceedings{kang_stories_2022,
	address = {New York, NY, USA},
	series = {{DIS} '22},
	title = {Stories from the {Frontline}: {Recuperating} {Essential} {Worker} {Accounts} of {AI} {Integration}},
	isbn = {978-1-4503-9358-4},
	shorttitle = {Stories from the {Frontline}},
	url = {https://dl.acm.org/doi/10.1145/3532106.3533564},
	doi = {10.1145/3532106.3533564},
	urldate = {2025-04-28},
	booktitle = {Proceedings of the 2022 {ACM} {Designing} {Interactive} {Systems} {Conference}},
	publisher = {Association for Computing Machinery},
	author = {Kang, Esther Y. and Fox, Sarah E.},
	month = jun,
	year = {2022},
	pages = {58--70},
}

@inproceedings{fox_worker-centered_2020,
	address = {New York, NY, USA},
	series = {{CHI} {EA} '20},
	title = {Worker-{Centered} {Design}: {Expanding} {HCI} {Methods} for {Supporting} {Labor}},
	isbn = {978-1-4503-6819-3},
	shorttitle = {Worker-{Centered} {Design}},
	url = {https://dl.acm.org/doi/10.1145/3334480.3375157},
	doi = {10.1145/3334480.3375157},
	urldate = {2025-04-28},
	booktitle = {Extended {Abstracts} of the 2020 {CHI} {Conference} on {Human} {Factors} in {Computing} {Systems}},
	publisher = {Association for Computing Machinery},
	author = {Fox, Sarah E. and Khovanskaya, Vera and Crivellaro, Clara and Salehi, Niloufar and Dombrowski, Lynn and Kulkarni, Chinmay and Irani, Lilly and Forlizzi, Jodi},
	month = apr,
	year = {2020},
	pages = {1--8},
}

@misc{logan_corporate_2025,
	title = {Corporate union busting in plain sight: {How} {Amazon}, {Starbucks}, and {Trader} {Joe}’s crushed dynamic grassroots worker organizing campaigns},
	shorttitle = {Corporate union busting in plain sight},
	url = {https://www.epi.org/publication/corporate-union-busting/},
	language = {en-US},
	urldate = {2025-03-31},
	journal = {Economic Policy Institute},
	author = {Logan, John},
	month = jan,
	year = {2025},
}

@article{levy_contexts_2015,
	title = {The {Contexts} of {Control}: {Information}, {Power}, and {Truck}-{Driving} {Work}},
	volume = {31},
	issn = {0197-2243},
	shorttitle = {The {Contexts} of {Control}},
	url = {https://doi.org/10.1080/01972243.2015.998105},
	doi = {10.1080/01972243.2015.998105},
	number = {2},
	urldate = {2025-03-27},
	journal = {The Information Society},
	publisher = {Routledge},
	author = {Levy, Karen E. C.},
	month = mar,
	year = {2015},
	note = {\_eprint: https://doi.org/10.1080/01972243.2015.998105},
	pages = {160--174},
}

@article{ajunwa_limitless_2017,
	title = {Limitless {Worker} {Surveillance}},
	volume = {105},
	issn = {0008-1221},
	url = {https://www.jstor.org/stable/44630759},
	number = {3},
	urldate = {2025-03-31},
	journal = {California Law Review},
	publisher = {California Law Review, Inc.},
	author = {Ajunwa, Ifeoma and Crawford, Kate and Schultz, Jason},
	year = {2017},
	pages = {735--776},
}

@misc{palmer_how_2020,
	title = {How {Amazon} keeps a close eye on employee activism to head off unions},
	url = {https://www.cnbc.com/2020/10/24/how-amazon-prevents-unions-by-surveilling-employee-activism.html},
	language = {en},
	urldate = {2025-03-03},
	journal = {CNBC},
	author = {Palmer, Annie},
	month = oct,
	year = {2020},
	note = {Section: Technology},
}

@inproceedings{guberek_keeping_2018,
	address = {New York, NY, USA},
	series = {{CHI} '18},
	title = {Keeping a {Low} {Profile}? {Technology}, {Risk} and {Privacy} among {Undocumented} {Immigrants}},
	isbn = {978-1-4503-5620-6},
	shorttitle = {Keeping a {Low} {Profile}?},
	url = {https://dl.acm.org/doi/10.1145/3173574.3173688},
	doi = {10.1145/3173574.3173688},
	urldate = {2025-02-28},
	booktitle = {Proceedings of the 2018 {CHI} {Conference} on {Human} {Factors} in {Computing} {Systems}},
	publisher = {Association for Computing Machinery},
	author = {Guberek, Tamy and McDonald, Allison and Simioni, Sylvia and Mhaidli, Abraham H. and Toyama, Kentaro and Schaub, Florian},
	month = apr,
	year = {2018},
	pages = {1--15},
}

@misc{calacci_fairfare_2025,
	title = {{FairFare}: {A} {Tool} for {Crowdsourcing} {Rideshare} {Data} to {Empower} {Labor} {Organizers}},
	shorttitle = {{FairFare}},
	url = {http://arxiv.org/abs/2502.11273},
	doi = {10.48550/arXiv.2502.11273},
	urldate = {2025-02-28},
	publisher = {arXiv},
	author = {Calacci, Dana and Rao, Varun Nagaraj and Dalal, Samantha and Di, Catherine and Pua, Kok-Wei and Schwartz, Andrew and Spitzberg, Danny and Monroy-Hernández, Andrés},
	month = feb,
	year = {2025},
	note = {arXiv:2502.11273 [cs]},
}

@misc{brown_hotel_2023,
	title = {Hotel {Workers} {Strike} against {Racist} {Hiring} and a {Scab} {App}},
	url = {https://labornotes.org/2023/08/hotel-workers-strike-against-racist-hiring-and-scab-app},
	language = {en},
	urldate = {2025-02-27},
	journal = {Labor Notes},
	author = {Brown, Jenny},
	month = aug,
	year = {2023},
}

@article{ball_workplace_2010,
	title = {Workplace surveillance: an overview},
	volume = {51},
	issn = {0023-656X},
	shorttitle = {Workplace surveillance},
	url = {https://doi.org/10.1080/00236561003654776},
	doi = {10.1080/00236561003654776},
	number = {1},
	urldate = {2025-02-25},
	journal = {Labor History},
	publisher = {Routledge},
	author = {Ball, Kirstie},
	month = feb,
	year = {2010},
	note = {\_eprint: https://doi.org/10.1080/00236561003654776},
	pages = {87--106},
}

@misc{brandom_amazon_2021,
	title = {Amazon changed traffic light timing during union drive, county officials say},
	url = {https://www.theverge.com/2021/2/17/22287191/amazon-alabama-warehouse-union-traffic-light-change-bessemer},
	language = {en-US},
	urldate = {2025-02-13},
	journal = {The Verge},
	author = {Brandom, Russell},
	month = feb,
	year = {2021},
}

@article{jaffe_itll_2021,
	title = {It’ll {Take} a {Movement}: {Organizing} at {Amazon} after {Bessemer}},
	volume = {30},
	issn = {1095-7960},
	shorttitle = {It’ll {Take} a {Movement}},
	url = {https://doi.org/10.1177/10957960211035077},
	doi = {10.1177/10957960211035077},
	language = {en},
	number = {3},
	urldate = {2025-02-12},
	journal = {New Labor Forum},
	publisher = {SAGE Publications Inc},
	author = {Jaffe, Sarah},
	month = sep,
	year = {2021},
	pages = {30--37},
}

@article{reese_teamsters_2022,
	title = {Teamsters {Confront} {Amazon}: {An} {Early} {Assessment}},
	volume = {31},
	issn = {1095-7960},
	shorttitle = {Teamsters {Confront} {Amazon}},
	url = {https://doi.org/10.1177/10957960221116835},
	doi = {10.1177/10957960221116835},
	language = {en},
	number = {3},
	urldate = {2025-02-12},
	journal = {New Labor Forum},
	publisher = {SAGE Publications Inc},
	author = {Reese, Ellen and Alimahomed-Wilson, Jake},
	month = sep,
	year = {2022},
	pages = {43--51},
}

@article{vgontzas_toward_2023,
	title = {Toward {Realignment}: {Big} {Tech}, {Organized} {Labor}, and the {Politics} of the {Future} of {Work}},
	volume = {48},
	issn = {0160-449X},
	shorttitle = {Toward {Realignment}},
	url = {https://doi.org/10.1177/0160449X231178772},
	doi = {10.1177/0160449X231178772},
	language = {EN},
	number = {3},
	urldate = {2025-03-25},
	journal = {Labor Studies Journal},
	publisher = {SAGE Publications Inc},
	author = {Vgontzas, Nantina},
	month = sep,
	year = {2023},
	pages = {265--275},
}

@article{fuchs_political_2013,
	title = {Political {Economy} and {Surveillance} {Theory}},
	volume = {39},
	issn = {0896-9205},
	url = {https://doi.org/10.1177/0896920511435710},
	doi = {10.1177/0896920511435710},
	language = {en},
	number = {5},
	urldate = {2025-01-17},
	journal = {Critical Sociology},
	publisher = {SAGE Publications Ltd},
	author = {Fuchs, Christian},
	month = sep,
	year = {2013},
	pages = {671--687},
}

@article{clawson_it_2017,
	title = {{IT} {Is} {Watching}: {Workplace} {Surveillance} and {Worker} {Resistance}},
	volume = {26},
	issn = {1095-7960},
	shorttitle = {{IT} {Is} {Watching}},
	url = {https://doi.org/10.1177/1095796017699811},
	doi = {10.1177/1095796017699811},
	language = {en},
	number = {2},
	urldate = {2025-02-25},
	journal = {New Labor Forum},
	publisher = {SAGE Publications Inc},
	author = {Clawson, Dan and Clawson, Mary Ann},
	month = may,
	year = {2017},
	pages = {62--69},
}

@article{logan_crushing_2021,
	title = {Crushing {Unions}, by {Any} {Means} {Necessary}: {How} {Amazon}’s {Blistering} {Anti}-{Union} {Campaign} {Won} in {Bessemer}, {Alabama}},
	volume = {30},
	issn = {1095-7960},
	shorttitle = {Crushing {Unions}, by {Any} {Means} {Necessary}},
	url = {https://doi.org/10.1177/10957960211035082},
	doi = {10.1177/10957960211035082},
	language = {en},
	number = {3},
	urldate = {2025-01-21},
	journal = {New Labor Forum},
	publisher = {SAGE Publications Inc},
	author = {Logan, John},
	month = sep,
	year = {2021},
	pages = {38--45},
}

@article{baptista_digital_2020,
	series = {Strategic {Perspectives} on {Digital} {Work} and {Organizational} {Transformation}},
	title = {Digital work and organisational transformation: {Emergent} {Digital}/{Human} work configurations in modern organisations},
	volume = {29},
	issn = {0963-8687},
	shorttitle = {Digital work and organisational transformation},
	url = {https://www.sciencedirect.com/science/article/pii/S0963868720300263},
	doi = {10.1016/j.jsis.2020.101618},
	number = {2},
	urldate = {2025-02-07},
	journal = {The Journal of Strategic Information Systems},
	author = {Baptista, João and Stein, Mari-Klara and Klein, Stefan and Watson-Manheim, Mary Beth and Lee, Jungwoo},
	month = jun,
	year = {2020},
	pages = {101618},
}

@article{marsh_digital_2022,
	title = {The digital workplace and its dark side: {An} integrative review},
	volume = {128},
	issn = {0747-5632},
	shorttitle = {The digital workplace and its dark side},
	url = {https://www.sciencedirect.com/science/article/pii/S0747563221004416},
	doi = {10.1016/j.chb.2021.107118},
	urldate = {2025-02-07},
	journal = {Computers in Human Behavior},
	author = {Marsh, Elizabeth and Vallejos, Elvira Perez and Spence, Alexa},
	month = mar,
	year = {2022},
	pages = {107118},
}

@inproceedings{ekbia_social_2016,
	address = {New York, NY, USA},
	series = {{CHI} '16},
	title = {Social {Inequality} and {HCI}: {The} {View} from {Political} {Economy}},
	isbn = {978-1-4503-3362-7},
	shorttitle = {Social {Inequality} and {HCI}},
	url = {https://dl.acm.org/doi/10.1145/2858036.2858343},
	doi = {10.1145/2858036.2858343},
	urldate = {2025-01-28},
	booktitle = {Proceedings of the 2016 {CHI} {Conference} on {Human} {Factors} in {Computing} {Systems}},
	publisher = {Association for Computing Machinery},
	author = {Ekbia, Hamid and Nardi, Bonnie},
	month = may,
	year = {2016},
	pages = {4997--5002},
}

@misc{ghaffary_real_2020,
	title = {The real cost of {Amazon}},
	url = {https://www.vox.com/recode/2020/6/29/21303643/amazon-coronavirus-warehouse-workers-protest-jeff-bezos-chris-smalls-boycott-pandemic},
	language = {en-US},
	urldate = {2025-01-28},
	journal = {Vox},
	author = {Ghaffary, Shirin},
	month = jun,
	year = {2020},
}

@misc{peterson_amazon-owned_2020,
	title = {Amazon-owned {Whole} {Foods} is quietly tracking its employees with a heat map tool that ranks which stores are most at risk of unionizing},
	url = {https://www.businessinsider.com/whole-foods-tracks-unionization-risk-with-heat-map-2020-1},
	language = {en-US},
	urldate = {2025-02-05},
	journal = {Business Insider},
	author = {Peterson, Hayley},
	month = apr,
	year = {2020},
}

@misc{kessler_companies_2020,
	title = {Companies {Are} {Using} {Employee} {Survey} {Data} to {Predict} — and {Squash} — {Union} {Organizing}},
	url = {https://onezero.medium.com/companies-are-using-employee-survey-data-to-predict-and-squash-union-organizing-a7e28a8c2158},
	language = {en},
	urldate = {2025-01-21},
	journal = {OneZero},
	author = {Kessler, Sarah},
	month = jul,
	year = {2020},
}

@article{mackenzie_marx_1984,
	title = {Marx and the {Machine}},
	volume = {25},
	issn = {0040-165X},
	url = {https://www.jstor.org/stable/3104202},
	doi = {10.2307/3104202},
	number = {3},
	urldate = {2025-01-21},
	journal = {Technology and Culture},
	publisher = {[The Johns Hopkins University Press, Society for the History of Technology]},
	author = {MacKenzie, Donald},
	year = {1984},
	pages = {473--502},
}

@misc{nlrb_office_of_public_affairs_union_2024,
	title = {Union {Petitions} {Filed} with {NLRB} {Double} {Since} {FY} 2021, {Up} 27\% {Since} {FY} 2023 {\textbar} {National} {Labor} {Relations} {Board}},
	url = {https://www.nlrb.gov/news-outreach/news-story/union-petitions-filed-with-nlrb-double-since-fy-2021-up-27-since-fy-2023},
	language = {en},
	urldate = {2025-01-21},
	author = {NLRB Office of Public Affairs},
	month = oct,
	year = {2024},
}

@misc{inc_labor_2024,
	title = {Labor {Unions}},
	url = {https://news.gallup.com/poll/12751/Labor-Unions.aspx},
	language = {en},
	urldate = {2025-01-21},
	journal = {Gallup.com},
	author = {Inc, Gallup},
	year = {2024},
	note = {Section: In Depth: Topics A to Z},
}

@article{kochan_worker_2019,
	title = {Worker {Voice} in {America}: {Is} {There} a {Gap} between {What} {Workers} {Expect} and {What} {They} {Experience}?},
	volume = {72},
	issn = {0019-7939},
	shorttitle = {Worker {Voice} in {America}},
	url = {https://doi.org/10.1177/0019793918806250},
	doi = {10.1177/0019793918806250},
	language = {en},
	number = {1},
	urldate = {2025-01-21},
	journal = {ILR Review},
	publisher = {SAGE Publications Inc},
	author = {Kochan, Thomas A. and Yang, Duanyi and Kimball, William T. and Kelly, Erin L.},
	month = jan,
	year = {2019},
	pages = {3--38},
}

@misc{sum_its_2024,
	title = {"{It}'s {Always} a {Losing} {Game}": {How} {Workers} {Understand} and {Resist} {Surveillance} {Technologies} on the {Job}},
	shorttitle = {"{It}'s {Always} a {Losing} {Game}"},
	url = {http://arxiv.org/abs/2412.06945},
	doi = {10.48550/arXiv.2412.06945},
	urldate = {2025-01-14},
	publisher = {arXiv},
	author = {Sum, Cella M. and Shi, Caroline and Fox, Sarah E.},
	month = dec,
	year = {2024},
	note = {arXiv:2412.06945 [cs]},
}

@article{bernhardt_data-driven_2023,
	title = {The {Data}-{Driven} {Workplace} and the {Case} for {Worker} {Technology} {Rights}},
	volume = {76},
	issn = {0019-7939},
	url = {https://doi.org/10.1177/00197939221131558},
	doi = {10.1177/00197939221131558},
	language = {en},
	number = {1},
	urldate = {2024-12-17},
	journal = {ILR Review},
	publisher = {SAGE Publications Inc},
	author = {Bernhardt, Annette and Kresge, Lisa and Suleiman, Reem},
	month = jan,
	year = {2023},
	pages = {3--29},
}

@inproceedings{vincent_data_2021,
	address = {New York, NY, USA},
	series = {{FAccT} '21},
	title = {Data {Leverage}: {A} {Framework} for {Empowering} the {Public} in its {Relationship} with {Technology} {Companies}},
	isbn = {978-1-4503-8309-7},
	shorttitle = {Data {Leverage}},
	url = {https://dl.acm.org/doi/10.1145/3442188.3445885},
	doi = {10.1145/3442188.3445885},
	urldate = {2024-12-11},
	booktitle = {Proceedings of the 2021 {ACM} {Conference} on {Fairness}, {Accountability}, and {Transparency}},
	publisher = {Association for Computing Machinery},
	author = {Vincent, Nicholas and Li, Hanlin and Tilly, Nicole and Chancellor, Stevie and Hecht, Brent},
	month = mar,
	year = {2021},
	pages = {215--227},
}

@article{doorn_at_2020,
	title = {At what price? {Labour} politics and calculative power struggles in on-demand food delivery},
	volume = {14},
	issn = {1745-641X},
	shorttitle = {At what price?},
	url = {https://www.scienceopen.com/hosted-document?doi=10.13169/workorgalaboglob.14.1.0136},
	doi = {10.13169/workorgalaboglob.14.1.0136},
	urldate = {2024-07-10},
	journal = {Work Organisation, Labour \& Globalisation},
	publisher = {Pluto Journals},
	author = {Doorn, Niels van},
	month = jan,
	year = {2020},
	pages = {136--149},
}

@inproceedings{irani_turkopticon_2013,
	address = {New York, NY, USA},
	series = {{CHI} '13},
	title = {Turkopticon: interrupting worker invisibility in amazon mechanical turk},
	isbn = {978-1-4503-1899-0},
	shorttitle = {Turkopticon},
	url = {https://dl.acm.org/doi/10.1145/2470654.2470742},
	doi = {10.1145/2470654.2470742},
	urldate = {2024-09-23},
	booktitle = {Proceedings of the {SIGCHI} {Conference} on {Human} {Factors} in {Computing} {Systems}},
	publisher = {Association for Computing Machinery},
	author = {Irani, Lilly C. and Silberman, M. Six},
	month = apr,
	year = {2013},
	pages = {611--620},
}

@article{khovanskaya_tools_2019,
	title = {The {Tools} of {Management}: {Adapting} {Historical} {Union} {Tactics} to {Platform}-{Mediated} {Labor}},
	volume = {3},
	shorttitle = {The {Tools} of {Management}},
	url = {https://doi.org/10.1145/3359310},
	doi = {10.1145/3359310},
	number = {CSCW},
	urldate = {2024-07-10},
	journal = {Proc. ACM Hum.-Comput. Interact.},
	author = {Khovanskaya, Vera and Dombrowski, Lynn and Rzeszotarski, Jeffrey and Sengers, Phoebe},
	month = nov,
	year = {2019},
	pages = {208:1--208:22},
}

@article{kapoor_weaving_2022,
	title = {Weaving {Privacy} and {Power}: {On} the {Privacy} {Practices} of {Labor} {Organizers} in the {U}.{S}. {Technology} {Industry}},
	volume = {6},
	shorttitle = {Weaving {Privacy} and {Power}},
	url = {https://doi.org/10.1145/3555574},
	doi = {10.1145/3555574},
	number = {CSCW2},
	urldate = {2024-07-10},
	journal = {Proc. ACM Hum.-Comput. Interact.},
	author = {Kapoor, Sayash and Sun, Matthew and Wang, Mona and Jazwinska, Klaudia and Watkins, Elizabeth Anne},
	month = nov,
	year = {2022},
	pages = {473:1--473:33},
}

@misc{hanley_eyes_2021,
	address = {Rochester, NY},
	type = {{SSRN} {Scholarly} {Paper}},
	title = {Eyes {Everywhere}: {Amazon}'s {Surveillance} {Infrastructure} and {Revitalizing} {Worker} {Power} - {An} {Update}},
	shorttitle = {Eyes {Everywhere}},
	url = {https://papers.ssrn.com/abstract=4089859},
	language = {en},
	urldate = {2024-07-10},
	author = {Hanley, Daniel},
	month = sep,
	year = {2021},
}

@article{nissim_future_2021,
	title = {The future of labor unions in the age of automation and at the dawn of {AI}},
	volume = {67},
	issn = {0160-791X},
	url = {https://www.sciencedirect.com/science/article/pii/S0160791X21002074},
	doi = {10.1016/j.techsoc.2021.101732},
	urldate = {2024-07-10},
	journal = {Technology in Society},
	author = {Nissim, Gadi and Simon, Tomer},
	month = nov,
	year = {2021},
	pages = {101732},
}

@inproceedings{lee_working_2015,
	address = {Seoul Republic of Korea},
	title = {Working with {Machines}: {The} {Impact} of {Algorithmic} and {Data}-{Driven} {Management} on {Human} {Workers}},
	isbn = {978-1-4503-3145-6},
	shorttitle = {Working with {Machines}},
	url = {https://dl.acm.org/doi/10.1145/2702123.2702548},
	doi = {10.1145/2702123.2702548},
	language = {en},
	urldate = {2024-07-10},
	booktitle = {Proceedings of the 33rd {Annual} {ACM} {Conference} on {Human} {Factors} in {Computing} {Systems}},
	publisher = {ACM},
	author = {Lee, Min Kyung and Kusbit, Daniel and Metsky, Evan and Dabbish, Laura},
	month = apr,
	year = {2015},
	pages = {1603--1612},
}


\newpage
\appendix
\sloppy
\section{Corpus Documents}
\subsection{Amazon}
\begin{enumerate}
\item AFL-CIO. \textit{How the PRO Act Would’ve Changed the Amazon Organizing Landscape}. Accessed: September 3, 2025. \url{https://aflcio.org/how-pro-act-wouldve-changed-amazon-organizing-landscape}

\item Alec MacGillis. \textit{Lessons From Bessemer: What Amazon’s Union Defeat Means for the American Labor Movement}. ProPublica. Accessed: September 3, 2025. \url{https://www.propublica.org/article/lessons-from-bessemer-what-amazons-union-defeat-means-for-the-american-labor-movement}

\item Alex N. Press. \textit{A Stunning New Chapter Begins for Amazon Warehouse Workers}. Jacobin. Accessed: September 3, 2025. \url{https://jacobin.com/2022/04/amazon-labor-union-victory-jfk8-staten-island-bessemer}

\item Alex N. Press. \textit{Amazon Workers Just Suffered a Defeat. But the Fight Is Far From Over.} Jacobin. Accessed: September 3, 2025. \url{https://jacobin.com/2022/05/amazon-staten-island-ldj5-sorting-center-union-vote-loss}

\item Alex N. Press. \textit{The Beating Heart of the US Labor Movement Was at Labor Notes}. Jacobin. Accessed: September 3, 2025. \url{https://jacobin.com/2022/06/labor-notes-unions-starbucks-amazon-teamsters}

\item Alex N. Press. \textit{The Union-Busting Crime Wave at Starbucks and Amazon Is Getting Worse}. Jacobin. Accessed: September 3, 2025. \url{https://jacobin.com/2022/05/union-busting-amazon-starbucks-alu-swu-organizing-nlrb}

\item Alina Selyukh. \textit{It’s A No: Amazon Warehouse Workers Vote Against Unionizing In Historic Election}. NPR. Accessed: September 3, 2025. \url{https://www.npr.org/2021/04/09/982139494/its-a-no-amazon-warehouse-workers-vote-against-unionizing-in-historic-election}

\item Alina Selyukh. \textit{Amazon Ordered to Let Workers Vote on Unionizing -- For the 3rd Time}. NPR. Accessed: September 3, 2025. \url{https://www.npr.org/2024/11/06/nx-s1-5087567/amazon-union-warehouse-bessemer-alabama}

\item Alvin Gaine, Brendan Radtke, Dylan D. Maraj, Ira Pollock, Luc Rene, and Ben Mabie. \textit{Advanced Detachments: Two Strikes Against Competition and Division at an Amazon Delivery Station}. Long-Haul Mag. Accessed: September 3, 2025. \url{https://longhaulmag.com/advanced-detachments-two-strikes/}

\item Annie Palmer. \textit{Amazon Moves Closer to Facing Its First Unionization Vote in Six Years}. CNBC. Accessed: September 3, 2025. \url{https://www.cnbc.com/2020/12/22/amazon-moves-closer-to-facing-its-first-unionization-vote-in-six-years.html}

\item Annie Palmer. \textit{How Amazon Keeps a Close Eye on Employee Activism to Head Off Unions}. CNBC. Accessed: September 3, 2025. \url{https://www.cnbc.com/2020/10/24/how-amazon-prevents-unions-by-surveilling-employee-activism.html}

\item Brad Porter. \textit{Response to Tim Bray’s Departure}. LinkedIn. Accessed: September 3, 2025. \url{https://www.linkedin.com/pulse/response-tim-brays-departure-brad-porter}

\item Brandon Magner. \textit{If the PRO Act Were Law, The Amazon Union Election Would Have Looked Very Different}. Labor Notes. Accessed: September 3, 2025. \url{https://labornotes.org/blogs/2021/03/if-pro-act-were-law-amazon-union-election-would-have-looked-very-different}

\item Caitlyn Clark. \textit{Countering Dangerous Work at Amazon}. Labor Notes. Accessed: September 3, 2025. \url{https://labornotes.org/2023/09/countering-dangerous-work-amazon}

\item Colin Lecher. \textit{How Amazon Automatically Tracks and Fires Warehouse Workers for ‘Productivity’}. The Verge. Accessed: September 3, 2025. \url{https://www.theverge.com/2019/4/25/18516004/amazon-warehouse-fulfillment-centers-productivity-firing-terminations}

\item Dave Jamieson. \textit{For The Starbucks Union Campaign, A Bruising Contract Fight Is Just Beginning}. HuffPost. Accessed: September 3, 2025. \url{https://www.huffpost.com/entry/starbucks-union-campaign-contract-fight_n_6259c871e4b066ecde1350f0}

\item Debbie Hanley and Samantha Hubbard. \textit{Eyes Everywhere: Amazon’s Surveillance Infrastructure and Revitalizing Worker Power}. Open Markets Institute, SSRN Scholarly Paper 4089862, Rochester, NY, September 2020. Accessed: September 3, 2025. \url{https://papers.ssrn.com/abstract=4089862}

\item Ellen Reese and Jake Alimahomed-Wilson. \textit{Teamsters Confront Amazon: An Early Assessment}. \textit{New Labor Forum}, vol. 31, no. 3, pp. 43--51, September 2022. \url{10.1177/10957960221116835}

\item Hayley Peterson. \textit{Amazon-owned Whole Foods Is Quietly Tracking Its Employees with a Heat Map Tool That Ranks Which Stores Are Most at Risk of Unionizing}. Business Insider. Accessed: September 3, 2025. \url{https://www.businessinsider.com/whole-foods-tracks-unionization-risk-with-heat-map-2020-1}

\item Jay Greene. \textit{Amazon Fights Aggressively to Defeat Union Drive in Alabama, Fearing a Coming Wave}. The Washington Post. Accessed: September 3, 2025. \url{https://www.washingtonpost.com/technology/2021/03/09/amazon-union-bessemer-history/}

\item Jason Del Rey and Shirin Ghaffary. \textit{Leaked: Confidential Amazon Memo Reveals New Software to Track Unions}. Vox. Accessed: September 3, 2025. \url{https://www.vox.com/recode/2020/10/6/21502639/amazon-union-busting-tracking-memo-spoc}

\item Jason Del Rey and Shirin Ghaffary. \textit{Amazon White-Collar Employees Are Fuming Over Management Targeting a Fired Warehouse Worker}. Vox. Accessed: September 3, 2025. \url{https://www.vox.com/recode/2020/4/5/21206385/amazon-fired-warehouse-worker-christian-smalls-employee-backlash-david-zapolsky-coronavirus}

\item Jeffrey Dastin. \textit{Amazon Warehouse Workers Protest Near Detroit, Days After NYC Walkout}. Reuters. Accessed: September 3, 2025. \url{https://www.reuters.com/article/business/amazon-warehouse-workers-protest-near-detroit-days-after-nyc-walkout-idUSKBN21J6IW/}

\item Joe DeManuelle-Hall. \textit{The BAmazon Loss and the Road Ahead}. Labor Notes. Accessed: September 3, 2025. \url{https://labornotes.org/2021/04/reflections-bamazon-loss-and-road-ahead}

\item John Logan. \textit{Corporate Union Busting in Plain Sight: How Amazon, Starbucks, and Trader Joe’s Crushed Dynamic Grassroots Worker Organizing Campaigns}. Economic Policy Institute. Accessed: September 3, 2025. \url{https://www.epi.org/publication/corporate-union-busting/}

\item John Logan. \textit{Crushing Unions, by Any Means Necessary: How Amazon’s Blistering Anti-Union Campaign Won in Bessemer, Alabama}. \textit{New Labor Forum}, vol. 30, no. 3, pp. 38--45, September 2021. \url{10.1177/10957960211035082}

\item Josh Dzieza. \textit{Robots Aren’t Taking Our Jobs — They’re Becoming Our Bosses}. The Verge. Accessed: September 3, 2025. \url{https://www.theverge.com/2020/2/27/21155254/automation-robots-unemployment-jobs-vs-human-google-amazon}

\item Josh Sherer. \textit{The Future of Work Depends on Stopping Amazon’s Union Busting: Shareholders and Policymakers Must All Play a Role in Protecting Amazon Workers’ Rights}. Economic Policy Institute. Accessed: September 3, 2025. \url{https://www.epi.org/blog/the-future-of-work-depends-on-stopping-amazons-union-busting-shareholders-and-policymakers-must-all-play-a-role-in-protecting-amazon-workers-rights/}

\item Labor Notes. \textit{Viewpoint: There’s More to Amazon Organizing Strategy Than Choke Points}. Accessed: September 3, 2025. \url{https://labornotes.org/blogs/2025/02/viewpoint-theres-more-strategy-choke-points}

\item Labor Notes. \textit{Amazon Stokes Racial Divides in Lead-Up to North Carolina Union Vote}. Accessed: September 3, 2025. \url{https://labornotes.org/2025/02/amazon-stokes-racial-divides-lead-north-carolina-union-vote}

\item LaborLab. \textit{Union-Busting Without Consequences: Amazon Thrives in Trump’s Anti-Worker America}. Accessed: September 3, 2025. \url{https://laborlab.us/union-busting-without-consequences-amazon-thrives-in-trumps-anti-worker-america/}

\item Lauren Kaori Gurley. \textit{Hundreds of Amazon Employees Are Out on COVID Leave at a Single Warehouse}. VICE. Accessed: September 3, 2025. \url{https://www.vice.com/en/article/hundreds-of-amazon-employees-are-out-on-covid-leave-at-a-single-warehouse/}

\item Lauren Kaori Gurley. \textit{Amazon Cracks Down on Organizing After Historic Union Win}. VICE. Accessed: September 3, 2025. \url{https://www.vice.com/en/article/amazon-cracks-down-on-organizing-after-historic-union-win/}

\item Lauren Kaori Gurley and Joseph Cox. \textit{Inside Amazon’s Secret Program to Spy On Workers’ Private Facebook Groups}. VICE. Accessed: September 3, 2025. \url{https://www.vice.com/en/article/amazon-is-spying-on-its-workers-in-closed-facebook-groups-internal-reports-show/}

\item Louise Matsakis. \textit{Amazon Came to the Bargaining Table—But Workers Want More}. Wired. Accessed: September 3, 2025. \url{https://www.wired.com/story/amazon-labor-protests-minnesota-nyc/}

\item Luis Feliz Leon. \textit{Amazon Workers on Staten Island Clinch a Historic Victory}. Labor Notes. Accessed: September 3, 2025. \url{https://labornotes.org/2022/04/amazon-workers-staten-island-clinch-historic-victory}

\item Lynn Rhinehart. \textit{How Amazon Gerrymandered the Union Vote—and Won}. Economic Policy Institute. Accessed: September 3, 2025. \url{https://www.epi.org/blog/how-amazon-gerrymandered-the-union-vote-and-won/}

\item Matthew Cunningham-Cook and Julia Rock. \textit{Amazon’s Union Busting Is Subsidized by the Government}. Jacobin. Accessed: September 3, 2025. \url{https://jacobin.com/2022/04/amazon-union-busting-corporate-subsidies-government-state-labor-law}

\item Natascha Elena Uhlmann. \textit{Amazon Workers Launch Largest Strike Yet: ‘It Doesn’t Feel Like a Job That Should Be Legal’}. Labor Notes. Accessed: September 3, 2025. \url{https://labornotes.org/2024/12/amazon-workers-launch-largest-strike-yet-it-doesnt-feel-job-should-be-legal}

\item Nikolas Vgontzas. \textit{A New Industrial Working Class? Challenges in Disrupting Amazon’s Fulfillment Process in Germany}. In J. Alimahomed-Wilson and E. Reese (Eds.), \textit{The Cost of Free Shipping: Amazon in the Global Economy}, Pluto Press, 2020, pp. 116--128. \url{10.2307/j.ctv16zjhcj.14}

\item Noam Scheiber. \textit{Amazon Is Everywhere. That’s What Makes It So Vulnerable.} The New York Times. Accessed: September 3, 2025. \url{https://www.nytimes.com/2023/05/19/business/amazon-union-choke-points.html}

\item Noam Scheiber and Santul Nerkar. \textit{Amazon Delivery Drivers at Seven Hubs Walk Out}. The New York Times. Accessed: September 3, 2025. \url{https://www.nytimes.com/2024/12/19/business/economy/amazon-teamsters-strike.html}

\item Orin Starn. \textit{Inside Amazon’s Union-Busting Tactics}. Sapiens. Accessed: September 3, 2025. \url{https://www.sapiens.org/culture/amazon-union-busting-anthropologist/}

\item Paul Blest. \textit{Leaked Amazon Memo Details Plan to Smear Fired Warehouse Organizer: ‘He’s Not Smart or Articulate’}. VICE. Accessed: September 3, 2025. \url{https://www.vice.com/en/article/leaked-amazon-memo-details-plan-to-smear-fired-warehouse-organizer-hes-not-smart-or-articulate/}

\item Peter Olney and Rand Wilson. \textit{BAmazon Union: Anticipating the Battle in Bessemer, Alabama}. Labor Notes. Accessed: September 3, 2025. \url{https://labornotes.org/2020/12/bamazon-union-anticipating-battle-bessemer-alabama}

\item Russell Brandom. \textit{Amazon Changed Traffic Light Timing During Union Drive, County Officials Say}. The Verge. Accessed: September 4, 2025. \url{https://www.theverge.com/2021/2/17/22287191/amazon-alabama-warehouse-union-traffic-light-change-bessemer}

\item Sam Gindin. \textit{Was the Teamsters’ Amazon Strike a Success?} Jacobin. Accessed: September 3, 2025. \url{https://jacobin.com/2025/03/teamsters-amazon-strike-union-strategy}

\item Santul Nerkar and Noam Scheiber. \textit{Amazon Warehouse Workers in New York City Join Protest}. The New York Times. Accessed: September 3, 2025. \url{https://www.nytimes.com/2024/12/21/business/amazon-warehouse-staten-island-strike.html}

\item Sarah Jaffe. \textit{It’ll Take a Movement: Organizing at Amazon After Bessemer}. \textit{New Labor Forum}, vol. 30, no. 3, September 2021. Accessed: September 4, 2025.

\item Shirin Ghaffary. \textit{Amazon Warehouse Workers urlng ‘Back-Breaking’ Work Walked Off the Job in Protest}. Vox. Accessed: September 3, 2025. \url{https://www.vox.com/recode/2019/12/10/21005098/amazon-warehouse-workers-sacramento}

\item Shirin Ghaffary and Jason Del Rey. \textit{The Real Cost of Amazon}. Vox. Accessed: September 3, 2025. \url{https://www.vox.com/recode/2020/6/29/21303643/amazon-coronavirus-warehouse-workers-protest-jeff-bezos-chris-smalls-boycott-pandemic}

\item Stefano Perago. \textit{Embracing Automation: AI, Robotics, and the Future of Work}. Accessed: September 4, 2025. \url{https://www.aboutamazon.eu/embracing-automation-ai-robotics-and-the-future-of-work}

\item Tamara Dowell and Josh Crowell. \textit{Amazon Goes into Union-Busting Overdrive to Fight Campaign at KCVG Air Hub}. Labor Notes. Accessed: September 3, 2025. \url{https://labornotes.org/blogs/2023/12/amazon-goes-union-busting-overdrive-fight-campaign-kcvg-air-hub}

\item The Economist. \textit{What Happens When Amazon Comes to Town}. Accessed: September 3, 2025. \url{https://www.economist.com/united-states/what-happens-when-amazon-comes-to-town/21808308}

\item Thomas Wiggin. \textit{Weaponizing the Workplace: How Algorithmic Management Shaped Amazon’s Antiunion Campaign in Bessemer, Alabama}. Socius, 11, 23780231251318389. August 2025. \url{10.1177/23780231251318389}

\item Tim Bray. \textit{Bye, Amazon}. Ongoing by Tim Bray. Accessed: September 3, 2025. \url{https://www.tbray.org/ongoing/When/202x/2020/04/29/Leaving-Amazon}

\item Toni Gilpin. \textit{Lessons from Labor History for Organizing Amazon}. Labor Notes. Accessed: September 3, 2025. \url{https://labornotes.org/blogs/2021/04/lessons-labor-history-organizing-amazon}

\item Will Evans. \textit{Ruthless Quotas at Amazon Are Maiming Employees}. The Atlantic. Accessed: September 3, 2025. \url{https://www.theatlantic.com/technology/archive/2019/11/amazon-warehouse-reports-show-worker-injuries/602530/}

\end{enumerate}

\subsection{Starbucks}
\begin{enumerate}
\item Alex N. Press. \textit{Starbucks Workers in Ithaca Say the Company Is Trying to Crush Them}. Jacobin. Accessed: September 3, 2025. \url{https://jacobin.com/2022/07/starbucks-workers-ithaca-new-york-union-busting-unfair-labor-practice-law}

\item Alex N. Press. \textit{The Starbucks Workers’ Union Has Finally Broken Through}. Jacobin. Accessed: September 3, 2025. \url{https://jacobin.com/2024/02/starbucks-workers-united-master-contract-bargaining}

\item Alina Selyukh. \textit{Starbucks union push spreads to 54 stores in 19 states}. NPR. Accessed: September 3, 2025. \url{https://www.npr.org/2022/01/31/1076978207/starbucks-union-push-spreads-to-54-stores-in-19-states}

\item Amelia Lucas and Kate Rogers. \textit{Starbucks will have at least one unionized cafe in Buffalo, New York — a U.S. first for the chain}. CNBC. Accessed: September 3, 2025. \url{https://www.cnbc.com/2021/12/09/starbucks-employees-at-a-buffalo-cafe-vote-to-unionize-a-first-for-the-coffee-chain-in-the-us.html}

\item Daniel Denvir, Christian Smalls, and Jaz Brisack. \textit{We’re Organizing Unions at Amazon and Starbucks. We Won’t Back Down.}. Jacobin. Accessed: September 3, 2025. \url{https://jacobin.com/2022/05/amazon-starbucks-labor-union-busting-nlrb}

\item Daniel Wiessner. \textit{Starbucks closed 23 stores to deter unionizing, US agency says}. Reuters. Accessed: September 3, 2025. \url{https://www.reuters.com/business/starbucks-closed-23-us-stores-deter-unionizing-agency-claims-2023-12-14/}

\item Danielle Wiener-Bronner. \textit{Starbucks union organizers wanted credit-card tipping. Now they’re being left out | CNN Business}. CNN. Accessed: September 3, 2025. \url{https://www.cnn.com/2022/12/16/business/starbucks-tipping}

\item Dave Jamieson. \textit{For The Starbucks Union Campaign, A Bruising Contract Fight Is Just Beginning}. HuffPost. Accessed: September 3, 2025. \url{https://www.huffpost.com/entry/starbucks-union-campaign-contract-fight_n_6259c871e4b066ecde1350f0}

\item Edward Ongweso Jr. \textit{Starbucks Is ‘Concerned’ White House Met With Starbucks Union}. VICE. Accessed: September 3, 2025. \url{https://www.vice.com/en/article/starbucks-is-concerned-white-house-met-with-starbucks-union/}

\item Faith Bennett. \textit{Starbucks Workers Are Organizing — and Management Is Worried}. Jacobin. Accessed: September 3, 2025. \url{https://jacobin.com/2021/10/starbucks-workers-united-buffalo-union-drive-organizing-coffee-shop-industry-labor}

\item Gino Canella. \textit{Networked Unionism: Reframing the Labour Movement and Starbucks Workers United’s Hybrid Organising Practices}. tripleC: Communication, Capitalism \& Critique, vol. 21, no. 1, pp. 1–17, Jan. 2023. 10.31269/triplec.v21i1.1358.

\item Hamilton Nolan. \textit{Buffalo Starbucks Workers Say They Will Unionize One Store At a Time}. In These Times. Accessed: September 3, 2025. \url{https://inthesetimes.com/article/buffalo-starbucks-workers-union-labor-india-walton-colectivo}

\item Jackie Kent. \textit{Starbucks workers hold picket line at Reserve Roastery, accuse company of union busting}. KOMO. Accessed: September 3, 2025. \url{https://komonews.com/news/local/starbucks-workers-hold-picket-line-at-reserve-roastery-accuse-company-of-union-busting}

\item Jenny Brown. \textit{Salts and Peppers Build a Union at Starbucks}. Labor Notes. Accessed: September 3, 2025. \url{https://www.labornotes.org/blogs/2025/08/salts-and-peppers-build-union-starbucks}

\item Joe Hernandez. \textit{NLRB sues Starbucks for retaliating against 3 workers involved in unionizing}. NPR. Accessed: September 3, 2025. \url{https://www.npr.org/2022/04/24/1094569620/nlrb-sues-starbucks-for-retaliating-against-3-workers-involved-in-unionizing}

\item Jodi Kantor. \textit{Working Anything but 9 to 5}. The New York Times. Accessed: September 3, 2025. \url{https://www.nytimes.com/interactive/2014/08/13/us/starbucks-workers-scheduling-hours.html}

\item John Logan. \textit{Starbucks’ Howard Schultz Should Win an Award for the Nation’s Most Flagrant Union Buster}. Jacobin. Accessed: September 3, 2025. \url{https://jacobin.com/2022/07/starbucks-howard-schultz-union-busting-firings-store-closures}

\item John Logan. \textit{Starbucks Is Breaking Ground as One of the Worst Union Busters in Recent Memory}. Jacobin. Accessed: September 3, 2025. \url{https://jacobin.com/2022/06/starbucks-union-busting-seattle-pike-place-stores}

\item John Logan. \textit{Starbucks Is Slowly Strangling Its Pro-Union Workers. Where Is ‘Pro-Union’ President Joe Biden?}. Jacobin. Accessed: September 3, 2025. \url{https://jacobin.com/2022/07/starbucks-strangling-pro-union-workers-joe-biden-howard-schultz-nlrb}

\item John Logan. \textit{Starbucks Is the Country’s Worst Labor Law Violator. Joe Biden Should Rein It In.}. Jacobin. Accessed: September 3, 2025. \url{https://jacobin.com/2022/05/starbucks-nlrb-biden-illegal-union-busting/}

\item John Logan. \textit{Starbucks’ caffeinated anti-union efforts may leave a bitter taste – but are they legal?}. The Conversation. Accessed: September 3, 2025. \url{http://theconversation.com/starbucks-caffeinated-anti-union-efforts-may-leave-a-bitter-taste-but-are-they-legal-182549}

\item John Logan. \textit{Starbucks’s Latest Union-Busting Tactic: Demand the Suspension of NLRB Elections Nationwide}. Jacobin. Accessed: September 3, 2025. \url{https://jacobin.com/2022/08/starbucks-union-busting-national-labor-relations-board-union-election-suspension}

\item Jordan Zakarin. \textit{EXCLUSIVE: Starbucks Illegally Withheld Raises \& Tips from Union Workers, NLRB Says}. More Perfect Union. Accessed: September 3, 2025. \url{https://perfectunion.us/starbucks-nlrb-credit-card-tip-complaint/}

\item Khaleda Rahman. \textit{Gen-Z Activists Flood Starbucks With Fake Job Applications Over Firings}. Newsweek. Accessed: September 3, 2025. \url{https://www.newsweek.com/gen-z-activists-flood-starbucks-fake-job-applications-over-firings-1681786}

\item Lauren Kaori Gurley. \textit{Starbucks Temporarily Closes 2 Stores That Are Trying to Unionize}. VICE. Accessed: September 4, 2025. \url{https://www.vice.com/en/article/starbucks-temporarily-closes-two-stores-that-are-trying-to-unionize/}

\item Lauren Kaori Gurley. \textit{The Starbucks Unionization Effort Is Now An ‘Unprecedented’ Nationwide Movement}. VICE. Accessed: September 3, 2025. \url{https://www.vice.com/en/article/the-starbucks-unionization-effort-is-now-an-unprecedented-nationwide-movement/}

\item Lauren Kaori Gurley. \textit{Starbucks Is Blocking Union Activists and Workers on Twitter}. VICE. Accessed: September 3, 2025. \url{https://www.vice.com/en/article/starbucks-is-blocking-union-activists-and-workers-on-twitter/}

\item Lauren Kaori Gurley. \textit{‘It’s Almost Comical:’ Starbucks Is Blatantly Trying to Crush Its Union}. VICE. Accessed: September 3, 2025. \url{https://www.vice.com/en/article/its-almost-comical-starbucks-is-blatantly-trying-to-crush-its-union/}

\item Lauren Kaori Gurley. \textit{Starbucks Baristas Declare Victory in Historic Union Election}. VICE. Accessed: September 3, 2025. \url{https://www.vice.com/en/article/buffalo-starbucks-union-election-victory/}

\item Lauren Kaori Gurley. \textit{The Starbucks Union Movement Is ‘Unstoppable’}. VICE. Accessed: September 3, 2025. \url{https://www.vice.com/en/article/the-starbucks-union-movement-is-unstoppable/}

\item Lauren McFerran, David M. Prouty, and Gwynne A. Wilcox. \textit{Starbucks Corporation d/b/a Starbucks Coffee Company and Philadelphia Baristas United and Echo Nowakowska and Tristan J. Bussiere}, 2023. \url{https://apps.nlrb.gov/link/document.aspx/09031d45839a5303}

\item Megan K. Stack. \textit{Opinion | Inside Starbucks’ Dirty War Against Organized Labor}. The New York Times. Accessed: September 3, 2025. \url{https://www.nytimes.com/2023/07/21/opinion/starbucks-union-strikes-labor-movement.html}

\item Michael Sainato. \textit{Starbucks creating ‘culture of fear’ as it fires dozens involved in union efforts}. The Guardian. Accessed: September 3, 2025. \url{https://www.theguardian.com/us-news/2022/aug/25/starbucks-union-employees-fired}

\item National Labor Relations Board. \textit{Summary of NLRB Decisions for Week of February 13 - 17, 2023}. NLRB. Accessed: September 3, 2025. \url{https://www.nlrb.gov/cases-decisions/weekly-summaries-decisions/summary-of-nlrb-decisions-for-week-of-february-13-17-0}

\item Nick Blumberg. \textit{Surveillance, Threats and Retaliation: Local Starbucks Workers Charge Company With Slew of Labor Law Violations}. WTTW. Accessed: September 3, 2025. \url{https://news.wttw.com/2022/07/18/surveillance-threats-and-retaliation-local-starbucks-workers-charge-company-slew-labor}

\item Noam Scheiber. \textit{Starbucks Faces Rare Union Challenge as Buffalo Workers Seek Vote}. The New York Times. Accessed: September 3, 2025. \url{https://www.nytimes.com/2021/08/30/business/starbucks-coffee-buffalo-union.html}

\item Noam Scheiber. \textit{Starbucks workers at three more Buffalo-area stores file for union elections}. The New York Times. Accessed: September 3, 2025. \url{https://www.nytimes.com/2021/11/09/business/economy/starbucks-workers-union-elections-buffalo.html}

\item Noam Scheiber. \textit{Starbucks workers at a Buffalo store unionize in a big symbolic win for labor}. The New York Times. Accessed: September 3, 2025. \url{https://www.nytimes.com/2021/12/09/business/economy/buffalo-starbucks-union.html}

\item Orit Naomi. \textit{Starbucks Adopts People-First Scheduling Strategy as It Reassesses Staffing Automation Investment}. Restaurant Technology News. Accessed: September 4, 2025. \url{https://restauranttechnologynews.com/2025/05/starbucks-adopts-people-first-scheduling-strategy-as-it-reassesses-staffing-automation-investment/}

\item Paul Blest. \textit{Starbucks Just Fired Yet Another Union Organizer}. VICE. Accessed: September 3, 2025. \url{https://www.vice.com/en/article/starbucks-fired-union-organizer/}

\item Paul Blest. \textit{Starbucks Just Fired a Union Organizer for Allegedly Breaking a Sink}. VICE. Accessed: September 3, 2025. \url{https://www.vice.com/en/article/starbucks-union-organizer-fired-raleigh/}

\item Paul Blest. \textit{Starbucks Fired a Bunch of Pro-Union Workers After They Went on TV}. VICE. Accessed: September 3, 2025. \url{https://www.vice.com/en/article/starbucks-fired-union-organizers-memphis/}

\item Paul Blest. \textit{Starbucks CEO Howard Schultz Says Companies Are Being ‘Assaulted’ by Unions}. VICE. Accessed: September 3, 2025. \url{https://www.vice.com/en/article/starbucks-union-howard-schultz/}

\item Paul Blest. \textit{Lateness, Cursing, a Broken Sink: Starbucks Keeps Firing Pro-Union Employees}. VICE. Accessed: September 3, 2025. \url{https://www.vice.com/en/article/alleged-starbucks-union-busting/}

\item Paul Blest. \textit{The Starbucks Union Push Is So Successful It Could Start a Turf War}. VICE. Accessed: September 3, 2025. \url{https://www.vice.com/en/article/wisconsin-starbucks-unions-ufcw/}

\item Peter Lucas. \textit{Starbucks Baristas Are Accusing the Company of Homophobic Worker Suppression}. Jacobin. Accessed: September 3, 2025. \url{https://jacobin.com/2023/06/starbucks-queer-pride-decorations-workers-united}

\item Saurav Sarkar. \textit{Across the Country, Starbucks’s Anti-Union Push Is Getting Worse}. Jacobin. Accessed: September 3, 2025. \url{https://jacobin.com/2022/09/starbucks-workers-united-anti-union-busting}

\item Saurav Sarkar. \textit{Boston Starbucks Workers Have Been on Strike for 3 Weeks}. Jacobin. Accessed: September 3, 2025. \url{https://jacobin.com/2022/08/boston-starbucks-ulp-strikes-union-busting}

\item Saurav Sarkar. \textit{Starbucks Workers Attempting to Unionize May Have Just Had Their Best Week Ever}. Jacobin. Accessed: September 3, 2025. \url{https://jacobin.com/2023/03/starbucks-unionization-white-collar-workers-nlrb-bernie-sanders-howard-schultz}

\item Steven Greenhouse. \textit{Will Starbucks’ Union-Busting Stifle a Union Rebirth in the United States?}. The Century Foundation. Accessed: September 3, 2025. \url{https://tcf.org/content/commentary/will-starbucks-union-busting-stifle-a-union-rebirth-in-the-united-states/}

\item Steven Greenhouse. \textit{Starbucks’ Aggressive Union-Busting Is a New Model for American Corporations}. Slate. Accessed: September 3, 2025. \url{https://slate.com/news-and-politics/2022/11/starbucks-union-busting-tactics-workers-labor-wave-nlrb.html}

\item Vanessa Romo. \textit{On Red Cup Day, thousands of Starbucks workers go on strike}. NPR. Accessed: September 3, 2025. \url{https://www.npr.org/2022/11/17/1137296597/starbucks-strike-red-cup-day}

\item WGRZ Staff. \textit{Starbucks workers urge customers to boycott Depew location over firing}. WGRZ2. Accessed: September 3, 2025. \url{https://www.wgrz.com/article/money/business/starbucks-workers-urge-customers-to-boycott-depew-location-over-firing/71-08345ce4-7791-4e62-80be-3a4e1400b1fb}

\item \textit{We are all Starbucks Partners}. One.Starbucks. Accessed: September 4, 2025. \url{https://one.starbucks.com/}

\end{enumerate}

\subsection{Boston University}
\begin{enumerate}
    \item Abigail Hassan. \textit{Boston University faculty, staff create successful petition against Provost’s strike pay policies}. The Daily Free Press. Accessed: September 4, 2025. \url{https://dailyfreepress.com/03/29/12/203314/boston-university-faculty-staff-create-successful-petition-against-provosts-strike-pay-policies/}

    \item Abigail Hassan and Maya Mitchell. \textit{Office of Provost changes payroll policy amid BUGWU strike}. The Daily Free Press. Accessed: September 3, 2025. \url{https://dailyfreepress.com/04/04/21/203500/office-of-provost-changes-payroll-policy-amid-bugwu-strike/}

    \item Adora Brown. \textit{BU faces two unfair labor charges amidst graduate student strike}. Boston.com. Accessed: September 3, 2025. \url{https://www.boston.com/news/local-news/2024/05/08/bu-faces-two-unfair-labor-charges-amidst-graduate-student-strike/}

    \item Angry Education Workers. \textit{Call to Action: Support the BUGWU Strike Fund!}. Angry Education Workers. Accessed: September 3, 2025. \url{https://www.angryeducationworkers.com/call-to-action-support-the-bugwu-strike-fund/}

    \item BU Office of the Provost. \textit{Archive: Negotiations Updates (July 2023--October 2024)}. BU Office of the Provost. Accessed: September 3, 2025. \url{https://www.bu.edu/provost/students/enrollment-student-life/bugwu-information/negotiations-updates/}

    \item BU Today Staff. \textit{Seven Month Strike Ends, as BU Graduate Workers Union Ratifies Contract}. BU Today. Accessed: September 3, 2025. \url{https://www.bu.edu/articles/2024/bu-graduate-workers-union-ratifies-contract/}

    \item Camille Bugayong. \textit{BUGWU calls undergraduates, faculty, staff to join walkout in response to Provost emails}. The Daily Free Press. Accessed: September 3, 2025. \url{https://dailyfreepress.com/05/02/14/204393/bugwu-calls-undergraduates-faculty-staff-to-join-walkout-in-response-to-provost-emails/}

    \item Charlotte Lawrence. \textit{BU Graduate Workers Union rejects BU’s voluntary recognition offer}. The Daily Free Press. Accessed: September 3, 2025. \url{https://dailyfreepress.com/11/08/00/192195/bu-graduate-workers-union-rejects-bus-voluntary-recognition-offer/}

    \item City Council. \textit{Council Supports Boston University Graduate Workers}. City of Boston. Accessed: September 3, 2025. \url{https://www.boston.gov/news/council-supports-boston-university-graduate-workers}

    \item Freddy Reiber. \textit{Opinion -- Reflections From an Organizer in the Longest Grad Student Strike in (Recent) History}. Working Mass. Accessed: September 3, 2025. \url{https://working-mass.com/2025/01/24/opinion-reflections-from-an-organizer-in-the-longest-grad-student-strike-in-recent-history/}

    \item Gloria Waters. \textit{Submitting Graduate Worker Pay Attestation for Previous Weeks of Unpaid Work}. October 29, 2024. Accessed: September 3, 2025. \url{https://www.bu.edu/provost/files/2024/11/Submitting-Graduate-Worker-Pay-Attestation-for-Previous-Weeks-of-Unpaid-Work-10-29-24.pdf}

    \item Jacksyn Bakeberg. \textit{Old Formulas, New Playbook: Building a Striking Department at Boston University}. Long-Haul Mag. Accessed: September 3, 2025. \url{https://longhaulmag.com/2025/06/30/old-formulas-new-playbook/}

    \item Jean Morrison. \textit{A Message to BU Graduate Students: Our Commitment to Progress Together}. BU Office of the Provost. Accessed: September 3, 2025. \url{https://www.bu.edu/provost/2022/09/23/a-message-to-bu-graduate-students-our-commitment-to-progress-together/}

    \item Johanna Alonso. \textit{Undergrads Suffer the Impact of BU’s Graduate Student Strike}. Inside Higher Ed. Accessed: September 3, 2025. \url{https://www.insidehighered.com/news/students/academics/2024/05/13/boston-u-strike-plagues-undergraduate-writing-students}

    \item Kayla Baltazar. \textit{Unions, student groups hold protests during BU presidential inauguration}. The Daily Free Press. Accessed: September 3, 2025. \url{https://dailyfreepress.com/09/27/23/205888/unions-student-groups-hold-protests-during-bu-presidential-inauguration/}

    \item Kayla Baltazar. \textit{BREAKING: BUGWU ends strike after seven months, ratifies first contract with University}. The Daily Free Press. Accessed: September 3, 2025. \url{https://dailyfreepress.com/10/16/20/206613/breaking-bugwu-ends-strike-after-seven-months-ratifies-first-contract-with-university/}

    \item Lauren Coffey. \textit{Boston University Denies It Would Use AI to Replace Striking Teaching Assistants}. Inside Higher Ed. Accessed: September 3, 2025. \url{https://www.insidehighered.com/news/quick-takes/2024/04/01/boston-university-denies-it-would-replace-striking-tas-ai}

    \item Maya Mitchell and Abigail Hassan. \textit{Students, administration uncertain about BUGWU strike as semester ends}. The Daily Free Press. Accessed: September 3, 2025. \url{https://dailyfreepress.com/04/29/02/204322/students-administration-uncertain-about-bugwu-strike-as-semester-ends/}

    \item Michael Knight. \textit{Faculty Begins Contract Strike at Boston University}. The New York Times. Accessed: September 3, 2025. \url{https://www.nytimes.com/1979/04/06/archives/faculty-begins-contract-strike-at-boston-university-campus-outrage.html}

    \item National Labor Relations Board. \textit{Trustees of Boston University | National Labor Relations Board}. NLRB. Accessed: September 3, 2025. \url{https://www.nlrb.gov/case/01-CA-315655}

    \item Nicolette Manglos-Weber. \textit{What’s behind the grad student strike at Boston University?}. The Christian Century. Accessed: September 3, 2025. \url{https://www.christiancentury.org/features/what-s-behind-grad-student-strike-boston-university}

    \item Office of the Provost. \textit{Strike Payroll | Office of the Provost}. Accessed: January 26, 2026. \url{https://web.archive.org/web/20240402153545/https://www.bu.edu/provost/students/enrollment-student-life/bugwu-information/strike-payroll/}

    \item Ryan Quinn. \textit{Boston U Grad Worker Strike Now Longest in a Decade}. Inside Higher Ed. Accessed: September 3, 2025. \url{https://www.insidehighered.com/news/faculty-issues/labor-unionization/2024/08/23/boston-u-grad-worker-strike-now-longest-decade}

    \item SAG-AFTRA. \textit{SAG-AFTRA Joins in Solidarity with BUGWU}. SAG-AFTRA. Accessed: September 3, 2025. \url{https://www.sagaftra.org/sag-aftra-joins-solidarity-bugwu}

    \item Samantha Genzer. \textit{BU community reacts to CAS dean suggestion to replace striking graduate teaching assistants with AI tools}. The Daily Free Press. Accessed: September 3, 2025. \url{https://dailyfreepress.com/04/05/20/203574/bu-community-reacts-to-cas-dean-suggestion-to-replace-striking-graduate-teaching-assistants-with-ai-tools/}

    \item SEIU Local 509. \textit{SEIU Local 509 and Boston University Graduate Workers announce campaign to form Graduate Workers Union}. Mass Insider. Accessed: September 3, 2025. \url{https://massinsider.net/press-releases/35634}

    \item Steve Gillis. \textit{Boston University Historic union win for grad workers}. Workers World. Accessed: September 3, 2025. \url{https://www.workers.org/2022/12/68283/}

    \item Tony Ho Tran. \textit{Boston University Suggests Replacing Striking Grad Students With AI}. Daily Beast. Accessed: September 3, 2025. \url{https://www.thedailybeast.com/boston-university-suggests-replacing-striking-grad-students-with-ai/}

    \item Truman Dickerson. \textit{BU graduate workers reflect on living conditions, union demands as strike carries on}. The Daily Free Press. Accessed: September 3, 2025. \url{https://dailyfreepress.com/04/10/22/203734/bu-graduate-workers-reflect-on-living-conditions-union-demands-as-strike-carries-on/}

    \item Vanessa Bartlett. \textit{Striking BU Graduate Workers Demand More for Parents and Families}. Working Mass. Accessed: September 3, 2025. \url{https://working-mass.com/2024/04/04/striking-bu-graduate-workers-demand-more-for-parents-and-families/}
\end{enumerate}

\end{document}